# Cold Atoms in Driven Optical Lattices

by

Muhammad Ayub

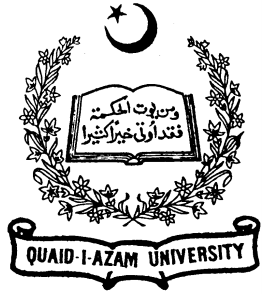

The work presented in this thesis is dedicated

to

**My father, mother (late), wife and kids**

# Certificate

It is certified that the work contained in this thesis entitled "Cold Atoms in Driven Optical Lattices" is carried out by **Mr. Muhammad Ayub** under the supervision of **Dr. Farhan Saif**.

Research Supervisor: ———————————

Dr. Farhan Saif

Pride of Performance

Associate Prof. Department of Electronics

Quaid-i-Azam University, Islamabad.

Chairman Department: ———————————

Dr. Farhan Saif

Associate Prof. Department of Electronics

Quaid-i-Azam University, Islamabad.

March 2012

# Abstract

The coherent control of matter waves in optical crystals is at the heart of many theoretical and experimental interests and has evolved as an active research area over the past two decades. Time-periodic modulations in matter-wave optics provides an additional handle to control coherence of matter waves.

In the thesis, the center of mass dynamics of cold atoms and the Bose-Einstein condensate in one dimensional optical lattice is considered both in the absence and in the presence of external forcing. We discuss three situations for matter waves: first, the cold atoms; second, sufficiently dilute condensate or the condensate for which the inter-particle interaction can be tuned to zero by exploiting Feshbach resonance, and condensate dynamics are governed by the single particle wave packet dynamics; third, strong interaction regime, where, inter-atomic interaction can no longer be ignored.

We show that wave packet evolution of the quantum particle probes parametric regimes in the optical lattices which support classical period, quantum mechanical revival and super revival phenomena. The analytical formalism developed for the two regimes, namely, deep optical lattice and shallow optical lattice. Parametric dependencies of energy spectrum and classical period, revival time and super revival are explained for the two regimes.

The inherent non-linearity of the condensate due to inter-atomic interactions and Bragg scattering of the matter-wave by an optical lattice play their role in the condensate dynamics. The condensate shows anomalous dispersion behavior at the edges of a Brillouin zone of the lattice and magnitude of this dispersion can be controlled by tuning the lattice depth. Therefore, the condensate spreading can either be controlled by actively controlling the lattice parameters or by utilizing the interaction between atoms.

The dynamics of condensate in driven optical lattice crystal are analyzed by studying dynamical stability of the condensate. The stability is determined by the dispersion behavior of the condensate excited in driven optical lattice for a variable range of modulation and interaction parameters.

The recurrence behavior of the condensate is analyzed as a function of time close to the nonlinear resonances occurring in the classical counterpart. Our mathematical formalism for the recurrence time scales is presented as: *delicate recurrences* which take place for instance when lattice is weakly perturbed; and *robust recurrences* which may manifest themself for sufficiently strong external driving force. The analysis is not only valid for dilute condensate but also applicable for strongly interacting homogeneous condensate provided, the external modulation causes no significant change in density profile of the condensate. We explain parametric dependence of the dynamical recurrence times which can easily be realized in laboratory experiments. In addition, we find a good agreement between the obtained analytical results and numerical calculations.

The stability of condensate is also explored in driven optical lattice numerically for rang of external parameters i.e. modulation strength, atom-atom interaction and lattice potential.

# Contents

# Chapter 1

# Introduction

The discovery of light pressure has a long history dating back to Kepler as he surmised that light exerts pressure on comets. Maxwell calculated radiation pressure in the framework of his own electromagnetic field theory in 1873. First time, the light force on a thin metallic plate was experimentally measured by Lebedev [Lebedev 1901] followed by similar experiment by Nicols [Nicols 1903]. Later, Lebedev [Lebedev 1910] tried to measure radiation pressure on gas molecules exerted by resonant field. Kapitza and Dirac [Kapitza and Dirac 1933] showed that a standing light field behave like a diffraction grating for electrons.

The interaction of the external degree of freedom of an atom with electromagnetic wave fields was observed first time in 1930, when Frisch measured the deflection of an atomic beam with resonant light from a sodium lamp [Frisch 1933]. The physical phenomenon responsible for mechanical action of radiation is the momentum and energy exchange between field and particle [Einstein 1917]. The deflection is due to the recoil momentum, an atom gains while absorbing or emitting a single photon of light. When an atom absorbs a photon from a beam of light, it gains momentum in the direction of the field. Since emission of photons is without preferred direction, the momentum acquired during the emission averages to zero over many cycles. This leads to a net force on the atom which is called the spontaneous force, or radiation pressure. The electric field induces a dipole moment in the atom. If the laser is detuned to the red side of the atomic transition then the induced dipole moment is in phase with the electric field. The interaction potential lowers in regions of high field and atoms experiences a force directing to those regions. On the other hand, if the laser is blue detuned to the atomic transition, the dipole moment consequently forces the atom to the regions of intensity minima.

The two forces make it possible to cool and trap atoms simultaneously. The spontaneous force cools the atoms while, dipole force is exploited to trap them [Cohen-Tannoudji et al 1992b]. Ashkin [Ashkin 1970] succeeded in trapping small particles with a pair of counter propagating laser beams, making use of both types of forces. The level of interest in atom-field interaction gained momentum when it was discovered that the radiation pressure force can be used to cool neutral atoms [Hansch 1975] and ions [Wineland 1975]. Recent developments in laser cooling and trapping techniques allow to cool and control the atoms with the dipole force alone [S. Chu 1998; J Holland et al 1999].

The experimental realization of Bose-Einstein condensation (BEC) [Anderson et al 1995; Davis and Blakie 2006; Bradley et al. 1995] and Fermi degeneracy [DeMarco et al 1999; Truscott et al 2001; Schreck et al 2001] in ultra-cold, dilute gases has opened new avenues in atomic and molecular physics, where we deal with the particle statistics and their interactions, rather than the study of single atom or photon. For many years, the main purpose in this area of research was to investigate the enriched phenomena associated with the coherent matter waves. The main achievements are the interference observation of two overlapping Bose-Einstein condensates [Andrews et al 1997], phase coherence with longe range [Bloch et al 2000] and molecular condensates of fermions with bound pairs [Greiner et al 2003; Jochim et al 2003; Zwierlein et al 2003]. The existence of a macroscopic coherent matter wave in an interacting many body system is common in all above phenomena, a much familiar concept in super-fluidity and super-conductivity.

In 1950, Ginzburg and Landau [Ginzburg and Landau 1950] coined the idea that a complex, macroscopic wave function $\psi(\mathbf{x}) = |\psi(\mathbf{x})| \exp[i\phi(\mathbf{x})]$ can describe the coherent many body state rather than a detailed microscopic understanding. Where, $|\psi(\mathbf{x})|^2$ gives the superfluid density, and $\phi(\mathbf{x})$, which is the phase of this wave function determines the superfluid velocity, $\hbar/M\nabla\phi(\mathbf{x})$, here, $M$ is mass of an atom and $\hbar$ is Planck's constant. This is similar [Langer 1968] to the description of laser light by a coherent state [Glauber 1963] and is equally valid to the standard bosonic condensates and weakly bound fermion pairs, building blocks of the Bardeen-Cooper-Schrieffer (BCS) picture of super-fluidity in fermionic systems. The macroscopic wave function description in dilute gases is more simple than the conventional super-fluids such as $^4$He, as the macroscopic wave function in later case provides only a phenomenological description of the superfluid degrees of freedom. In the presence of weak interactions, the BEC's of dilute gases are essentially pure condensates sufficiently below the transition temperature. The macroscopic description of wave function thus, has direct connection with the microscopic degrees of freedom and provides a complete picture of static and time dependent phenomena in nonlinear Schrödinger wave equation, known as Gross-Pitaevskii equation [Gross 1961; Pitaevskii 1961]. Hence, the many body effects of a BEC in dilute gases can effectively be described by single-particle wave packet and interactions can be understood as an additional potential which is proportional to the local particle density. An addition of small fluctuations around the zeroth order solutions develops a theory of weakly interacting Bose gases known as Bogoliubov theory. Many body problem then completely be solved as a set of non-interacting quasi-particles like the closely related BCS-superfluid of weakly interacting fermions. Ultracold dilute gases serve as a reliable tool to understand the many body physics and many characteristic effects have indeed been verified quantitatively [Bloch et al 2008].

In the last few years, two major experimental breakthrough were observed: first, the capability to control interaction strength in cold gases through magnetic [Inouye et al 1998; Courteille et al 1998] or optical [Theis et al 1998] Feshbach resonances [Chin et al 2010] both for bosons or fermions; second, the ability to change the dimensionality with optical lattices, particularly to produce periodic potentials through the optical lattices.

These have enhanced the range of exploration of the physical phenomena with ultra-cold gases [Greiner et al 2002a]. These two achievements either independently or collectively, enable us to explore regimes where interactions can no longer be ignored. Such enriched phenomena are the characteristics exist due to strongly correlations. Earlier, this research field was only confined to the strongly interacting quantum liquids or nuclear physics and gases were never considered as a candidate to exhibit strong correlations.

It was proposed that a strongly correlated regime can be realized by loading a condensate in an optical crystal which is nearly an ideal experimental realization of the Bose-Hubbard model [Jaksch 1998]. A quantum phase transition from a superfluid to an insulator state was predicted when interaction is increased [Fisher 1989] and this Mott insulator transition in a 3-D lattice was experimentally realized by M. Greiner and coworkers [Greiner et al 2002a]. Moreover, there have been performed many experiments to demonstrate the loss of quantum coherence in strongly correlated regime by measuring number squeezing [Orzel et al 2001] or by studying the collapse and revival of coherence in matter waves [Greiner 2002]. Further advancements were done following the suggestions [Olshanii 1998; Petrov et al 2000] to observe a Tonks-Girardeau gas with BEC's confined in 1-D lattice crystal and to understand physics of quantum Hall effect in fast rotating gases [Wilkin and Gunn 2000].

The first experimental milestone was reached [Cornish et al 2000] when strong coupling regime was explored in dilute gases using Feshbach resonances for bosonic atoms. However, an increase in the scattering length $a_s$ faces a strong decrease in the lifetime of the condensate caused by three body losses and these losses are proportional to $a_s^4$ [Fedichev et al 1996; Petrov 2004]. The set back was encountered in strong correlated regime by using a different technique [Greiner et al 2002a] to avoid the problem of condensate lifetime. The quantum phase transition from a superfluid to a Mott-insulating phase was observed even in the regime where the average inter-particle spacing is longer than the scattering length. On the other hand, the strong confinement option with optical lattices opened the opportunity to study the low dimensional systems with the possibility of the emergence of new phases. The first realization of a bosonic Luttinger liquid was reported with the observation of a (Tonks-Girardeau) hard core Bose gas in one dimension [Paredes et al 2004; Kinoshita and Weiss 2004]. The fast rotating BEC's [Schweikhard et al 2004; Bretin et al 2004] give an access to the physics of strongly correlated systems in the lowest Landau level, where, the melting of vortex lattice due to quantum fluctuations was predicted. For the atoms with larger permanent magnetic moment, i.e., $^{52}$Cr, BEC's with strong dipolar interactions have been realized [Griesmaier et al 2005]. In short, combination of Feshbach resonances [Chin et al 2010] and dimensionality control opens the ways to tune the nature and range of the interaction [Lahaye et al 2007], which enables to reach novel many body states accessible even in the fractional Quantum Hall effect.

In Fermi gases, three body losses are strongly suppressed by Pauli exclusion principle. The rate of suppression decreases with increasing scattering length [Petrov et al 2004a] and Feshbach resonances enable to explore the strong coupling regime in ultra-cold Fermi gases [OHara et al 2002; Bourdel et al]. Especially, there exist stable molecular states of weakly bounded fermion pairs in highly excited vibrational states [Strecker et al 2003] [Cubizolles et al 2003]. The extra-ordinary stability of fermions near Feshbach resonances explore the transition phenomenon from a molecular BEC to a BCS-superfluid of weakly bound Cooper-pairs [Regal et al 2004; Zwierlein 2004]. Particularly, the presence of pairing due to many body effects has been probed by spectroscopy of the gap [Chin et al 2004], or the closed channel fraction [Partridge et al 2005] while super-fluidity has been verified by the observation of quantized vortices [Zwierlein et al 2005]. These studies have also been extended to Fermi gases for the spin down and spin up components [Zwierlein et al 2006; Partridge et al 2006], where, the difference in the respective Fermi energies suppresses the pairing.

The fermions with repulsive interaction and confined in an optical lattice, can be realized as an ideal and tunable version of the Hubbard model, a paradigm to understand strong correlation problems in condensed matter physics. Experimentally, some of the fundamental properties of degenerate fermions in optical lattices like the Fermi surface existence and a band insulator with unit filling [Köhl et al 2005] have been observed. Since, it is difficult to cool fermions to the temperature range much below the optical lattice band width. However, in the deep optical lattices these experiments have raised the hope to access unconventional superconducting or magnetically ordered phases with cold fermionic gases. The tunability and ultimate control on interaction strength in these systems provide a ground to study many body physics especially, to enter the regimes which have never been explored in nuclear and condensed matter physics.

The system of cold atoms in an optical lattice, isolated from the environment leads to the long coherence time which is used for quantum information processing [Nielsen and Chang 2000; Enrico Fermi 2002], and many schemes have been proposed for entanglement engineering. Generally, those long lived qbit states are considered which are internal states of atoms localized at each site in a deep optical lattice [Jaksch and Zoller 2005]. This can be a large array of atoms, with a single atom at each lattice site, either for Fermions [Viverit et al 2004] or for Bosons (Mott Insulator regime [Jaksch 1998]). In this system, dipole [Brennen et al 1999] or collisional interactions [Jaksch et al 1999; Calarco et al 2000] between the atoms can be used to perform gate operations. Experimental realization of entanglement for a large array of atoms using controlled collisions has already been demonstrated [Mandel et al 2003] by producing a one dimensional cluster state required for quantum computing [Raussendorf and Briegel 2001]. Many other schemes have also been proposed for gate operations, based on, tunneling of atoms between neighboring lattice sites [Mompart et al 2003; Pachos and Knight 2003], strong dipole-dipole interactions between Rydberg atoms [Jaksch et al 2000] or the external degrees of freedom of atoms [Charron et al 2002]. There are several possibilities which can be used individually to address the atoms at particular lattice site [Calarco et al 2004]. The most immediate application in quantum computing using cold atoms in optical lattices is a quantum simulator [Jane et al 2003]. There are many other important subjects including Bose-Fermi mixtures, many species condensate, spinor gases, and quantum spin systems in optical lattices, which are discussed in Ref: [Lewenstein et al 2007].

In short, ultra-cold atoms and Bose-Einstein condensates in the presence of optical lattice provide an wonderful tool for engineering simple quantum systems with tunable parameters and serve as "quantum simulator" [Feynman 1982] to understand physics of different systems.

The wave packet dynamics of cold atoms in optical lattices closely resembles to the electron dynamics in solid state crystals. In many way, cold atom dynamics in optical lattice are experimentally more accessible than the dynamics of electrons in crystal lattices. Such as: i) Initial momentum distribution of cold atoms can be tailored according to the desire; ii) In these experiments, potential depth, lattice period and acceleration of atoms can be controlled at will with high precision of experimental control [Peil et al 2003; Blair et al 2001]; iii) The impurity defects and phonon vibrations are absent in optical lattices; iv) Dynamics of cold atoms in optical lattices occur on a time scale, many orders longer than the dynamics of electrons in crystal lattices. Due to all these experimental advantages, dynamics of cold atoms in optical lattices can demonstrate all those quantum effects which can only be realized indirectly in crystal lattice. When ultra-cold bosons are loaded in shallow optical crystals, the system is in the weakly interacting regime. Beautiful experiments have been performed in this regime and have provided strong evidences of band structure [Greiner 2001; Ferlaino et al 2002], coherent matter wave interferometry [Anderson et al 1998; Morsch 2002; Cronin et al 2009], super-fluidity [Burger et al 2001], Bloch oscillations and diffraction [Morsch 2001].

Observation of the quantum phase transitions in the gases of ultra-cold atoms confined by lattice potential has enhanced the theoretical and experimental study in this field [Morsch 2006; Lewenstein et al 2007; Bloch et al 2008]. This field emerged as a major area of contemporary research and is driven by the promise of simulating complex dynamics of condensed-matter and obtaining insight into phenomena which are not understood so far, such as high-temperature superconductivity and nonlinear dynamics of condensate in optical lattice crystals.

In experiments with dilute condensates in optical lattices, or condensates for which the inter-particle $s$-wave scattering length can be tuned close to zero with the advent of a Feshbach resonance [Chin et al 2010], the effects caused by inter-particle interaction can be ignored, and one observes single-particle phenomena, such as ordinary or *super* Bloch oscillations, with condensates [Ivanov et al 2008; Gustavsson et al 2008; Alberti et al 2009] [Halleret al 2010].

In recent days, cold atoms in driven optical lattices, with an aim to coherently control the matter waves, is emerging as a new area of research both in theoretical and experimental matter wave optics. We observe various dynamical modes in a system modulated by time periodic forcing. In the corresponding classical systems, the stable nonlinear resonances are immersed in stochastic sea and the system may display global stochasticity beyond a critical value of coupling or modulation strength [Lichtenberg 1992; Lichtenberg 2004; Ott 1993]. In the case of spatially periodic potentials driven by periodic force, the classical counterpart of the dynamical system displays dominant regular dynamics and dominant stochastic dynamics, one after the other, as a function of increasing modulation amplitude [Raizen 1999]. Due to spatial and temporal periodicity in the driven optical crystal, the quantum dynamics of a particle inside a nonlinear resonance is effectively mapped on the Mathieu equation. For a general quantum system driven by a periodically time dependent external force, the Hamiltonian operator is time periodic $H(t) = H(t + T)$ where, $T$ is period of external deriving force. The discrete time translation $t \rightarrow t + T$ symmetry validate the Floquet formalism [Floquet 1883]. Floquet theory is an elegant formalism to study periodically driven system. In periodically driven systems, Floquet analysis gives quasi-eigen states and quasi-energy eigen values.

In such systems the role of the stationary states is taken over by the Floquet states [Zeldovich 1966; Ritus 1966]. If the operator

$$\mathcal{H} := H(t) - i\hbar\partial_t \, , \tag{1.1}$$

acting on an extended Hilbert space of $T$-periodic functions [Sambe 1973], has $T$-periodic eigenfunctions, i.e., if the eigenvalue equation

$$[H(t) - i\hbar\partial_t]u_\alpha(t) = \varepsilon_\alpha u_\alpha(t), \tag{1.2}$$

can be solved with periodic boundary conditions in time, $u_\alpha(t) = u_\alpha(t + T)$, then the Floquet states

$$\psi_\alpha(t) := u_\alpha(t) \exp(-i\varepsilon_\alpha t/\hbar), \tag{1.3}$$

are solutions to the time-dependent Schrödinger equation. The eigenvalue Eq. (1.2), now plays a role analogous to that of the stationary Schrödinger equation for time-independent systems, and the objective of a semiclassical theory is to approximate computation of the Floquet eigen-functions $u_\alpha(t)$ and the quasi-energies $\varepsilon_\alpha$, starting again from invariant objects in the classical phase space.

However, driven quantum systems pose some mathematical difficulties that are rarely met in the time-independent case. Consider a Hamiltonian operator of the form $H(t) = H_0 + \lambda H_1(t)$, where only $H_1(t) = H_1(t + T)$ carries the time dependence, and $\lambda$ is a dimensionless coupling constant. Assume further that $H_0$ possesses a discrete spectrum of eigen-values $E_n$ ($n = 1, 2, \ldots, \infty$), with corresponding eigen-functions $\varphi_n$. Then for $\lambda = 0$ the Floquet states can be written as

$$\psi_{(n,m)}(t) = \left(\varphi_n e^{im\omega_m t}\right) \exp[-i(E_n + m\hbar\omega_m)t/\hbar], \tag{1.4}$$

with $\omega_m = 2\pi/T$. If $m$ is an integer number, the Floquet function $u_{(n,m)}(t) = \varphi_n e^{im\omega_m t}$ is $T$-periodic, as required and corresponding quasi-energies are $\varepsilon_{(n,m)} = E_n + m\hbar\omega_m$. The index $\alpha$ in Eq. (1.2) thus becomes a double index, $\alpha = (n, m)$, and the quasi energy spectrum is given by the energy eigenvalues *modulo* $\hbar\omega_m$ which can be mapped into first Brillouin zone of width $\hbar\omega_m$. For the Hermitian operator $\mathcal{H}(x, t)$, it is convenient to introduce the composite Hilbert space $R \otimes \tau$ composed of the Hilbert space $R$ of square integrable function on configuration space and space $\tau$ of functions periodic in time with period $T = 2\pi/\omega_m$ [Sambe 1973]. For the spatial part, the inner product is defined as

$$< \phi_n \mid \phi_m > = \int dx \phi_n^*(x) \phi_m(x) = \delta_{n,m}. \tag{1.5}$$

Whereas, the temporal part is spanned by the orthogonal set of Fourier vectors $< t \mid n > \equiv \exp(im\omega_m t)$, where, $n = 0, \pm 1, \pm 2, \pm 3, ......$ and inner product in $\tau$ is read as

$$\frac{1}{T} \int_0^T dx \phi_n^*(x) \phi_m(x) = \delta_{n,m}. \tag{1.6}$$

Thus the eigen vectors of $\mathcal{H}$ obey the ortho-normality condition in the composite Hilbert space too.

Floquet state formalism has been applied to a number of time-dependent problems: from coherent states of driven Rydberg atoms [Vela-Arevalo 2005], chaotic quantum ratchets [Hur 2005], electron transmission in semiconductor hetero-structures [Zhang 2006], selectively suppressing of tunneling in quantum-dot array [Villas-Bôas et al 2004] to frequency-comb laser fields [Son and Chu 2008].

Time periodic modulation in matter wave optics have given birth both to hybrid nano-opto-mechanical systems [Steinke et al 2011] and driven billiards [Leonel et al]. It was pointed out earlier that a metal-insulator transition of ultra-cold atoms in quasi-periodic optical lattices is controllable by adjusting the amplitude of a sinusoidal external forcing [Eckardt 2005]. The experimental work in this direction gained the momentum recently, with the novel observation of dynamical suppression of tunneling, and even reversal of the sign of the tunneling matrix element induced by shaken optical lattices [Lignier et al 2007; Eckardt et al 2009]. An analog of photon-assisted tunneling with Bose-Einstein condensates in driven optical crystals has been observed [Sias et al 2008]. Experimental demonstration of coherent control of superfluid to Mott-insulator transition of the matter wave in an optical lattice [Zenesini et al 2009], is also verified the theoretical proposal [Eckardt 2005]. The same principle reveals that this form of coherent control has recently been explored successfully for frustrated magnetism in driven triangular optical lattices [Struck et al 2011]. Moreover, demonstration of control over correlated tunneling is first time realized experimentally in sinusoidally driven optical lattices [Chen et al 2011].

Apart from experimental achievements of coherent control of matter waves, many theoretical advancements such as, the possibility of bound-pair transport control [Kudo et al 2009; Zhang 2010; Dahan et al 1996], quantum chaos [Hensinger 2001; Saif 2005], controlled wave-packet manipulation [Arlinghaus et al 2012], dynamical localization [Eckardt et al 2009; Eckardt 2010], entanglement [Creffield 2007a], precision measurement of gravitational acceleration [Poli et al 2011], extracting nearest-neighbor spin correlations in a fermionic Mott insulator [Greif et al 2011], coherent acceleration of matter wave [Pötting 2001; Benjamin et al 2007] and spectroscopic study of cold atoms in phase-modulated optical lattices [Tokuno et al 2011] is predicted. Furthermore, it is proved that sinusoidal forcing may even convert the repulsive inter-particle interaction to attractive one, creating a possibly to study an effectively attractive Hubbard model below the superconducting transition temperature [Tsuji et al 2011].

Moreover, apart from external modulation (phase modulation or amplitude modulation), dynamics of a condensate can also be controlled by inducing time-dependent nonlinearity [Bludov et al 2010] using Feshbach resonances [Chin et al 2010]. Many clean experiments like observation of stable Bloch oscillations [Gual et al 2009; Gual et al 2011], Dynamical localization of gap solitons [Bludov et al 2009] and probing the collective excitation of trapped condensate [Pollack et al 2010; Vidanović et al 2011] have been performed by controlling time-dependent interactions.

On the other hand, atoms and molecules interacting with specifically tailored laser pulses in optical lattices, suggests more available control options to the new emerging field of mesoscopic matter waves [Judson et al 1992; Rice 1992; Assion et al 1998; Baumert 2011]. Condensate exposed to a sequence of laser pulses can also be used to engineer different momentum states [Xiong et al 2011]. We get insight from simplified model systems [Holthaus 2001], that lead to a considerable progress with the help of advanced numerical simulations [Poletti et al 2011]. **Outline:** In this thesis, first, we discuss center of mass dynamics of single particle wave packet in optical crystal. This situation is attained in experiments with dilute condensates, or condensates for which the inter-particle $s$-wave scattering length can be tuned close to zero with the advent of a Feshbach resonance. The effects caused by inter-particle interaction in this situation can be ignored and one observes single-particle phenomena. The energy spectrum of single particle in optical crystal is a base for further discussion on role of atom-atom interaction and modulation in the dynamics. Later, we study the dynamics of condensate in driven lattice in the presence of strong correlations.

In **chapter 2**, center of mass dynamics of single particle wave packet are discussed in the absence of external modulation discussed. This chapter mostly consists of our published work in Ref: [Ayub et al 2009].

In **chapter 3**, for the sake of completeness, a tour d'horizon of the role of interaction in the dynamics of the condensate is given from the already published work. With numerical simulation, it is also shown that how we can explain the nonlinear phenomena like solitons, truncated Bloch waves (self-trapped states) and solitonic trains by studying spatiotemporal behavior, position and momentum dispersion, and wave packet revivals of condensate in deep and shallow lattice in different nonlinear regimes.

The main results of the thesis are discussed in **chapter 4**, and **chapter 5**.

In **chapter 4**, single particle wave packet dynamics is discussed in the presence of external modulation. This chapter mostly consists of our published work in Ref: [Ayub et al 2011]. In this chapter, we find analytical expression for wave packet revivals in deep and shallow case. The spatio-temporal dynamics, position and momentum dispersion and auto-correlation confirm our analytical results numerically.

In **chapter 5**, we consider nonlinear dynamics of Bose-Einstein condensate in optical lattices in the presence of periodic external drive. Stability of condensate in driven optical lattice is studied versus nonlinear interaction and modulation numerically. Later, spatio-temporal behavior of the condensate in optical lattice is studied. Chapter 5 consists of our published [Ayub and Saif 2012] and work in process of publication [Ayub and Saif 2012a].

Results are discussed and concluded in **chapter 6**.

# Chapter 2

# Cold atoms in optical lattices

## 2.1 Introduction

The physical phenomenon responsible of mechanical action of radiation is the momentum and energy exchange between field and particle [Einstein 1917]. The deflection is due to the recoil momentum, an atom gains when absorbing or emitting a single photon of light. When an atom absorbs a photon from a beam of light, it gains momentum in the direction of the field beam. Since emission of photons is without preferred direction, the momentum acquired during the emission averages to zero over many cycles. This leads to a net force on the atom which is called the spontaneous force, or radiation pressure. The spontaneous force scales with the scattering rate and for large detuning falls off quadratically with the detuning $\delta_L$, between the atomic transition frequency and the field frequency [Cohen-Tannoudji et al 1992b]

$$F_{spont} \propto \frac{I}{\delta_L^2},$$ (2.1)

here, $I$ is laser intensity. There is another force named dipole force, which is based on the coherent atom photon interaction. The electric field of light induces a dipole moment in the atom. If the induced dipole moment is in phase with the electric field, the interaction potential lowers in regions of high field and the atoms experiences a force directing to those regions. If dipole moment is out of phase by $\pi$, a force pointing away from regions of high field is experienced by the atom, consequently forced to the regions of low field intensity. Dipole force falls inversely with the detuning, $\delta_L$, from the atomic resonance in the limit of large detuning, such that

$$F_{dipole} \propto \frac{\nabla I}{\delta_L}. \tag{2.2}$$

The role of two forces is very important in manipulating and confining neutral atoms. The spontaneous force $F_{spont}$, used to cool atomic sample [J Holland et al 1999] while, dipole force is exploited to trap the atoms. The expressions of dipole and spontaneous forces show that with sufficient detuning the spontaneous force can be made negligibly very small while still keeping an appreciable dipole force.

## 2.2 Interaction Hamiltonian

In this section, we derive the effective Hamiltonian for a two-level atoms interacting with a far-detuned classical monochromatic standing wave field, $\hat{E}(x,t)$. In the semi-classical derivation, we treat the electromagnetic field classically, whereas, the atom is treated quantum mechanically with a ground state $|g\rangle$, and an excited state $|e\rangle$, separated in energy by $\hbar\omega_o$. The interaction of the atom with laser field $\hat{E}(x,t)$ is governed by the Hamiltonian which consists of three contributions, that is,

$$\hat{H} = \hat{H}_{cm} + \hat{H}_{internal} + \hat{H}_{interaction}, \tag{2.3}$$

where,

$$\hat{H}_{cm} = \frac{\hat{p}_x^2}{2M}, \tag{2.4}$$

$$\hat{H}_{internal} = \frac{1}{2}\hbar\omega_o\sigma_z, \tag{2.5}$$

$$\hat{H}_{interaction} = -\hat{d}\cdot\hat{E}(x,t),$$
$$= -(\langle e|\hat{d}.\hat{E}|g\rangle\sigma_+ + \langle g|\hat{d}.\hat{E}|e\rangle\sigma_-). \tag{2.6}$$

Here, $\sigma_\pm$ are atomic raising and lowering operators, $\sigma_z$ is the Pauli spin matrix and $\hat{d}$ is dipole moment. For linear polarization of electromagnetic field in y-direction, resonant Rabi frequency is defined as

$$\Omega = -\frac{\langle e|\hat{d}.\hat{E}|g\rangle}{\hbar} = -\frac{\langle g|\hat{d}.\hat{E}|e\rangle}{\hbar} = -\frac{\langle e|d.\hat{e}_y|g\rangle}{\hbar}E. \tag{2.7}$$

Here, $\hat{e}_y$, is the unit vector along y-direction and we have applied dipole approximation which implies that amplitude, $|\hat{E}(x,t)|$ varies slowly on atomic size scale.

## 2.3 Optical Lattice

Counter propagating laser beams overlap spatially to create an optical lattice. We consider two linearly polarized counter propagating beams having same wave number $k_L$ and their polarization vectors are parallel. The electric filed for the optical lattice is

$$\hat{E}(x,t) = \hat{e}_y[\varepsilon_o \cos(k_L x)e^{-i\omega_L t} + c.c.], \tag{2.8}$$

where, $\omega_L$, $\varepsilon_o$ are frequency and amplitude of laser field, respectively.

The interaction Hamiltonian is modified as

$$H_{interaction} = \hbar\Omega\varepsilon_o cos(k_L x)e^{-i\omega_L t}\sigma_+ + H.c. \tag{2.9}$$

In last expression, we have used rotating wave approximation to eliminate the counter rotating terms $\sigma_+ e^{i\omega_L t}$ and $\sigma_- e^{-i\omega_L t}$ [Loudon 1983]. To separate the centre of mass motion of atom and their internal states, we write the atomic state as

$$|\Psi(x,t)\rangle = \Psi_g(x,t)|g\rangle + \Psi_e(x,t)e^{-i\omega_L t}|e\rangle. \tag{2.10}$$

The Schrödinger equation for the physical system is

$$-i\hbar\frac{\partial}{\partial t}|\Psi(x,t)\rangle = \hat{H}|\Psi(x,t)\rangle. \tag{2.11}$$

By substituting the Hamiltonian, $\hat{H}$, in Eq. (2.3), and $|\Psi\rangle$ in above equation and later taking projection onto internal states $|e\rangle$ and $|g\rangle$, we get two coupled equations, viz.,

$$i\hbar\frac{\partial\Psi_g}{\partial t} = -\frac{\hbar^2}{2M}\frac{\partial^2\Psi_g}{\partial x^2} - \frac{\hbar\Omega}{2}\cos(k_L x)\Psi_e,$$

and

$$i\hbar\frac{\partial\Psi_e}{\partial t} = -\frac{\hbar^2}{2M}\frac{\partial^2\Psi_e}{\partial x^2} + \hbar\delta_L\Psi_e - \frac{\hbar\Omega}{2}\cos(k_L x)\Psi_g. \tag{2.12}$$

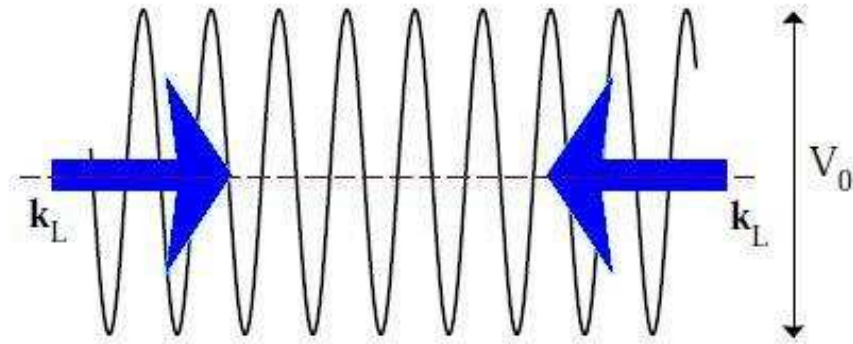

Figure 2.1: An optical lattice: standing wave field formed by two counter propagating laser beams.

In the presence of sufficiently large detuning between the atomic transition frequency and the field frequency, i.e., $\delta_L = \omega_0 - \omega_L$, spontaneous emission can be neglected [Graham 1992]. Furthermore, the probability to find an atom in the excited state is negligible as a consequence of large detuning, the evolution properties of the atom in the field are completely determined by the ground state amplitude which effectively describes the evolution of the atom in the electromagnetic field in its ground state,

$$i\hbar \frac{\partial \Psi_g}{\partial t} = -\frac{\hbar^2}{2M} \frac{\partial^2 \Psi_g}{\partial x^2} - \frac{\hbar \Omega_{eff}}{4} \cos^2(k_L x) \Psi_g, \qquad (2.13)$$

where, $\Omega_{eff} = \Omega^2 / \delta_L$ is the effective Rabi frequency. The dynamics of an atom in its ground state, with energy shift $\frac{\hbar \Omega_{eff}}{8}$, are governed by the effective Hamiltonian, that is,

$$H = \frac{p_x^2}{2M} - \frac{V_0}{2} \cos(2k_L x),$$

where, $V_0 = \frac{\hbar \Omega_{eff}}{4}$ is an effective lattice potential amplitude. When the laser field is red detuned to the atomic transition ($\delta_L > 0$), atoms are attracted towards the locations of maximal laser intensity. While, in the case of blue detuning, ($\delta_L < 0$), the atoms are repelled towards the intensity minima.

The Rabi frequency $\Omega$, which can be expressed in terms of laser intensity, $I$, and the lattice depth, $V_0$, is expressed as [Dahan et al 1996; Morsch 2006] as

$$V_0 = \tilde{\xi}\hbar\frac{I}{I_s}\frac{\Gamma_s^2}{\delta_L},\qquad(2.14)$$

where, $\tilde{\xi}$ is a correction related to level structure, $I_s$ is saturation intensity and $\Gamma_s$ is photon scattering rate [Grimm 2000].

## 2.4 Schrödinger Equation for Optical Lattice and Mathieu Solutions

The center of mass (CM) dynamics of an atomic wave packet in the presence of a standing wave field for sufficiently large detuning is effectively controlled by the time independent Schrödinger wave equation,

$$-\frac{\hbar^2}{2M}\frac{\partial^2\psi(x)}{\partial x^2} - \frac{V_0}{2}\cos(2k_Lx)\psi(x) = E_n\psi(x),$$

here, for simplicity, we have taken $\langle x|\Psi_g(x,t)\rangle = e^{-iEt}\psi(x)$. Substituting $\bar{x} = k_Lx - \pi$ in above equation we find

$$-\frac{\partial^2\psi(\bar{x})}{\partial x^2} + \frac{V_0}{2E_R}\cos(2\bar{x})\psi(\bar{x}) = \frac{E_n}{E_R}\psi(\bar{x}).$$

where, $E_R = \frac{\hbar^2 k_L^2}{2M}$ is recoil energy gained by the atom as it annihilates or emits a photon. The above equation is standard Mathieu equation, that is,

$$\frac{\partial^2\psi_n(x)}{\partial x^2} + [a_n - 2q\cos(2x)]\psi_n(x) = 0.\qquad(2.15)$$

Here, and onward, for simplicity, the bar on variable $x$ is dropped. We find Mathieu characteristic parameters[1]

$$q_0 = \frac{V_0}{4E_R},\qquad(2.16)$$

----

[1]Here, Mathieu characteristic parameter $q_0$ and $a_n$ are scale by recoil energy and scaling is different than in [Ayub et al 2009]. We use symbol $q_0$ for stationary optical lattice and symbol $q$ is reserved for driven optical lattice.

and

$$a_n = \frac{E_n}{E_R}. \tag{2.17}$$

Mathieu characteristic parameter $q_0$ is scaled optical lattice potential and $a_n$ is scaled energy of $n^{th}$ lattice band both expressed in the units of recoil energy.

One of the most important characteristics of periodic function is emergence of band structure. The periodicity of potential dictates that eigen states $\psi_n(x)$ can be chosen to have the form [Ashcroft and Mermin 1967]

$$\psi_n(x) = e^{ikx}\mu_n(x), \tag{2.18}$$

with periodicity condition,

$$\psi_n(x) = \psi_n(x + d), \tag{2.19}$$

where, $k$ is quasi momentum, $d = \frac{1}{2}\frac{2\pi}{k_L} = \frac{\lambda_L}{2}$ is periodicity of optical lattice.

Energy bands, and band gaps as a function of Mathieu characteristic parameters $q_0$ (scaled potential depth) are shown in Fig-2.2. Even Mathieu solutions, $a_{2m}$ are blue curves and odd Mathieu solutions, $b_{2m}$ are red curves mostly overridden by blue curves. The dotted lines correspond to characteristic parameter $a_n = \pm 2q_0$ i.e., $E = \pm V_0/E_R$. Note that for a given value of $q_0$ where $q_0 \gg a_n$, the gap between the lowest energy state (lowest solid curve) is roughly one-half of the spacing between solid and dashed curves, corresponding to the zero-point energy in the oscillator like limit.

Eqs. (2.15-2.17) state that for a fixed value of $q_0$, there are countably infinite number of solutions, labeled by $n$. However, only for specific values of the parameter $a_n$, the solutions will be periodic, with periods $\pi$ or $2\pi$ in the variable $x$, which are denoted by $a_n$, and $b_n$, respectively for the even and odd solutions. Because of the intrinsic parity of the potential, the solutions can be characterized as being even, $ce_n(x, q_0)$ or cosine-like for integral values of $n$, with $n \geq 0$. Whereas, they are odd, $se_n(x, q_0)$ or sine-like for integral values of $n$, with $n \geq 1$. Limiting cases i.e., $q_0 = 0$, $q_0 \lesssim 1$ and $q_0 \gg 1$ for quantum pendulum are discussed in detail in reference [Doncheski 2003; Ayub et al 2009].

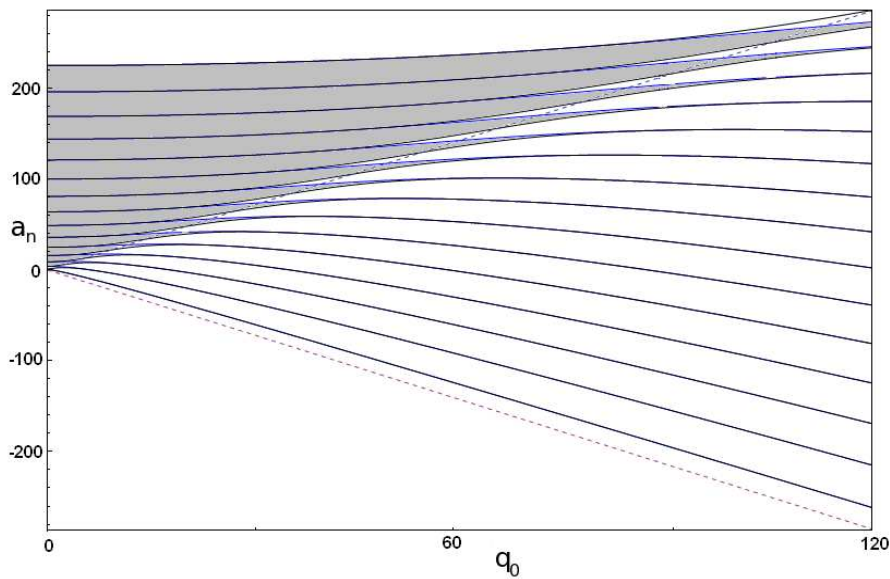

Figure 2.2: Characteristic values for the Mathieu equation, $a_{2m}$ (for even solutions, gray (on-line colour blue) curves) and $b_{2m}$ (for odd solutions, dark (on-line colour red) curves, mostly overridden by blue curves) versus $q_0$ for the quantum pendulum: The dotted lines correspond to characteristic parameter $a_n = \pm 2q_0$ i.e., $E = \pm V_0/E_R$. Note that for a given value of $q_0$ where $q_0 \gg a_n$, the gap between the lowest energy state (lowest solid curve) is roughly one-half of the spacing between solid and dashed curves, corresponding to the zero-point energy in the oscillator like limit.

Atom in an optical lattice may observe deep or shallow potential depths corresponding to its energy. The atom observes a deep optical lattice potential if effective lattice depth $V_0$ is of the order of a few hundred single photon recoil energies and temperature of the atom is around recoil temperature. In this case the dynamics of the quantum particle in the individual well is independent and one obtains multiple realization of an-harmonic oscillators. On the other hand when the depth of the potential is just few recoil energies and atom is at about recoil temperatures, it sees a shallow potential. In this case, the quantum mechanical effects caused by spatial periodicity of optical lattices such as formation of Bloch waves become important. Furthermore, the levels are broadened into bands due to resonant tunneling between adja-cent wells [Drese et al 1997]. Tunneling in the low-lying bands is suppressed as the lattice potential is increased and particle motion is dominated by single-well dynamics as it is discussed for large $V_0$. A moderate values of $V_0$, with an effective Plank's constant of order unity, indicates the deep quantum regime. The semi-classical dynamics of the atom in the standing wave field are observed however for large values of $V_0$ and here, we find several tightly bound energy bands. Quantum mechanical effects become important once the atomic de Broglie wavelength $\frac{2\pi\hbar}{P}$ significantly exceeds the lattice constant $d$.

Approximate expressions for the values of $a_n$ and $b_n$ both in $q_0 \lesssim 1$ and in $q_0 \gg 1$ limits are provided by references [Abramowitz 1970; McLachlan 1947]. In next sections, we discuss them separately.

## 2.5   Tight Binding Limit

The regime in which the gap energy between lowest and first excited band is larger than the width of the lowest band is usually referred as "tight bind regime". As linear Bloch waves are strongly localized in deep lattice potential (see Fig-2.4). In the tight binding limit $E << V_0$, analytical results can easily be derived. Following the Wannier states description, the time-dependent wave function for a state in the lowest energy band is written as

$$\psi(x,t) = \sum_{j=-\infty}^{\infty} \phi_j(t)\psi_j(x), \tag{2.20}$$

where, $\psi_j(x) = \psi_j(x - jd)$ is the ground state of the $j^{th}$ lattice site and

$$\phi_j(x) = \sqrt{n_j(t)}e^{-i\varphi_j t}, \tag{2.21}$$

is a complex function describing the amplitude, $\sqrt{n_j}$, and the phase, $\varphi_j$, associated to the wave function in the $j^{th}$ site. It is safe to assume for deep lattice potential that the Wannier functions almost coincide with the ground state of the single potential well. The functions $\psi_j(x)$ are well localized at the lattice sites, with a very small overlap among functions $\psi_j(x)$ and $\psi_{j+1}(x)$ located at neighboring sites. With the assumption that next to nearest neighboring tunneling is negligible, a set of discrete Schrödinger equations for $\phi_j(t)$ is written as

$$i\hbar \frac{d}{dt}\phi_j = -J(\phi_{j-1} + \phi_{j+1}), \tag{2.22}$$

where,

$$J = \frac{8\sqrt{2}}{\sqrt{\pi}}(q_o)^{\frac{3}{4}}\exp(-4\sqrt{q_o}), \tag{2.23}$$

is hoping matrix element. We see that $J$ is a function of the potential height, as the states $\psi_j(x)$ depend on the shape of the potential. Exact solutions of Eq. (2.22) are the Bloch waves, in which the complex functions $\phi_j(t) = e^{jkd - iEt/\hbar}$ replaces the plane waves. In this case the phase difference between neighboring sites $\Delta\phi = \phi_{j+1} - \phi_j = kd$ is constant across the entire lattice and dependent on the quasi-momentum $k$. Introducing this Bloch ansatz in Eq. (2.22), we find an analytical expression for the shape of the lowest energy band [Kevrekidis 2009]

$$E = \epsilon_0 - 2J\cos(kd), \tag{2.24}$$

where, $\epsilon_0$ is a constant energy. In Fig-2.3, lowest energy bands are shown for $V_0 = 2E_r$ (a) and $V_0 = 8E_r$ (b). Width of first excited band is indicated by blue arrows. Band gap between lowest and first excited band for $V_0 = 2E_r$ is shown in red arrows. The band width of lowest band, band gaps between lowest and first excited band and hopping matrix elements are shown in Table:2.1. From Fig-2.3 and Table-2.1, we see that as the lattice potential is increased, band width of lower band and hopping matrix elements decreases, while, band gap between ground and first excited band increases.

## 2.6   Dynamics of Cold Atoms in Deep Optical Lattices

In this section, we discuss the energy spectrum and wave packet dynamics of a single particle in deep optical lattice. An optical lattice is considered deep when the hopping matrix element $J$, for next to nearest lattice sites is negligible. In table-2.1 band width of lowest band, band gap between lowest band and first excited band, nearest neighboring hopping matrix elements $c_1$ and next to nearest neighboring hopping matrix elements, $c_2$, are shown for different lattice depths. We note that band width is four times of hopping matrix elements for nearest neighbor lattice sites i.e., *band width* $= 4J$. From Table-2.1, we observe that for $V_0 = 6E_r$ and above, $c_2$ is negligible.

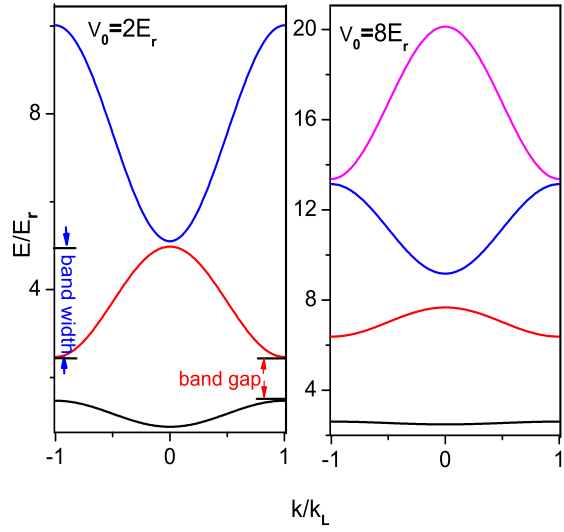

Figure 2.3: Band characteristics of an optical lattice for $V_0 = 2E_r$ (a) and $V_0 = 8E_r$ (b).

### 2.6.1   Energy Spectrum

In limiting case, $q_0 \gg 1$, the spectrum is given as

$$a_n \approx b_{n+1} \approx -2q_0 + 2s\sqrt{q_0} - \frac{s^2+1}{2^3} - \frac{s^3+3s}{2^7\sqrt{q_0}} - ..........,  \qquad (2.25)$$

where, $s = 2n + 1$. It has, thus, $(n + \frac{1}{2})\hbar\omega_h$ dependence in lower order, which resembles harmonic oscillator energy for $\omega_h = 2\sqrt{V_0}$. Here, in the deep optical lattice limit, the band width is defined as [Abramowitz 1970]

$$b_{n+1} - a_n \simeq \frac{2^{4n+5}\sqrt{\frac{2}{\pi}}q_0^{\frac{n}{2}+\frac{3}{4}}\exp(-4\sqrt{q_0})}{n!},  \qquad (2.26)$$

| $V_0(E_r)$ | band width $(E_r)$ | band gap $(E_r)$ | $c_1$ | $c_2$ |
|:---:|:---:|:---:|:---:|:---:|
| 2.0 | 0.59242 | 0.99611 | $-0.14276$ | $2.04 \cdot 10^{-2}$ |
| 4.0 | 0.34489 | 1.96936 | $-0.08549$ | $6.15 \cdot 10^{-3}$ |
| 6.0 | 0.20355 | 2.89921 | $-0.05077$ | $1.91 \cdot 10^{-3}$ |
| 8.0 | 0.12328 | 3.76987 | $-0.03080$ | $6.35 \cdot 10^{-4}$ |
| 10.0 | 0.07675 | 4.57226 | $-0.01918$ | $2.27 \cdot 10^{-4}$ |
| 12.0 | 0.04901 | 5.30442 | $-0.01225$ | $8.66 \cdot 10^{-5}$ |
| 14.0 | 0.03202 | 5.97026 | $-0.00800$ | $3.49 \cdot 10^{-5}$ |
| 16.0 | 0.02134 | 6.57719 | $-0.00533$ | $1.47 \cdot 10^{-5}$ |
| 18.0 | 0.01447 | 7.13379 | $-0.00362$ | $6.48 \cdot 10^{-6}$ |
| 20.0 | 0.00997 | 7.64827 | $-0.00249$ | $2.95 \cdot 10^{-6}$ |

Table 2.1: Band width of lowest energy band, band gap between first excited and lowest band, hopping matrix elements for nearest neighbor site $c_1$ and next to nearest neighbor site $c_2$.

which shows that in the deep optical lattice ($q_0 \gg 1$ limit) energy bands are realized as degenerate energy levels as the band width is negligible. The band structure of optical lattice is shown in Fig-2.2, for large $q_0$ near the bottom of the lattice, thin bands are seen, band width increases and band gap decreases for excited bands with increasing band index. The hoping matrix element, $J$, is related to band width in deep optical lattice case as *band width* $= 4J$, which explains the tunneling between adjacent sites for deep optical lattice. Eqs. (2.26) and (2.23) show that the width of the bands corresponds to tunneling of the atom from one lattice site to the other. In the deep lattice potentials, this tunneling probability will be exponentially small and band width therefore, reduces exponentially as a function of lattice potential.

### 2.6.2 Wave Packet Revivals

The time evolution of the particle, initially in state $\psi(x,0)$, is obtained by time evolution operator $\hat{U}$, such that,

$$\psi(x,t) = \hat{U}\psi(x,0) = \sum_{n=0}^{\infty} c_n \phi_n(x) \exp(-i\frac{E_n}{\hbar}t), \qquad (2.27)$$

where, $E_n$, and $\phi_n(x)$, are energy eigen values and eigen states corresponding to quantum number, $n$. The probability amplitudes, $c_n$, are defined as, $\langle \phi_n(x) | \psi(x, 0) \rangle$. The quantum particle wave packet in optical potential, narrowly peaked around a mean quantum number $\bar{n}$, displays quantum recurrences at different time scales, defined as

$$T_{(j)} = \frac{2\pi}{(j!\hbar)^{-1} E_n^j |_{n=\bar{n}}},$$ (2.28)

where, $E_n^j$ denotes the $j^{th}$ derivative of $E_n$ with respect to $n$. The time scale, $T_{(1)}$, is termed as classical period as it provides a time at which the particle completes its evolution following the classical trajectory and reshapes itself. Whereas, at $T_{(2)}$ the particle reshapes itself as a consequence of quantum interference in a nonlinear energy spectrum, which is purely a quantum phenomenon and thus named as quantum revival time. In the parametric regime of a changing non-linearity with respect to quantum number, $n$, we find, the super revival time $T_{(3)}$ for the quantum particle [Leichtle et al 1996].

Around the minima of lattice sites, harmonic evolution prevails and in the presence of the higher order terms it gradually modified to the original potential. Microscopic investigation of the atom-optical system, using term by term contribution of cosine potential expansion reveals the dominant role of the system's particular parametric regime in the formation of eigen-states and eigen energies. This leads to simplified analytical solutions around the potential minima in the system, as discussed below.

### 2.6.3 Harmonic Oscillator Like Limit

In the deep lattice limit, the potential near the minima can be approximated as quadratic. Thus the particle placed near the minima of the lattice potential, experiences a harmonic potential. The energy in this regime is obtained as, $E_n^{(0)} = (2n + 1)\hbar\sqrt{V_0} - V_0$, which can be identified in Eq. (2.25), by ignoring square and higher powers in $s$. The eigen states of quadratic potential are given as $\phi_n(x) = \sqrt{\frac{\beta}{2^n n! \sqrt{\pi}}} H_n(\beta x) \exp(\frac{-\beta^2 x^2}{2})$, where, $H_n(\beta x)$ are Hermite polynomials and $\beta = (\frac{2\sqrt{V_0}}{\hbar})^{\frac{1}{2}} = \sqrt{2} q_0^{\frac{1}{4}}$.

We study the time evolution of the material wave packet using square of the autocorrelation function [Nauenberg 1990],

$$|A(t)|^2 = \sum_{n=0}^{\infty} |c_n|^4 + 2 \sum_{n \neq m}^{\infty} |c_n|^2 |c_m|^2 \cos[(E_n - E_m)\frac{t}{\hbar}].$$

In the present parametric regime the square of the autocorrelation function is written as

$$|A(t)|^2 = \sum_{n=0}^{\infty} |c_n|^4 + 2 \sum_{n \neq m}^{\infty} |c_n|^2 |c_m|^2 \cos((n-m)2\sqrt{V_0}t), \qquad (2.29)$$

where, $\sum_{n=0}^{\infty} |c_n|^4$ is independent of time, and defines the interference free, averaged value of $|A(t)|^2$. The most dominant contribution to the $|A(t)|^2$ comes from $n-m=1$ in the second part at the right side of Eq. (2.29). Other terms with, $m-n>1$, have negligible role because their oscillation frequency is an integral multiple of fundamental frequency $2\sqrt{V_0}$, and are averaged out to zero. For the reason the square of the autocorrelation function in the present case oscillates following a cosine law with a frequency $2\sqrt{V_0}$ which leads to a classical time period $T_{cl}^{(0)} = \frac{\pi}{\sqrt{V_0}}$ as shown in Fig-2.4. Here, zero in the superscript of $T_{cl}^{(0)}$ defines the system's classical period in the absence of perturbation. At the integral multiples of classical period, $|A(t)|^2$ is unity, whereas, at time, which is odd integral multiple of the half of the classical period

$$|A(t)|^2 = \sum_{n=0}^{\infty} |c_n|^4 + 2 \sum_{n \neq m}^{\infty} |c_n|^2 |c_m|^2 \cos[(n-m)\pi]. \qquad (2.30)$$

Here, $\cos[(n-m)\pi]$ has alternatively values $+1$ and $-1$, when $n-m$ becomes even or odd, respectively. Thus, after the cancellation of positive terms with the negative ones, second summation reduces to minimum value and $|A(t)|^2$ attains its minima. We may write

$$\psi(x,t) = \exp[-i(2\sqrt{V_0} - \frac{V_0}{\hbar})t] \sum_{n=0}^{\infty} c_n \phi_n(x) \exp(-i2n\sqrt{V_0}t), \qquad (2.31)$$

it is notable that the eigen states $\phi_n(x)$ have parity $(-1)^n$. In case of even parity, only the even terms $c_{2n}$ are non vanishing and $n$-dependent exponent factor oscillates two times faster than the general case. At half of classical period the wave packet reappears towards other turning point of the well. In case the wave packet is initially placed at the centre of the cosine well, it reappears at half classical period at the same position but out of phase by $\pi$. In this case we see classical revivals of the initial atomic wave packet and quantum revivals take place at infinitely long time. Hence, we find revival of the atomic wave packet after each classical period only. Experimentally we may realize the situation by placing very few recoil energy atoms deep in the cosine potential well. This reveals the information about the level spacing around the bottom of the cosine potential. Interestingly we find an equal spacing between the energy levels from Fig-2.2, for large $q_0$ and small value of $n$. The spatio-temporal behavior of the wave packet in the quadratic potential, as shown in Fig-2.4, confirms the above results.

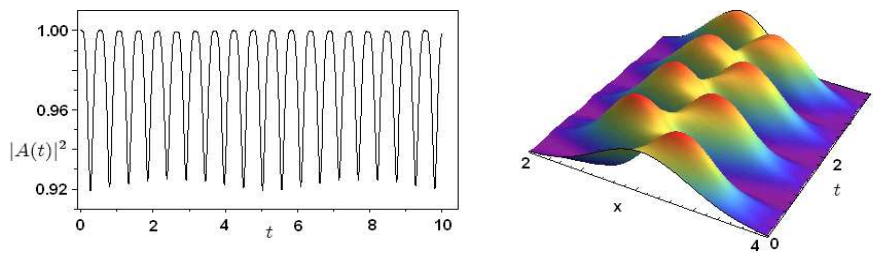

Figure 2.4: Time evolution of particle wave packet placed at the bottom of the cosine potential. The dimensions of the wave packet are $\hbar = 0.5$, $\Delta p = 0.5$ with $V_0 = 10$. We show autocorrelation function vs time (right side) and spatio-temporal behavior of the material wave packet (left side). The wave packet see equally spaced energy levels and rebuilds after every classical period. Analytically calculated value of classical period and numerical results show an excellent agreement.

### 2.6.4  Quartic Oscillator Limit

Beyond harmonic oscillator limit, we find oscillator with non-linearity and energy level spacing different from a constant value. The correction to the energy of harmonic oscillator comes from the first order perturbation (see

Appendix-A for energy corrections), that is, $E_n^{(4)} = -\frac{\hbar^2}{8}(2n^2 + 2n + 1)$, which again can be identified in Eq. (2.25) by ignoring cubic and higher order powers in $n$. The atoms with little higher energy, which is equivalent to several recoil energies observe another time scale in which it reconstructs itself beyond classical period, i.e., quantum revival time. The behavior of auto-correlation function for the wave packet exactly placed in this region, where only first order correction is sufficient, is shown in Fig-2.5. From this figure, we note that the wave packet displays revivals at quantum revival time. Thus, little above from the bottom of the of an optical lattice, the wave packet dynamics are modified due to nonlinear energy spacing. The classical time is modified as $T_{cl}^{(1)} = \alpha^{(1)} T_{cl}^{(0)}$, where $\alpha^{(1)} = 1 + \frac{\bar{s}}{8\sqrt{q_0}}$, and classical periodicity for a particle in present situation is related to potential height and mean quantum number, here, $\bar{s} = 2\bar{n} + 1$. As $\bar{n}$ increases, classical revival time also increases, and the ratio, $\frac{\bar{s}}{8\sqrt{q_0}}$, is always less than unity in the region where first correction is sufficient. The quantum revival time, $T_{rev}^{(1)} = \frac{8\pi}{\hbar}$, is independent of mean quantum number $\bar{n}$, whereas super revival time in this case is infinite.

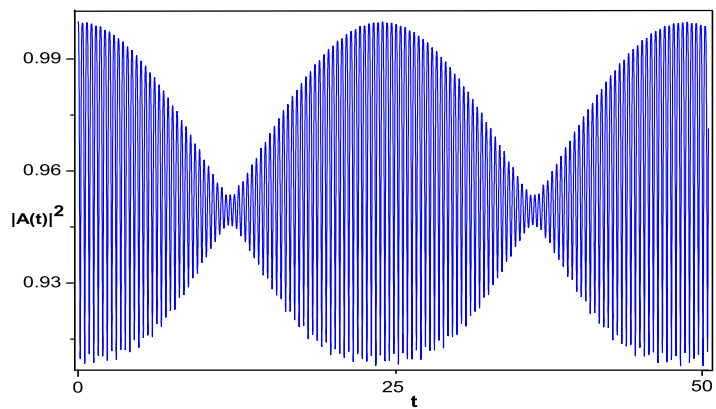

Figure 2.5: Autocorrelation function for a particle undergoing quantum revival evolution in time. The parameters are same as in Fig-2.4. The wave packet was placed close to the bottom of the potential well in the regime where first order correction is sufficient, it observes quantum revivals after many classical periods.

The eigen states of quartic oscillator due to first order perturbation are given as $\phi_n^{q_0}(x) = \phi_n(x) + \phi_n^{(1a)}(x)$, where, $\phi_n^{(1a)}(x)$ is first order correction to the harmonic oscillator wave function and is defined as

$$\phi_n^{(1a)} = D_1(\eta_1\phi_{n-4} + \eta_2\phi_{n-2} - \eta_3\phi_{n+2} - \eta_4\phi_{n+4}). \qquad (2.32)$$

Here, $D_1$, $\eta_1$, $\eta_2$, $\eta_3$ and $\eta_4$ are constants and defined in Appendix-A.

In Fig-2.6, eigen states of quadratic, quartic, sixtic and octic oscillators are mapped on numerically obtained eigen states of cosine potential. In this figure for $V_0 = 10$ and $\hbar = 1$, first order correction to the eigen states of unperturbed system matches with harmonic oscillator eigen states up to $n = 3$, and mapping of quartic oscillator with exact solution is much improved compare to harmonic oscillator. Similarly mapping of sixtic oscillator is better than quartic oscillator and is quite good for octic oscillator for all bands with energy less than potential depth. In this case eight bands exist in side the potential. From Fig-2.2, we note that as $q_0$ increases, number of bound bands also increases.

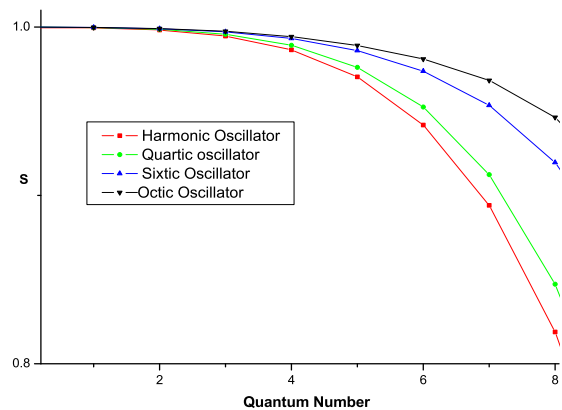

Figure 2.6: comparison of cosine potential with the simplified potentials is made by calculating the projection, S, of the eigen states of the cosine potential on the eigen states of the simplified potentials . For first few quantum numbers, the cosine potential very much resembles to the harmonic potential, however, for higher quantum numbers the higher order corrections to the harmonic oscillator are needed to make the resemblance.

The square of autocorrelation function in this case takes the form

$$|A(t)|^2 = \sum_{n=0}^{\infty} |c_n|^4 + 2 \sum_{n \neq m}^{\infty} |c_n|^2 |c_m|^2 \cos[(n-m)2\sqrt{V_0}t$$
$$+ (m-n)(n+m+1)\frac{\hbar}{4}t]. \quad (2.33)$$

Here, nonlinear dependence of energy eigen values on quantum number, $n$, makes the argument of cosine function non-linear. The nonlinear term $(m-n)(n+m+1)$ removes a degeneracy present for the harmonic case between the cosine waves corresponding to nearest neighboring off diagonal terms and beyond. Hence, there overall evolution display a gradual dispersion leading to collapse which latter transforms in revival as the dispersion in waves disappears. The values of $|A(t)|^2$ in this regime at $T_{rev} = \frac{8\pi}{\hbar}$ is simplified as

$$|A(t)|^2 = \sum_{n=0}^{\infty} |c_n|^4 + 2 \sum_{n \neq m}^{\infty} |c_n|^2 |c_m|^2 \cos[16\pi\sqrt{q_0}(n-m)]. \quad (2.34)$$

From the Eq. (2.34), we infer that $|A(t)|^2 = 1$, when $\sqrt{q_0}$ is an integral multiple of $\frac{1}{8}$. In case $\sqrt{q_0}$ is not an integral multiple of $\frac{1}{8}$, the wave packet revival occurs little earlier than the revival time $T_{rev} = \frac{8\pi}{\hbar}$ and also $|A(t)|^2$ approaches to unity little earlier than $\frac{T_{rev}}{2}$. Here, reconstruction of the wave packet at $\frac{T_{rev}}{2}$ is out of phase by $\pi$, i.e., at that time all the waves are moving exactly in opposite directions compare to initial direction. But at $T_{rev}$ they are all moving in the same direction and each wave is in phase not only with there initial states but also with each other.

Fig-2.7 shows spatio-temporal behavior of the wave packet in lattice potential. We see that wave packet spreads and oscillates in the lattice well, and after some time the original wave packet is divided into sub-wave-packets Fig-2.7-b. At a quantum revival time these sub-wave-packets constructively interfere and the wave packet gains the original shape at the same initial position Fig-2.7-c. At quantum revival time, the evolution is the same as at the start of the time evolution shown in Fig-2.7-a.

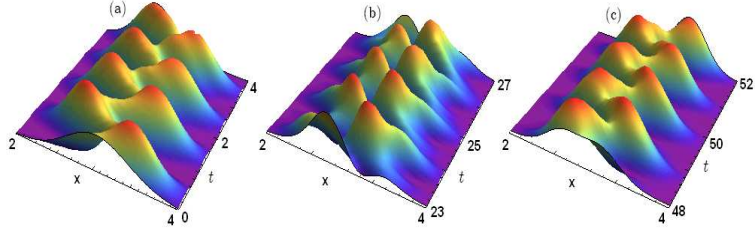

Figure 2.7: The time evolution of a wave packet in a cosine potential is displayed for the same parameters as in Fig-2.4. For a short time, the wave packet initially shows restructuring at classical periods, as shown in the inset (a), whereas, (b) we find sub wave packets in long time dynamics and (c) later constructive interfere between these sub wave packets leads to reconstruction at quantum revival time $T_{rev}$.

### 2.6.5 Sixtic Oscillator Limit: Existence of Super Revivals and Beyond

Higher order non-linearity in the energy spectrum of the quantum pendulum show up beyond quartic limit. In the presence of the second correction, the energy is modified by the term $E_n^{(6)} = -\frac{\hbar^3}{\sqrt{V_0}} \frac{(2n^3 + 3n^2 + 3n + 1)}{32}$. Second correction to the energy modifies the time scales. The classical time period is now $T_{cl}^{(2)} = \alpha^{(2)} T_{cl}^{(0)}$, whereas the quantum revival time is modified as $T_{rev}^{(2)} = |\beta^{(1)}| T_{rev}^{(1)}$. Here, $\alpha^{(2)} = \alpha^{(1)} + \frac{3(\tilde{s}^2 + 1)}{2^8 q_0}$, $\beta^{(1)} = \frac{3\tilde{s}}{16 q_0} - 1$ are constants. In addition system shows another time scale, i.e., super revival time $T_{spr}^{(2)} = \frac{64\pi \sqrt{V_0}}{\hbar^2}$, at which reconstruction of original wave packet takes place. The other quantum revival times occur at infinity.

The energy eigen states in this regime are given as $\phi_n^{(s)} = \phi_n + \phi_n^{(1,a)} + \phi_n^{(1,b)} + \phi_n^{(2,a)}$, where, $\phi_n^{(2,a)}$ is perturbation in eigen states due to the $H^{(6)}$ term and $\phi_n^{(1,b)}$ is second order perturbation caused by $H^{(4)}$ term [Doncheski 2003]. The expressions $\phi_n^{(1,b)}$ and $\phi_n^{(2,a)}$ are given as

$$\phi_n^{(1,b)} = D_2[\delta_1 \phi_{n-8} + \delta_2 \phi_{n-6} + \delta_3 \phi_{n-4} + \delta_4 \phi_{n-2} + \delta_5 \phi_{n+2} + \delta_6 \phi_{n+4} + \delta_7 \phi_{n+6} + \delta_8 \phi_{n+8}], \qquad (2.35)$$

and

$$\phi_n^{(2,a)} = D_6[\chi_1\phi_{n-6} + \chi_2\phi_{n-4} + \chi_3\phi_{n-2} + \chi_4\phi_{n+2} + \chi_5\phi_{n+4} + \chi_6\phi_{n+6}].$$

The constants $D_2, D_6, \delta_j's$ and $\chi_j's$ are calculated in Appendix-A, where, $j$ takes integral values.

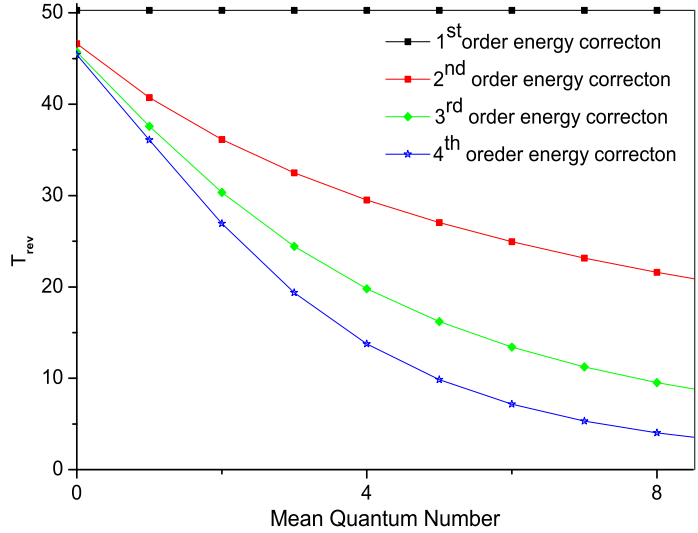

Figure 2.8: The quantum revival time vs mean quantum number is shown for simplified potentials. In the presence of first order correction, i.e., for quartic potential, the quantum revival time is a constant. For higher order corrections it decreases with increasing mean quantum number.

Again in this region classical time increases by increasing $\bar{n}$ and increases faster than as it was in quartic limit. However, in the present regime, the quantum revival time is not constant and decreases as $\bar{n}$ increases. The $\bar{n}$ dependence of quantum revival time is shown in Fig-2.8. The dependence of revival time on mean quantum number is associated with the non-linearity in the energy spectrum. For sufficiently deep lattice potential, near the lattice depth, non-linearity is almost zero and wave packet observes a linear spectrum which implies a quantum revival at infinity. However, as we move away from lowest energy band to the higher one, non-linearity emerges and defines the revival time. Since, revival time is inversely proportional to non-linearity, the quantum revival time decreases as $\bar{n}$ increases. This behavior of wave packet revival can be seen if we are away from the top of the lattice potential as well, where, continuum in energy is seen due to overlap of energy bands and classical limit in reached at which recurrence behavior is expected to vanish. The super revival time is independent of mean quantum number. It is directly proportional to square root of potential height and inversely proportional to the square of scaled Plank's constant. The temporal behavior of an atom in optical potential placed in this regime shows three time scales: classical periods enveloped in quantum revivals and quantum revivals enveloped in super revivals as shown in Fig-2.9. At each super revival time, the atomic wave packet reconstructed.

Similarly third correction in energy modifies the energy so that $E_n^{(8)} = -\frac{\hbar^4}{V_0} \frac{(5n^4 + 10n^3 + 16n^2 + 11n + 3)}{2^8}$. The time scales in this case are $T_{cl}^{(3)} = \alpha^{(3)} T_{cl}^{(0)}$ and $T_{rev}^{(3)} = |\beta^{(2)}| T_{rev}^{(1)}$. Where, $\alpha^{(3)} = \alpha^{(2)} + \frac{(5\bar{s}^3 + 17\bar{s})}{2^{11} q_0^{\frac{3}{2}}}$ and $\beta^{(2)} = \beta^{(1)} + \frac{15\bar{s}^2 + 17}{2^8 q_0}$ and super revival time is $T_{spr}^{(3)} = |\gamma| T_{spr}^{(2)}$ where $\gamma = \frac{5\bar{s}}{8\sqrt{q_0}} - 1$. Furthermore, the super quartic revival time, $T_4$, is independent of $\bar{n}$. The higher order corrections in energy show that other time scales do exist in the system, but their times of recurrence are too large to consider them finite. In the presence of third correction to energy, classical time increases as $\bar{n}$ increases but increases little faster than in the cases of quartic and sixtic corrections, whereas, the quantum revival time decreases faster as $\bar{n}$ increases compared to the case of sixtic correction. The super revival time is not constant but decreases as $\bar{n}$ increases.

Energy corrections increase anharmonicity in the system. We discussed that the first order correction to harmonic potential energy led to the quantum revivals, the second order energy correction led to super revivals and fourth correction led to quartic revival time. Comparison of revival times for different energy corrections is shown in Fig-2.8. For first order energy correction quantum revival time is constant, but for higher order corrections, revival time may decrease with increasing mean quantum number of the wave packet.

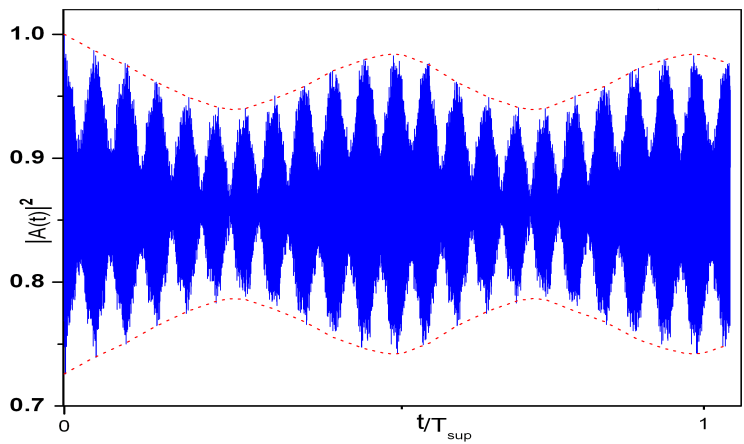

Figure 2.9: The wave packet dynamics in sixtic potential displays three time scales, the classical periods (making the dense region), the quantum revival times (making the peaks in the dense region) and the super revival times (making the peaks in the envelop of the quantum revivals). The parametric values are the same as in Fig-2.4.

In Fig-2.6, we show the projection of numerically calculated eigen states of lattice potential on the eigen states of quadratic, quartic (first correction to quadratic potential), sixtic (second correction to quadratic potential) and octic (third correction to quadratic potential) potentials. We note that the eigen states of all above mentioned potentials match to the eigen states of quadratic potential near the bottom of lattice potential. For little large quantum numbers, we see that projection of quadratic potential falls sharply and improves with higher order potentials. This correction is quite good for octic potential when $q_0 = 40$ justifying $q_0 >> 1$ condition. Eigen states and eigen energies of quartic, sixtic and octic potentials are analytically calculated using the method given in Appendix-A.

Here, for deep lattice case, asymptotic expansion Eq. (2.25) is quite good for Mathieu characteristic exponent $n$ satisfying the condition $n << \sqrt{q_0}$ and for $n >> \sqrt{q_0}$, Eq. (2.36), satisfy the numerically obtained energy spectrum. The intermediate range ($n \sim \sqrt{q_0}$), where, energy spectrum changes its character from lower to high value is estimated as $n_c \approx 2 \parallel \sqrt{\frac{q_0}{2}} \parallel$ [Rey 2005], where, $\parallel y \parallel$ denotes the closest integer to $y$.

## 2.7 Dynamics of Cold Atoms in Shallow Optical Lattices

In shallow optical lattice the condition $q_0 \lesssim 1$, is satisfied and next to nearest hopping matrix elements also needed to take into account. In this section, we discuss energy spectrum and wave packet dynamics of shallow optical lattice.

### 2.7.1 Energy Spectrum

In shallow optical lattices energy spectrum is given as [Abramowitz 1970; McLachlan 1947]

$$a_0(q_0) = -\frac{q_0^2}{2} + \frac{7q_0^4}{128} + \dots,$$

$$a_1(-q_0) \approx b_1(q_0) = 1 - q_0 - \frac{q_0^2}{8} - \frac{q_0^4}{1436} \dots,$$

$$a_2(q_0) = 4 + \frac{5q_0^2}{12} - \frac{763q_0^4}{13824} + \dots,$$

$$b_2(q_0) = 4 - \frac{q_0^2}{12} + \frac{5q_0^4}{13824} + \dots,$$

$$a_3(-q_0) \approx b_3(q_0) = 9 + \frac{q_0^2}{16} - \frac{q_0^3}{64} + \dots,$$

$$a_4(q_0) = 16 + \frac{q_0^2}{30} + \frac{433q_0^4}{864000} + \dots,$$

$$b_4(q_0) = 16 + \frac{q_0^2}{30} - \frac{317q_0^4}{864000} + \dots,$$

$$a_n \simeq b_n = n^2 + \frac{q_0^2}{2(n^2 - 1)} + \dots. \text{ for } n \geqslant 5. \tag{2.36}$$

The above expression is not limited to integral value of $n$ and is a very good approximation when $n$ is of the form, $m + \frac{1}{2}$. In case of integral value of $n = m$, the series holds only up to the terms not involving $n^2 - m^2$ in the denominator. The difference between the characteristic values for even and odd solutions satisfy the relation

$$a_n - b_n = O(\frac{q_0^n}{n^{n-1}}) \qquad \text{as } n \to \infty.$$

### 2.7.2 Wave packet Revivals

In shallow lattice potential limit, i.e., $q_0 \lesssim 1$, neglecting the higher order terms in $q_0$, the classical frequency and non-linearity are given by the expressions $\omega = 2n\{1 - \frac{q_0^2}{2(n^2-1)^2}\}$, $\zeta = 2 + \frac{q_0^2}{2}\frac{3n^2+1}{(n^2-1)^3}$ respectively. In the case of shallow optical potential classical period is

$$T_0^{(cl)} = \{1 + \frac{q_0^2}{2(n^2-1)^2}\}\frac{\pi}{n}, \tag{2.37}$$

where, $n$ is band index of lattice. The quantum revival time is

$$T_0^{(rev)} = 2\pi\{1 - \frac{q_0^2}{2}\frac{(3n^2+1)}{(n^2-1)^3}\}, \tag{2.38}$$

and super revival time is expressed as

$$T_0^{(spr)} = \frac{\pi(n^2-1)^4}{q_0^2 n(n^2+1)}, \tag{2.39}$$

Numerically dynamics are studied by evolving initially localized Gaussian wave packet in the lattice well. Fig-2.10, shows the square of auto-correlation function for Gaussian wave packet with $\Delta x = \Delta p = 0.5$, placed in an optical lattice with $V_0 = 2E_r$. Wave packet revivals are clearly seen but revival amplitude decrease with the passage of time due to tunneling to the adjacent lattice sites. While, this behavior of decreasing amplitude is not present in the case of deep optical lattice as tunneling is suppressed due to strong confinement. The spatio-temporal behavior of the same wave packet is shown in the Fig-2.11. Revival phenomenon of wave packet along with tunneling is also seen. In this regime next to nearest tunneling is not negligible as given in Table-2.1. Spatio-temporal behavior behavior explains the decrease in the amplitude with each revival in Fig-2.10.

## 2.8 Discussion

In this chapter, we have extended the understanding of eigen energy levels and eigen states in optical lattices. Two asymptotic cases namely deep optical lattice and shallow lattice are focused. We note that solutions obtained through perturbation theory and Mathieu solution are showing similar results and are in very good agreement with exact numerical solutions. In deep lattice case, the energy bands collapses to a single level as tunneling is suppressed and near the bottom of lattice, these energy levels are equally spaced as in the case of quadratic potential. Increasing the lattice potential depth, the number of equally spaced energy levels can be increased. A wave packet placed in this region revives after each classical period. From Figure 2.6, it is also noted that all potentials discussed have equally spaced eigen levels near the bottom of the potential well as their overlap with harmonic oscillator is unity (see 2.6).

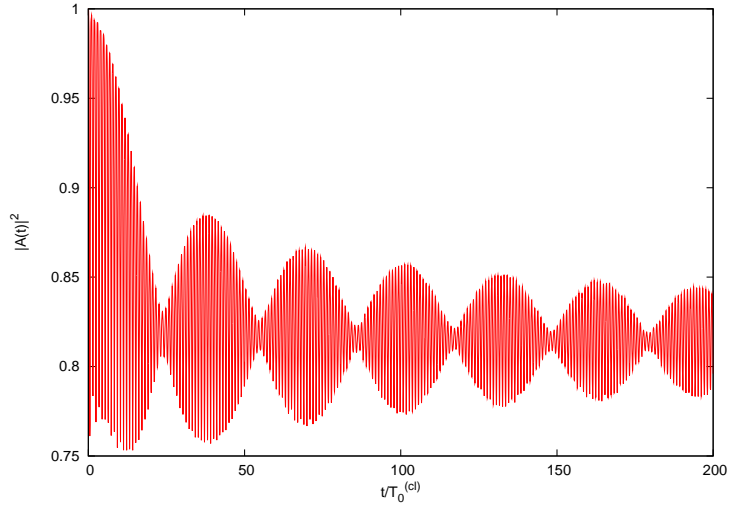

Figure 2.10: Autocorrelation function vs scaled time for wave packet initially placed in shallow optical lattice near the bottom of potential well. $V_0 = 2E_r$, $\Delta p = 0.5$.

Beyond linear regime, energy dependence is quadratic and any wave packet evolved in this region shows complete quantum revivals enveloping many classical periods. Interestingly in this regime quantum revival time ($T_{rev}^{(1)} = \frac{8\pi}{\hbar}$) is independent of potential height, however has inverse proportionality with Plank's constant, $\hbar$. We show that for deep optical lattice, there is a region in which revivals are independent of lattice depth and this region expands with the increase in lattice depth, however beyond this re-gion quantum revival time is no longer constant but decreases as $\bar{n}$ increases. The higher order time scale, super revival time $T_{spr}^{(2)} = \frac{64\pi\sqrt{V_0}}{\hbar^2}$ exists in this region and is independent of $\bar{n}$. Again this region expands as potential height is increased but super revival time in this region is directly proportional to square root of potential height. Above this region other time scales also exist where, super revival time is $\bar{n}$ dependent and decreases as $\bar{n}$ increases but these time scales are too long to consider them finite.

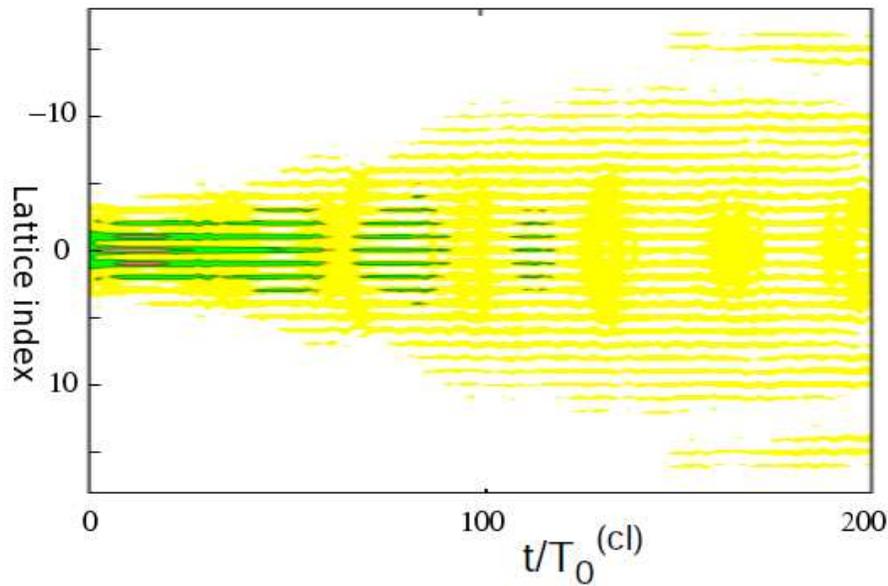

Figure 2.11: Spatio-temporal behavior of atomic wave packet initially placed in shallow lattice with $V_0 = 2E_r$. Other parameters are same as in Fig-2.10. Dark regions represent maximum probability.

In shallow optical lattices, quantum tunneling plays dominant role and we encounter wide energy bands. Two energy scales, band gap and band width evolve in the system. If the band gap is larger than band width, tight binding approximations can be applied in which next to nearest tunneling can safely be ignored. But in situations where, lattice is very shallow, next to nearest tunneling can't be ignored and wave packet disperses in position space with time causing in decrease in revival amplitude.

# Chapter 3

# BEC in Optical Lattices

## 3.1 Bose-Einstein condensation

Bose-Einstein condensation (BEC) is a pure quantum phenomenon consisting of the macroscopic occupation of a single-particle state by an ensemble of identical bosons in thermal equilibrium at finite temperature. In this chapter we provide a review on the dynamics of BEC in optical lattices. In addition we also explain our results related to BEC in shallow and deep optical lattices. Satyendra Nath Bose worked on the statistics of indistinguishable mass-less particles, that is, photons [Bose 1924], whereas, Albert Einstein extended the work to a gas of non-interacting mass particles and concluded that, below a critical temperature, a fraction of the total number of particles would occupy the lowest energy single-particle state [Einstein 1924]. Such a system undergoes a phase transition to form a BEC, which contains a macroscopic number of atoms occupying the lowest energy state. Bose-Einstein condensation in ultracold trapped atomic clouds for rubidium atoms was observed first time in 1995 [Anderson et al 1995]. The most fascinating application of these systems is the possibility to observe the quantum phenomena at a macroscopic scales.

When an ideal quantum gas of particles with mass $M$, adopts Bose-Einstein statistics at temperature $T$, a phase transition occurs and the *de Broglie* wavelength, $\lambda_{dB} = \sqrt{2\pi\hbar^2/Mk_BT}$, becomes comparable to the mean inter-particle separation (see Fig-3.1), $r = n_d^{-1/3}$, where, $k_B$ is the Boltzmann constant, and $n_d$ is the atom number density. The particles density at the centre of a condensed atomic sample is of the order of $10^{13}$-$10^{15} cm^{-3}$, much lower than the density of room-temperature molecules in air ($10^{19} cm^{-3}$), density of atoms in liquids or solids ($10^{22} cm^{-3}$) and density of nucleons in atomic nuclei ($10^{38} cm^{-3}$) [Pethick and Smith 2001]. Quantum phenomena at such a low density corresponds to the temperature of the order of $10^{-5} K$ or less.

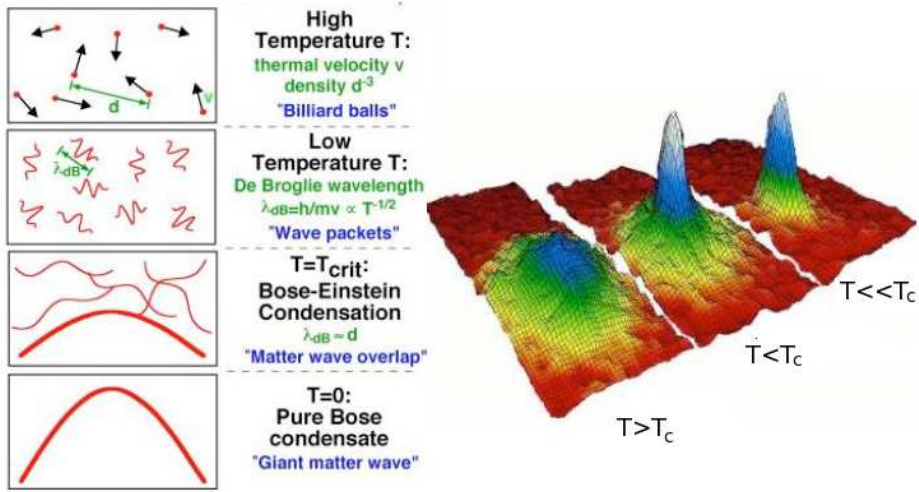

Figure 3.1: Illustration of Bose-Einstein condensation by Ketterle (left) [Anderson et al 1995]. Momentum distribution of BEC (right), first time realized by the group of Wieman and Cornell. Taken from [Ketterle 2002].

Rapid progress in laser cooling techniques [Cohen-Tannoudji 1992] provide macroscopic population of atoms in the ground state at such a low temperature. A realistic Bose gas shows some level of inter particle interaction. Therefore, the system of an ideal Bose gas of non-interacting particles is a fictitious system. However, this model provides the simplest example for the realization of BEC.

## 3.2   Bose-Einstein condensation in Optical Lattices

In this section we discuss theoretical description of BEC in a periodic potential. Turning on an optical field introduces fragmentation of the wave function of the continuous BEC into local wave functions centred around the potential minima. The fragmentation of BEC leads to a crystal like structure in which fragments mutually interact. Properly controlling the parameters of the lattice, macroscopic quantum interference phenomenon for vertical BEC array in gravitational field is observed [Anderson et al 1998].

In quantum mechanics, time evolution of any finite wave packet in free space shows dispersion, i.e., it expands with a velocity inversely proportional to its initial spread. The presence of the periodic potential modifies the dispersion. The dispersion relation (energy momentum relation $E(k)$) can sufficiently be approximated with a parabola. In the linear case, $E(k)$ relation is equivalent to the chemical potential $\bar{\mu}$ (generally the chemical potential includes mean-field interaction energy). As the dispersion relation in free space is quadratic in momentum, the parabola curvature can be associated with an effective mass description. This mass can be negative or positive resulting in either anomalous or normal dispersion. In both cases the dispersion leads to the spreading of the condensate, however, in the negative mass regions, the effective time arrow is reversed (due to the symmetry of the Schrödinger wave equation in free space: $M < 0$, $t > 0 \rightarrow M > 0$, $t < 0$). The two regimes were experimentally explored and it was shown that the dispersion of a BEC in an optical lattice can effectively be controlled [Eiermann et al 2003; Fallani et al 2003].

So far only linear regime is considered which is theoretically defined by band structures completely. Many-body interaction in the system gives rise to a non-linear term in the wave equation for the condensate order parameter. The presence of the interaction can be simplified significantly, as interaction between the particles at low energies is due to binary collisions. Nonlinear dynamics of BEC in optical lattices is expressed by mean-field Gross-Pitaevskii equation (GPE) [Pitaevskii et al 2008; Dalfovo et all 1999], that is,

$$i\hbar\frac{\partial}{\partial t}\psi(r,t) = \left[-\frac{\hbar^2}{2M}\bigtriangledown^2 - V(r)\right]\psi(r,t) + g_{3D}|\psi|^2\psi(r,t). \quad (3.1)$$

Here, $g_{3D} = 4\pi\hbar^2 a_s/M$ and $a_s$ is the s-wave scattering length which is independent of the details of two-body interaction potential and characterizes the collisions. GPE is an appropriate description of condensate dynamics given that there is negligible depletion out of the condensed mode and $n_d|a_s|^3 \ll 1$ where, $n_d$ is the density of condensate atom. The trapping potential is

$$V(r) = V_0[\sin^2(k_L x) + \sin^2(k_L y) + \sin^2(k_L z)] + \frac{M}{2}(\omega_x^2 x^2 + \omega_y^2 y^2 + \omega_z^2 z^2), \quad (3.2)$$

where, $\omega_x$, $\omega_y$ and $\omega_z$ are harmonic trapping frequencies. We focus our attention on one dimensional (1-D) optical lattice potential. Eq. (3.1) can be expressed in dimensionless parameters where, characteristic length, $a_L = d/\pi$, characteristic energy, $E_L = 2E_r$ and time with a scaling factor $2\omega_r$. With this scaling, interaction term is $g_{3D} = 4\pi(a_s/a_L)$ and lattice potential is measured in the unit of lattice recoil energy $E_r$.

For a condensate trapped in 1-D optical lattice along x-axis, with $\omega_x$ and $\omega_y$, $\omega_z \equiv \omega_\perp$ as the trapping frequencies, Gross-Pitaevskii Eq. (3.1) model can be reduced to a 1-D GPE with a periodic potential. The condition for the 1-D optical lattice is given only when the linear density is smaller than the critical density $n_{1D} < \frac{1}{2a_s}$. For $^{87}Rb$ atoms $n_{1D} < 100$ atoms $\mu m^{-1}$. The transverse wave function can be given by the ground state of a 2-D, radially symmetric quantum version of harmonic oscillator. The normalization condition is given as, $\int_{-\infty}^{\infty}|\Phi|^2 dx dy = 1$. The three dimensional wave function then can be separated as $\psi(r,t) = \Phi(r)\psi(x,t)$, to eliminate transverse dimensions [Pérez-García et al 1998], thus, yielding the 1-D GPE,

$$i\frac{\partial}{\partial t}\psi(z,t) = \left[-\frac{1}{2}\frac{\partial^2}{\partial z^2} - V_0\cos(2z)\right]\psi(z,t) + g_{1D}|\psi|^2\psi(z,t). \quad (3.3)$$

Here, $g_{1D} = (a_s/d)(\omega_\perp/\omega_r)$ and $z = k_L x$. We have neglected the contribution of harmonic potential in the direction of the optical lattice and constant energy shift in energy band. The stationary states of condensate in one dimensional optical lattice can be described by the solutions of Eq. (3.3) such that $\psi(z,t) = \phi(z)Exp[-i\bar{\mu}t]$, here, $\bar{\mu}$ is the chemical potential.

The non-linear interaction term introduces an additional energy scale in the system which enriches the linear stationary states. These states can be categorized as spatially extended non-linear Bloch waves, self-trapped states (truncated Bloch waves with an arbitrary localization length) or gap solitons (strongly localized matter-waves or non-spreading condensates).

## 3.3 Non-linear Bloch Spectrum

We have discussed the energy spectrum of single particle wave packet in section 2.6.1 for deep optical lattice and for shallow optical lattice in section 2.7.1. In this section, we analytically calculate the non-linear energy spectrum for well localized condensate in the presence of mean field interaction term, which shows how energy spectrum is modified in the presence of interaction term. Later, non-linear dynamics of the condensate initially localized to many lattice sites are studied numerically. The Bloch spectrum for condensate in optical lattice can be represented as

$$E_{nk} = \int_{-L/2}^{L/2} \varphi_{nk}^*(z) \; H_{NLS}[\varphi_{nk}] \; \varphi_{nk}(z) \; dz \; ,$$
(3.4)

where, $H_{NLS}$ is Hamiltonian (3.3) can be written as

$$H_{NLS}[\varphi] = H_L(z) + g_{1D}|\varphi|^2 \; ,$$
(3.5)

here,

$$H_L(z) \equiv -\frac{1}{2} \frac{\partial^2}{\partial z^2} \; + \; \frac{V_0}{2} \cos(2z),$$
(3.6)

is the linear lattice term (lattice Hamiltonian when interaction is absent). The chemical potential of condensate,

$$\bar{\mu}_0 = \lim_{k \to 0} E_{0k} \; ,$$
(3.7)

equals the system chemical potential, $\bar{\mu}_0 = \bar{\mu}$. The particle group velocity, $v_{nk}$, and the effective mass, $M_{nk}^*$ [Yukalov 2009], are respectively defined as

$$v_{nk} \equiv \frac{\partial E_{nk}}{\partial k}, \tag{3.8}$$

$$\frac{1}{M_{nk}^*} \equiv \frac{\partial^2 E_{nk}}{\partial k^2}. \tag{3.9}$$

In tight-binding approximation, only the nearest-neighbor hoping matrix elements contribute to the tunneling. For the linear term Eq. (3.6), two types of matrix elements over Wannier functions exist [Yukalov 2009], that is the on-site integral

$$h_0 \equiv \int_{-L/2}^{L/2} w(z) H_L(z) w(z) \, dz, \tag{3.10}$$

and, the integral representing the nearest-neighbor overlap

$$h_1 \equiv \int_{-L/2}^{L/2} w(z) H_L(z) w(z-d) \, dz \, . \tag{3.11}$$

The matrix element for non-linear term in Eq. (3.5) are generally proportional to the integral

$$I_{j_1 j_2 j_3 j_4} \equiv \int_{-L/2}^{L/2} w_{j_1}(z) w_{j_2}(z) w_{j_3}(z) w_{j_4}(z) \, dz \, . \tag{3.12}$$

Considering tight-binding approximation, three kinds of integrals are encountered. The on-site integral is

$$I_0 \equiv \int_{-L/2}^{L/2} w^4(z) \, dz \, , \tag{3.13}$$

the first-order overlap integral is

$$I_1 \equiv \int_{-L/2}^{L/2} w^3(z) w(z-d) \, dz \, , \tag{3.14}$$

and second-order overlap integral is given as

$$I_2 \equiv \int_{-L/2}^{L/2} w^2(z) w^2(z-d) \, dz \, . \tag{3.15}$$

Considering only the first order overlap integrals, the on-site term in Bloch spectrum,

$$\varepsilon_0 \equiv h_0 + g_{1D}I_0, \tag{3.16}$$

tunneling parameter defined by first-order overlap integrals,

$$J \equiv -h_1 - 4I_1g_{1D}, \tag{3.17}$$

and the Bloch spectrum Eq. (3.4) defined as

$$E_k = \varepsilon_0 - 2J\cos(kd). \tag{3.18}$$

When the second-order overlap integrals Eq. (3.15) are considered, an additional term with $\cos(2kd)$ appears.

In this case, the chemical potential Eq. (3.7) is

$$\bar{\mu}_0 \equiv \lim_{k \to 0} E_k = \varepsilon_0 - 2J, \tag{3.19}$$

provided, only lowest band is considered. Substituting the Eqs. (3.16) and (3.17) in the above equation gives the expression of chemical potential as

$$\bar{\mu}_0 = h_0 + 2h_1 + g_{1D}(I_0 + 8I_1). \tag{3.20}$$

The isothermal compressibility, $\kappa_T = \frac{d}{\bar{\nu}^2}\frac{\partial\bar{\nu}}{\partial\bar{\mu}} = \frac{a}{\bar{\nu}^2}(\frac{\partial\bar{\mu}}{\partial\bar{\nu}})^{-1}$, can be defined by considering only the dependence of the coupling parameter $g_{1D}$ on the filling factor $\bar{\nu}$, and neglecting other dependencies on $\bar{\nu}$ of the Wannier functions [Yukalov 2009]. Then

$$\kappa_T = \frac{Md\,l_\perp^2}{2a_s\bar{\nu}^2(I_0 + 8I_1)} = \frac{1}{\rho(I_0 + 8I_1)g_{1D}}, \tag{3.21}$$

where, $l_\perp$ is transverse localization length. We note that, bosons in a lattice can be stable only in the presence of non-zero repulsive interactions. When atomic interactions are attractive (i.e., $a_s < 0$), the compressibility Eq. (3.21) would be negative. In case of ideal Bose gas (i.e., $g_{1D} \to 0$), $\kappa_T$ would be infinite. In these two cases, the system would be unstable [Courteille et al 2001; Yukalov 2005a; Yukalov 2005b]. In the case of attractive interactions, ( i.e., $a_s < 0$), a trapping potential in all directions can restore the system stability for finite number of atoms [Courteille et al 2001; Yukalov 1997; Gammal 2002; Yukalov 2005c; Holz 2006].

Considering the Bloch spectrum Eq. (3.18), now the group velocity Eq. (3.8) and the effective mass, given in Eq. (3.9), are

$$v_k = 2Jd\sin(kd),\tag{3.22}$$

$$M_k^* = \frac{1}{2Jd^2\cos(kd)}.\tag{3.23}$$

The group velocity and the effective mass in the limit $k \to 0$, are simplified as

$$M^* \equiv M_0^* = \frac{1}{2Jd^2}\qquad (k=0),\tag{3.24}$$

$$v_k \simeq 2Jd^2 k\tag{3.25}$$

For tight-bind limit, the localized Wannier orbitals can be approximated by a Gaussian function. However, the Gaussian ansatz neglects the small oscillations characteristic in the tail of the Wannier function, the tunneling matrix element $J$ is then underestimated by almost an order of magnitude. The corresponding Wannier functions for a one-dimensional lattice is

$$w(z-z_0) = \frac{1}{(\sqrt{2\pi}\,\Delta z_0)^{1/2}}\,\exp(-\frac{(z-z_0)^2}{4(\Delta z_0)^2}).\tag{3.26}$$

Using Eq. (3.26), the on-site integral given in Eq. (3.13) becomes

$$I_0 = \frac{1}{2\Delta z_0\,\sqrt{\pi}},\tag{3.27}$$

and the first order integral given in Eq. (3.14) becomes

$$I_1 = I_0\tilde{\gamma}^{\frac{3}{4}},\tag{3.28}$$

where, $\tilde{\gamma} = \exp[-\frac{d^2}{4(\Delta z_0)^2}]$. The second order overlap integral in Eq. (3.15) is

$$I_2 = I_0\tilde{\gamma} = I_1\tilde{\gamma}^{\frac{1}{4}}.\tag{3.29}$$

From above three expressions we note that

$$I_2 \ll I_1 \ll I_0\qquad (\frac{\Delta z_0}{d} \ll 1),\tag{3.30}$$

which shows that second-order overlap integrals can be neglected Eq. (3.29). Now, The on-site integral Eq. (3.10) is

$$h_0 = -\frac{1}{8(\Delta z_0)^2} + \frac{V_0}{2}\exp[-4(\Delta z_0)^2],$$ (3.31)

and the overlap integral Eq. (3.11), we have

$$h_1 = \tilde{\gamma}^{\frac{1}{2}}\left[-\frac{d^2 - 4(\Delta z_0)^4}{32(\Delta z_0)^4} + \frac{V_0}{2}\exp[-2(\Delta z_0)^2]\cos(d)\right].$$ (3.32)

The local energy term (3.16) becomes

$$\varepsilon_0 = \frac{1}{2\Delta z_0}\left[-\frac{1}{4\Delta z_0} + V_0\exp[-4(\Delta z_0)^2] + \frac{g_{1D}}{\sqrt{\pi}}\right].$$ (3.33)

This leads to the nearest-neighbour hopping parameter in Eq. (3.17), as,

$$J = \tilde{\gamma}^{\frac{1}{2}}\left[\frac{d^2}{32(\Delta z_0)^4} - \frac{1}{8} - \frac{V_0}{2}\exp[-2(\Delta z_0)^2]\cos(d) - \frac{2g_{1D}\tilde{\gamma}^{\frac{1}{4}}}{\Delta z_0\sqrt{\pi}}\right].$$ (3.34)

Hence the chemical potential defined in Eq. (3.20) reads as

$$\bar{\mu}_0 = 2\tilde{\gamma}^{\frac{1}{2}}\left[-\frac{d^2 - 4(\Delta z_0)^4}{32(\Delta z_0)^4} + \frac{V_0}{2}\exp[-2(\Delta z_0)^2]\cos(d)\right]$$
$$+ g_{1D}\frac{1}{2\Delta z_0\ \sqrt{\pi}}\left[1 - \tilde{\gamma}^{\frac{3}{4}}\right] - \frac{1}{8(\Delta z_0)^2} + \frac{V_0}{2}\exp[-4(\Delta z_0)^2].$$ (3.35)

As a consequence the compressibility expression (3.21) gets the form

$$\kappa_T = \frac{2\Delta z_0\sqrt{\pi}}{\rho g_{1D}[1 + 8\tilde{\gamma}^{\frac{3}{4}}]}.$$ (3.36)

In the presence of non-linear atom-atom interaction, non-linear Bloch waves (spatially extended Bloch waves) exist with the periodicity of the lattice potential. For the case of weak non-linearity, the non-linear Bloch states and linear ones are qualitatively similar. As the non-linearity (local condensate density) increases, the chemical potential corresponding to the Bloch state near the linear band edges *shifts* into the linear spectrum band gap [Ostrovskaya et al 2008]. The magnitude of this shift is directly proportional to the non-linearity and its direction is determined by the scattering length sign as shown in Fig-3.2.

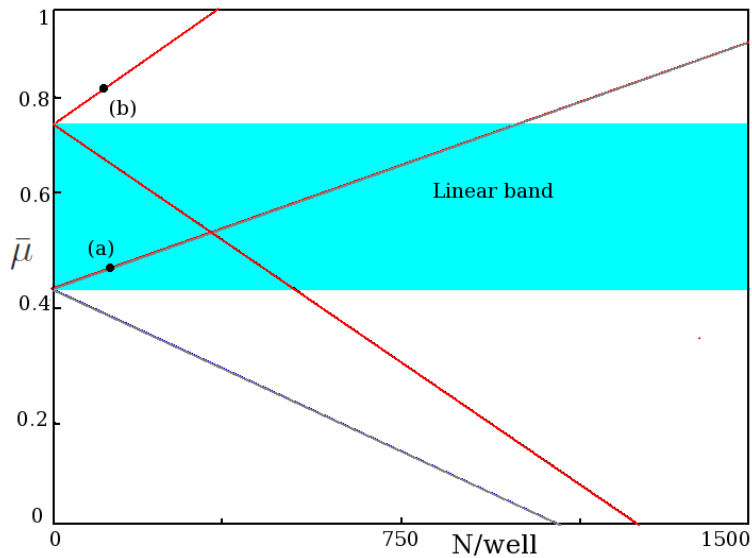

Figure 3.2: Nonlinear shift of the band edges corresponding to the Bloch states at quasi-momenta $k = 0$ (Grey) and $k = 1$ (red) for a repulsive [lines marked with points (a) and (b)], $g_{1D} = 0.001$, and attractive (unmarked lines), $g_{1D} = -0.001$, condensate in a shallow 1-D OL ($V_0 = 1.0$), shown as a dependence of chemical potential on the number of atoms per lattice site [Ostrovskaya et al 2008].

The non-linear Bloch waves show different stability properties for different quasi-momenta values. In case of repulsive BEC, Bloch states corresponding to the bottom of the lowest linear band ($k = 0$) are stable. On the other hand, non-linear Bloch states representing the top of linear ground band ($k = 1$) are dynamically unstable [Konotop and Salerno 2002]. This instability arises due to small fluctuations in condensate density and results in exponential growth in Bogoliubov excitations. Experimentally dynamical instability was detected with BEC of large number of atoms, adiabatically driven to upper edge of the ground state, or non-adiabatically loading of BEC in $k = 1$ state [Fallani et al 2004; Sarlo et al 2005]. Dynamical instabilities lead to the loss of condensate atoms due to spatial fragmentation of the BEC density.

We discuss roll of the non-linearity in the dynamics of condensate for three different regimes namely: i) Non-linearity is the smallest energy scale, it is easy to realize experiments in this regime; ii) Non-linearity is larger than the band width but smaller than the band gap energy, achieved by increasing the transverse trapping frequency or increasing the atoms per lattice well; iii) Non-linearity is the largest energy scale. Following, we discuss the condensate dynamics in each regime one after the other for shallow and deep optical lattices as mathematical formalism is different for the two.

To explore the dynamics of condensate in optical lattice, we study spatio-temporal behavior, spatial and momentum dispersion numerically. For shallow lattice, we consider an optical lattice with lattice potential $V_0 = 2E_r$. In this case the width of lowest band is $0.592Er$ and band gap between first excited band and lowest band is $0.996E_r$. For shallow lattice case, dynamics are explored for the cases i.e. in the absence of non-linearity ($g_{1D} = 0$), for non-linearity $g_{1D} = 0.3$ (interaction term is the smallest energy scale), for $g_{1D} = 0.7$ (interaction energy is in intermediate regime) and $g_{1D} = 1.5$, (interaction energy is dominant). For deep lattice, an optical lattice with potential $V_0 = 8E_r$ is considered. In this case the width of lowest band is $0.12Er$ and band gap between first excited band and lowest band is $3.77E_r$. We study dynamics of condensate in deep lattice for non-linearity $g_{1D} = 0.1$, (smallest non-linearity energy scale), $g_{1D} = 1.5$ (non-linearity in intermediate range) and $g_{1D} = 4$ (non-linearity is dominant energy scale).

### 3.3.1 Smallest Non-linear Energy

In this regime, only small changes are expected in the single particle spectrum due to the presence of non-linear term. Even though, stationary solutions have no significant differences from the linear case, the dynamics in this case is totally different. Formation of bright solitons when interaction is repulsive (when quasi-momentum is in the regime of negative mass) and dynamical instability are two important examples in this regard.

**Shallow lattice potential**

The condensate in the lowest band with a small momentum distribution centred around a mean momentum $k_0$ can be efficiently described by a slowly varying amplitude (on the single lattice scale), $A(z,t)$, times Bloch state corresponding to quasi-momentum $k_0$,

$$\psi(z,t) = A(z,t)\phi_{n=0,k_0}(z)e^{-\frac{i}{\hbar}\tilde{\mu}(k_0)t}. \tag{3.37}$$

In the weakly interacting case, a differential equation for slowly varying envelope, $A(z,t)$, can be derived which has the same form as the Gross-Pitaevskii equation with a modified interaction energy and dispersion,

$$i\hbar(\frac{\partial}{\partial t} + v_g\frac{\partial}{\partial z})A(z,t) = [-\frac{\hbar^2 k_L^2}{2M_{eff}}\frac{\partial^2}{\partial z^2} + V(z,t) + g_{1D}\alpha_{n1}|A|^2]A(z,t). \tag{3.38}$$

Here, $\alpha_{n1} = (1/d)\int_{-d/2}^{d/2} dz|\phi_{k_0}|^4 \sim 1-2$, re-normalizes the interaction energy as it increases with stronger localization in the lattice. The stationary solutions of above Eq. (3.38) do not show significant difference from linear case but dynamics shows some different characteristics i.e., formation of bright solitons (termed as gap-solitons by [Steel and Zhang 1998]) even for repulsive interaction with the condition that the mean quasi-momentum is in negative effective mass regime. Another very interesting phenomenon arising in the presence of non-linearity is dynamical instability, that is, exponentially fast growth of small perturbation of condensate wave function [Konotop and Salerno 2002].

A condensate with Gaussian profile is adiabatically loaded in the optical lattice by ramping up potential slowly. The adiabatic loading prepares the condensate in positive mass regime (initially occupying the lower edge of the ground band). Four parameters characterize the Gaussian condensate: (a) the width of the condensate, (b) the center-of-mass position, (c) the linear phase which describes the group velocity of the wave packet, and (d) the quadratic phase over the condensate which describes the linear evolution of the condensate. The quadratic dispersion in momentum space directly translates into a quadratic phase in real space. Moreover, the non-linear energy term also leads to a quadratic phase in first approximation since the density is quadratic near the Gaussian maximum.

In Fig-3.3 and Fig-3.4, we show spatio-temporal dynamics of attractive and repulsive condensate respectively, for non-linearity $g_{1D} = 0$, $g_{1D} = 0.3$ (interaction term is the smallest energy scale), for $g_{1D} = 0.7$ (interaction energy is in intermediate regime) and $g_{1D} = 1.5$, (interaction energy is dominant) for shallow lattice. Here and in rest of the figures, dark shades represent the maximum density for spatio-temporal structures. Spatial dispersion of the condensate both for attractive and repulsive case in shallow lattice is shown in Fig-3.5a and Fig-3.5b, respectively. Momentum dispersion for both attractive and repulsive condensate is shown in Fig-3.6a and Fig-3.6b, respectively.

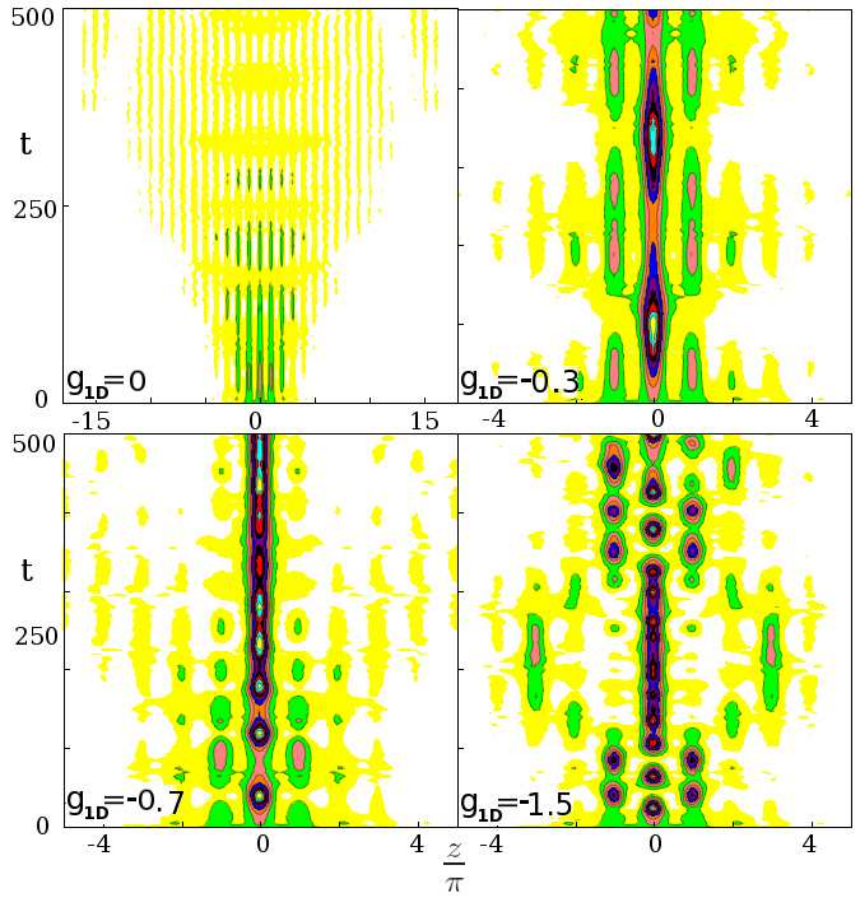

Figure 3.3: Spatio-temporal dynamics of attractive condensate with initial momentum dispersion $\Delta p_0 = 0.1$ for non-linearity, $g_{1D} = 0$, $g_{1D} = -0.3$, $g_{1D} = -0.7$ and $g_{1D} = -1.5$. Lattice potential is $V_0 = 2E_r$ and condensate is adiabatically loaded in the lattice. In this figure dark colours regions show density maxima.

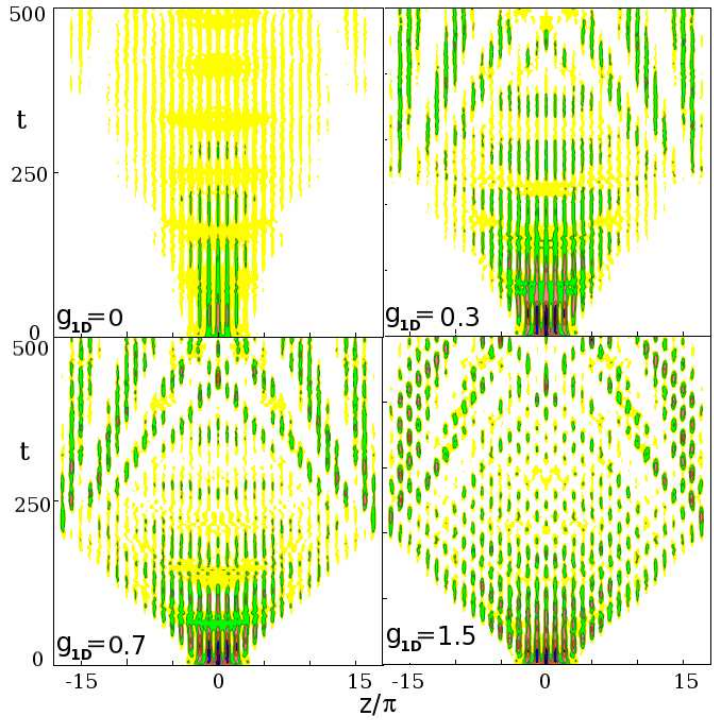

Figure 3.4: Spatio-temporal dynamics of repulsive condensate with initial momentum dispersion $\Delta p_0 = 0.1$ for non-linearity, $g_{1D} = 0$, $g_{1D} = 0.3$, $g_{1D} = 0.7$ and $g_{1D} = 1.5$. Lattice potential is $V_0 = 2E_r$ and condensate is adiabatically loaded in the lattice. In this figure dark colours regions show density maxima.

From Fig-3.3 and Fig-3.4, we note that the dynamics for shallow lattice in the absence of non-linearity shows diffusion to the neighboring lattice sites. The spreading of condensate is due to Josephson like tunneling from one lattice site to other. While, with the introduction of small non-linearity $g_{1D} = -0.3$, (attractive case), dispersion decreases and condensate is confined in few lattice sites only. On the other hand, for repulsive case ($g_{1D} = 0.3$), initially, the rate of diffusion is faster compared to liner case. However, due to the presence of non-linearity, cloud not only expands in momentum space but also explodes i.e. momentum distribution expands too [Ostrovskaya et al 2008]. As cloud reaches to the top edge of spectral band (Ground band with $n = 0$ and $k = 1$ in this case), the effective mass of matter-wave becomes negative which effectively leads to the focusing of the matter-wave. A balance between the repulsive interaction and negative dispersion near the top edge of the spectral band leads to the spatially localized non-spreading wave packets with zero group velocity termed as *matter-wave solitons*. The chemical potential corresponding to such localized waves lies in the gaps of the linear Bloch-wave spectrum and termed as *gap solitons*. In initially published work [Pelinovsky et al 2004], it is shown that generally two type of solitons are bifurcated from each band edge. These are bright solitons centred on the minimum (on-site) of the lattice potential and maximum (off-site) of the lattice potential. The gap solitons are well approximated by a sech-like broad envelope $A(z)$, with low-amplitude of the corresponding Bloch wave, centered either off or on-site. The solitons are comparatively weakly localized and have low density (number of particles). That is why, preparation of a very small condensate was required to observe a one dimensional gap solitons. On the other hand low density also reduces the chance of transverse excitations. In the dynamics, the distinction between the on-site and off-site states does not play significant role because the off-site states undergoes symmetry-breaking instability and transforms into a stable on-site states [Eiermann et al 2004].

The time evolution of spatial dispersion in Fig-3.5a and Fig-3.5b, shows that dispersion also increases with time in the absence of non-linearity. However, as the non-linearity ($g_{1D}$) is introduced, the spatial dispersion is modified. Initially, it increases, later, in attractive case it decreases while, for repulsive case, it keeps on increasing with time and it settles around a constant value. This suppression is due to formation of weakly localized solitons. These solitons are clearly seen in Fig-3.4 for $g_{1D} = 0.3$. Beyond $t = 250$, well localized soliton matter-waves appear as a balance between diffusion and focusing (due to negative effective mass) is established.

The momentum dispersion for the case of zero non-linearity ($g_{1D} = 0$) initially, quickly increases and then fluctuates around a mean value showing collapse and revivals in momentum dispersion. With the introduction of non-linearity, $g_{1D}$, the behavior of momentum dispersion is modified. Although the $\Delta p$ fluctuations around the same mean as in the case for zero non-linearity but its collapse and revival behavior is changed for both attractive (Fig-3.6a) and repulsive (Fig-3.6b) condensate.

**Deep Optical Lattice**

In the limit of deep periodic potential, condensate wave functions can be described by Wannier wave functions localized at the potential minima. The localization of condensate by deep optical lattice increases the on-site atom-atom interaction. The condensate wave function can be described with localized Wannier states associated with the lowest band [Chiofalo et al 2000]. It is notable that the strong localization modifies the linear Wannier states due to increase in atomic densities which enhance atom-atom interaction. The dynamics of BEC is governed by discrete non-linear Schrödinger equation

$$i\hbar \frac{d}{dt}\psi_n = J(\psi_{n-1} + \psi_{n+1}) + g_{1D}|\psi_n|^2\psi_n + E_n\psi_n. \qquad (3.39)$$

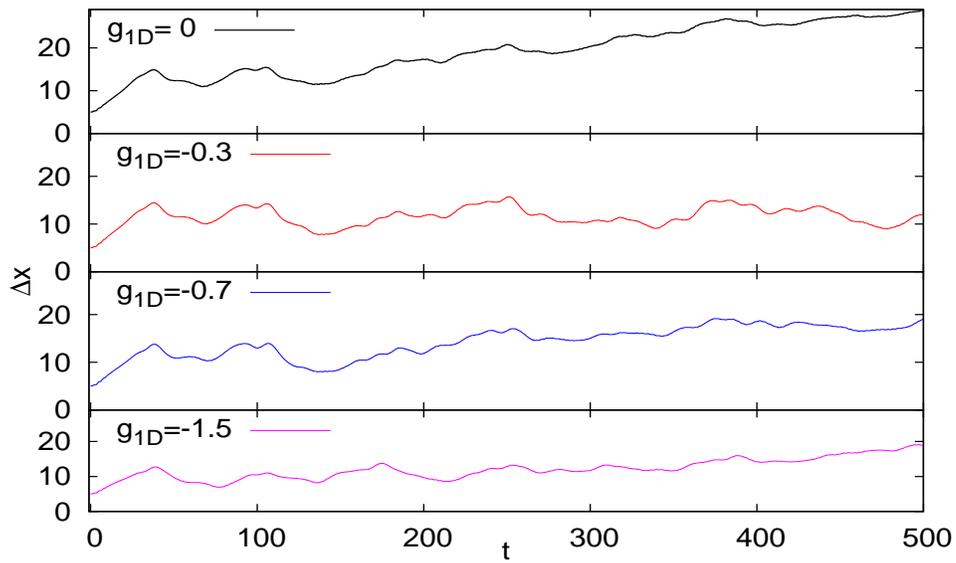

(a) Spatial dispersion vs time with lattice potential $V_0 = 2E_r$ for $g_{1D} = 0$, $g_{1D} = -0.3$, $g_{1D} = -0.7$ and $g_{1D} = -1.5$. Other parameters are $\Delta z_0 = 5$ and condensate is adiabatically loaded in the lattice.

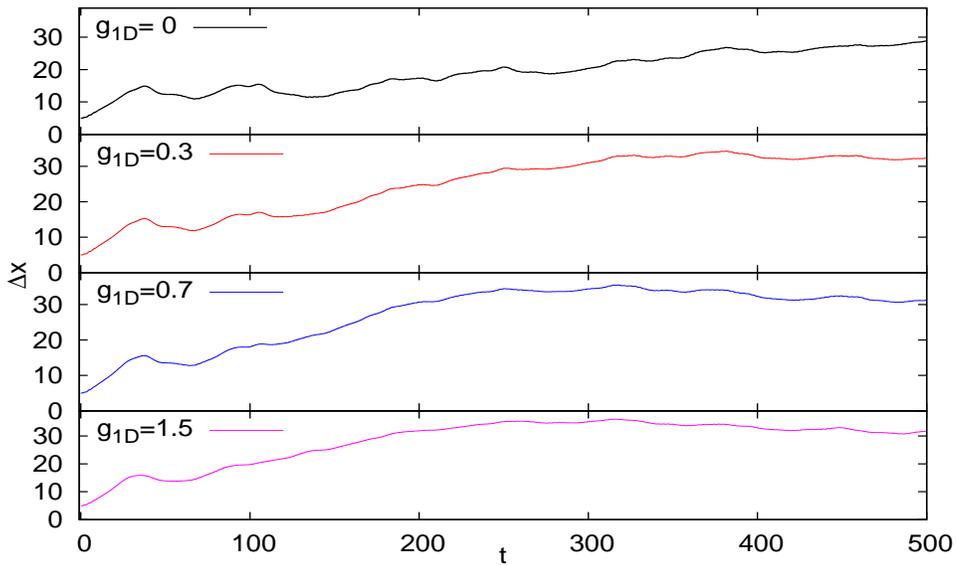

(b) Spatial dispersion vs time with lattice potential $V_0 = 2E_r$ for $g_{1D} = 0$, $g_{1D} = 0.3$, $g_{1D} = 0.7$ and $g_{1D} = 1.5$. Other parameters are $\Delta z_0 = 5$ and condensate is adiabatically loaded in the lattice.

Figure 3.5: Spatial dispersion vs time for shallow lattice.

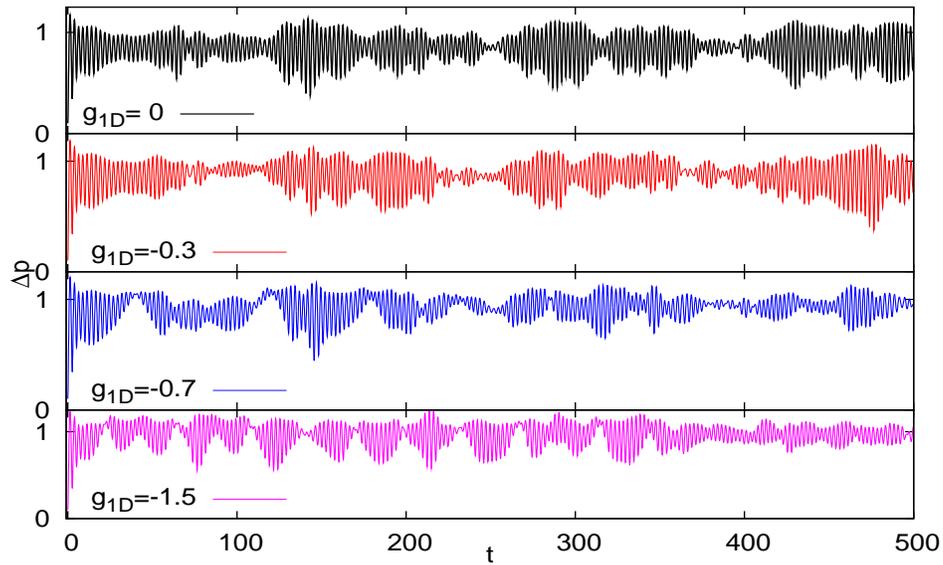

(a) Momentum dispersion vs time with lattice potential $V_0 = 2E_r$ for $g_{1D} = 0$, $g_{1D} = -0.3$, $g_{1D} = -0.7$ and $g_{1D} = -1.5$. Other parameters are , $\Delta p_0 = 0.1$ and condensate is adiabatically loaded in the lattice.

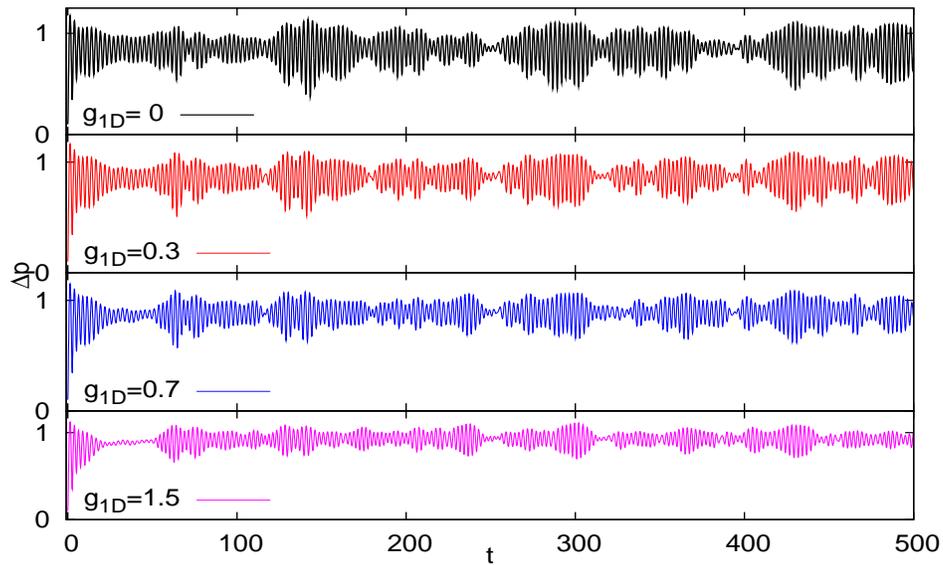

(b) Momentum dispersion vs time with lattice potential $V_0 = 2E_r$ for $g_{1D} = 0$, $g_{1D} = 0.3$, $g_{1D} = 0.7$ and $g_{1D} = 1.5$. Other parameters are , $\Delta p_0 = 0.1$ and condensate is adiabatically loaded in the lattice.

Figure 3.6: Momentum dispersion vs time for shallow lattice.

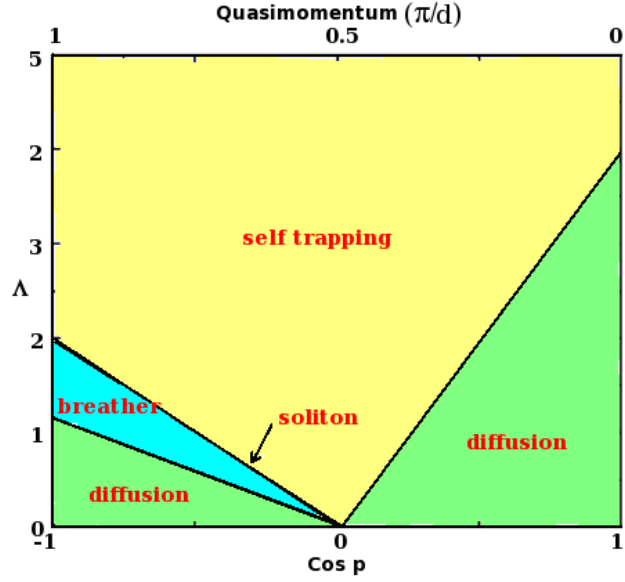

Figure 3.7: Classification of the non-linear propagation in the presence of deep periodic potentials (from Trombettoni et al 2001). Gaussian condensate initially centred around $k_0$ is propagated in a periodic potential depends on the quasi-momentum $k_0$ and total atom number $\Lambda \propto N_t$. After some time, the condensate expansion stops for large non-linearity and this behavior is independent of the initial quasi-momentum (self-trapping regime). For small non-linearity (small atom number), the condensate expands indefinitely (diffusive regime). The solitonic propagation or breathers are only possible for quasi-momenta with negative mass regime. Since this excitation relies on a delicate balance between linear spreading and non-linearity, it only appears for very well-defined atom numbers.

The dynamics is characterized by two basic parameters $\Lambda$ and $\cos p$, where, $\Lambda = g_{1D}/2J$ and $p = kd$ is quasi-momentum. In Fig-3.7, we show the propagation characteristics which depend on these two parameters and the resulting evolution can be characterized by diffusion, self-trapping and solitonic propagation. The solitonic evolution is found by applying the condition that neither the quadratic phase nor the width is time dependent and $\Lambda_{sol} = 2\sqrt{\pi}|\cos p_0| \exp\{-\frac{1}{2\Delta z_0}\}/\Delta z_0$. Here, $\Delta z_0$ is initial width of Gaussian condensate in units of the lattice constant. The number of atoms in a soliton is inversely proportional to the width of the soliton. These solitons very closely resemble the solitons in weak potential [Steel and Zhang 1998; Zobay 1999], populate only a few lattice sites and show a reduced mobility compare to the gap solitons in the weak potential.

In Fig-3.8 and Fig-3.9, we show spatio-temporal evolution of condensate in deep optical lattice for attractive and repulsive condensate respectively. Spatial dispersion for attractive and repulsive condensate is shown in Fig-3.10a and Fig-3.10b respectively. While momentum dispersion for attractive and repulsive condensate is plotted in Fig-3.11a and Fig-3.11b respectively.

The spatio-temporal dynamics of condensate in deep lattice for the absence of non-linearity is almost non-dispersive as tunneling have been suppressed due to strong localization by lattice field. An introduction of small non-linearity, $g_{1D} = 0.1$, whether, attractive or repulsive causes no significant difference. However, collapse and revival behavior of small fragments of condensate trapped on individual lattice sites is modified.

The spatial dispersion does not show a significant difference with the introduction of small non linearity $g_{1D} = -0.1$. The spatial dispersion in this case is suppressed earlier compared to shallow lattice even for small non-linearity, for both attractive and repulsion condensates, as shown in Fig-3.10b and Fig-3.10a. The instability arises due to non-linearity that prompts non-linear Bloch states which decay into localized soliton trains [Ostrovskaya et al 2008].

The momentum dispersion for the case of zero non-linearity ($g_{1D} = 0$) quickly increases initially, and then fluctuates around a mean value showing collapse and revivals in dispersion. With the introduction of non-linearity, the behavior of momentum dispersion is not modified significantly. Although $\Delta p$ fluctuates around the same mean as in the case of zero non-linearity but its collapse and revival behavior is slightly changed for both attractive (Fig-3.11a) and repulsive (Fig-3.11b) condensates, as shown for $g_{1D} = 0.1$.

### 3.3.2  Intermediate range non-linear Energy

The dynamics with intermediate non-linearity is an approximation which is good as long as the non-linearity is such that $4J < g_{1D} < E_{gap}$.

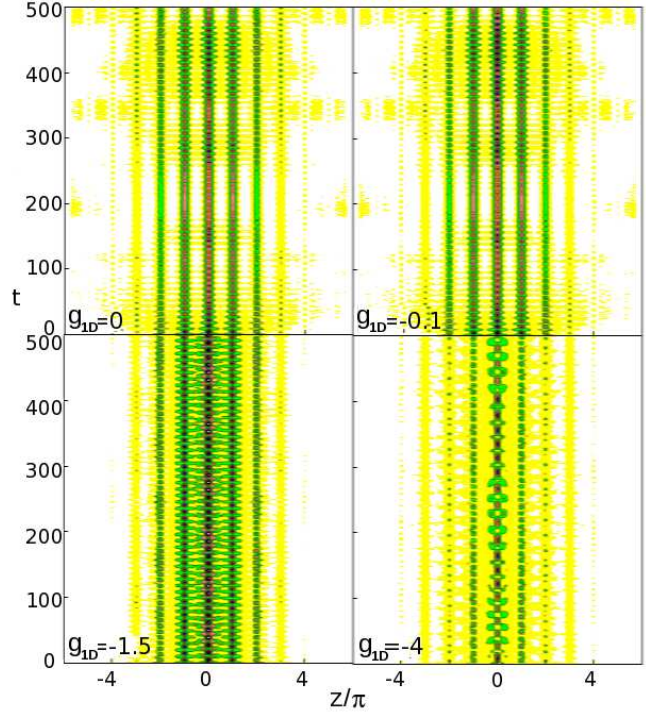

Figure 3.8: Spatio-temporal dynamics of attractive condensate with initial momentum dispersion $\Delta p_0 = 0.1$, non-linearity, $g_{1D} = 0$, $g_{1D} = -0.1$, $g_{1D} = -1.5$ and $g_{1D} = -4$. Lattice potential is $V_0 = 8E_r$ and condensate is adiabatically loaded in the lattice. The dark regions represent the maximum population.

The self-trapping region shown in Fig-3.7 is explored when the width of the condensate remains finite for very very long time. The critical value of $\Lambda$ is $\Lambda_c = 2\sqrt{\pi}\Delta z_0 |\cos p_0| (1 - exp\{-\frac{1}{2\Delta z_0^2}\})$. When $\Lambda < \Lambda_c$, on-site interaction energy per particle is the smallest energy scale and the evolution is qualitatively described with the assumption that the condensate having a mean quasi-momentum. In this limit the condensate is in the diffusive regime as shown in Fig-3.7. In the case, when $\Lambda > \Lambda_c$, the band width is smaller than the on-site interaction energy and the description of a wave packet based on a single central quasi-momentum is failed. The spreading of condensate is suppressed due to local dynamics at the band edges. Solitonic solutions exist and show structures on the length scale of the periodicity, moreover,

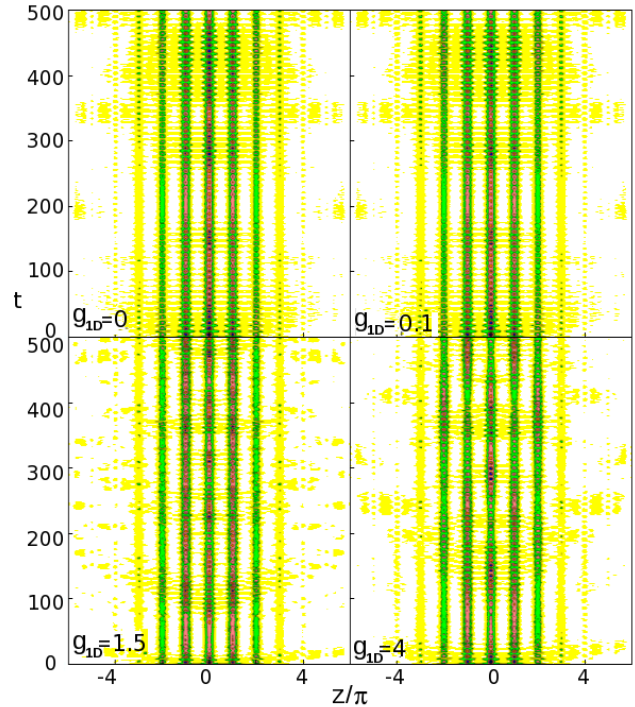

Figure 3.9: Spatio-temporal dynamics of repulsive condensate with initial momentum dispersion $\Delta p_0 = 0.1$, non-linearity, $g_{1D} = 0$, $g_{1D} = 0.1$, $g_{1D} = 1.5$ and $g_{1D} = 4$. Lattice potential is $V_0 = 8E_r$ and condensate is adiabatically loaded in the lattice. The darker regions represent the maximum energy density.

the solitonic solutions can be classified in terms of their symmetry with respect to the potential minima [Louis et al 2003]. Spatio-temporal dynamics (Fig-3.4) for repulsive condensate in shallow lattice shows that due to the stronger repulsion, cloud explodes to the band edge faster as compared to the case where non-linearity is the smallest energy. Moreover, spatial dispersion curve saturated earlier as compared to previous case. Fig-3.12a shows the normalized population of the condensate at different times. From this figure, we note that as the time passes steep edges appear in the density profile of the condensate. At $t = 400$ the steep edges are developed for the case of repulsive condensate with $g_{1D} = 0.7$ and these steep edges are responsible of self-trapping. The corresponding spatial dispersion is also suppressed at this

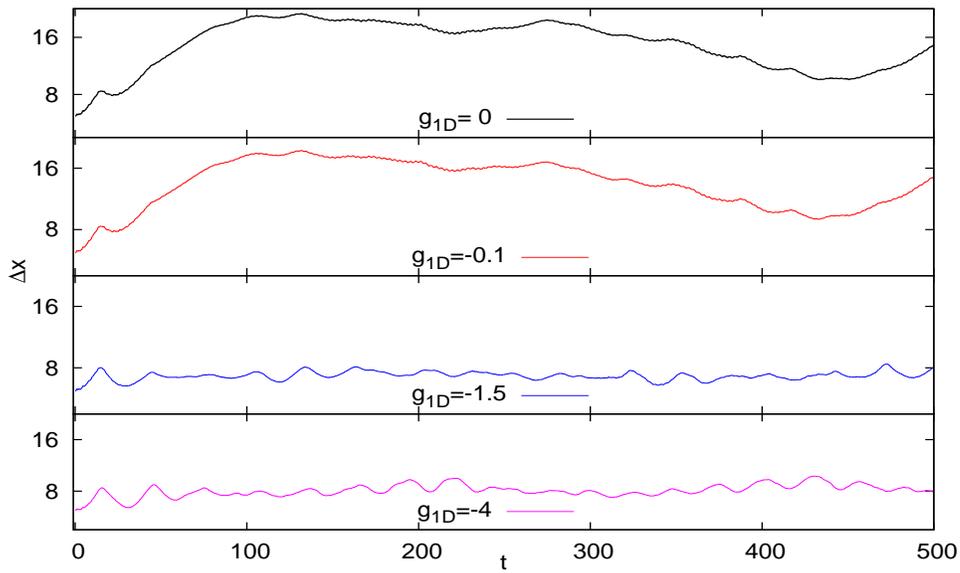

(a) Spatial dispersion vs time for lattice potential $V_0 = 8E_r$, non-linearity, $g_{1D} = 0$, $g_{1D} = -0.1$, $g_{1D} = -1.5$ and $g_{1D} = -4$. Other parameters are $\Delta p_0 = 0.1$ and repulsive condensate is adiabatically loaded in the lattice.

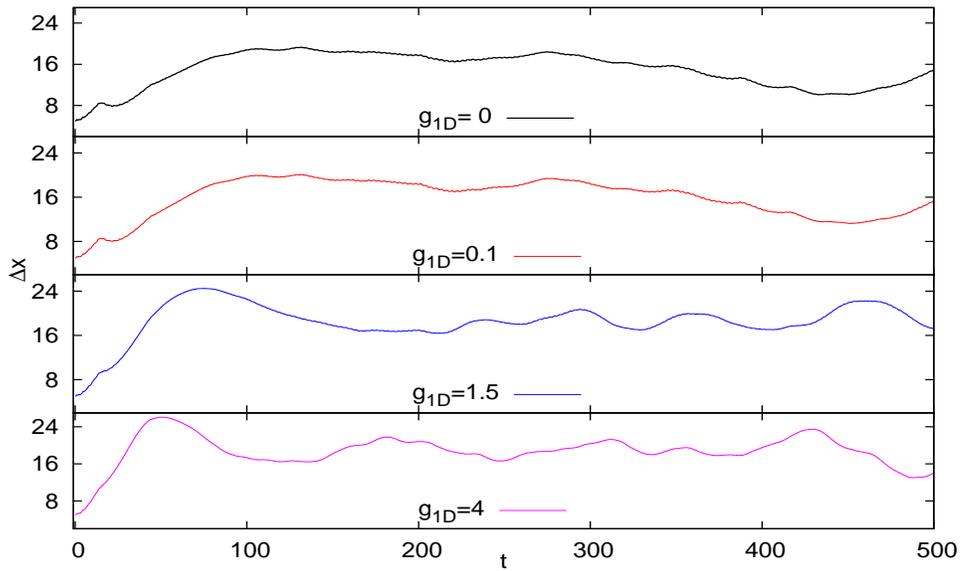

(b) Spatial dispersion vs time for lattice potential $V_0 = 8E_r$, non-linearity, $g_{1D} = 0$, $g_{1D} = 0.1$, $g_{1D} = 1.5$ and $g_{1D} = 4$. Other parameters are $\Delta p_0 = 0.1$ and repulsive condensate is adiabatically loaded in the lattice.

Figure 3.10: Spatial dispersion vs time for deep lattice.

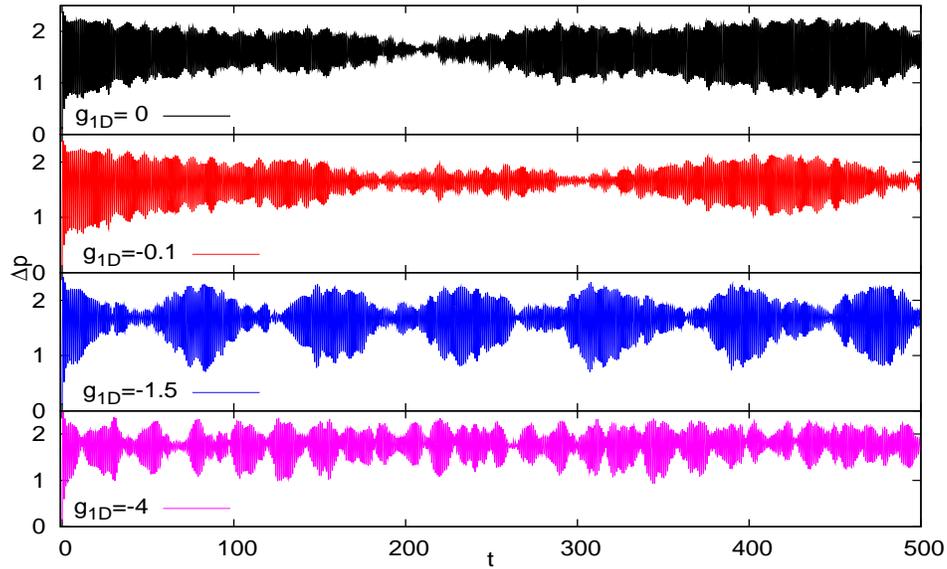

(a) Momentum dispersion vs time for $V_0 = 8E_r$, non-linearity, $g_{1D} = 0$, $g_{1D} = -0.1$, $g_{1D} = -1.5$ and $g_{1D} = -4$. Other parameters are $\Delta p_0 = 0.1$ and attractive condensate is adiabatically loaded in the lattice.

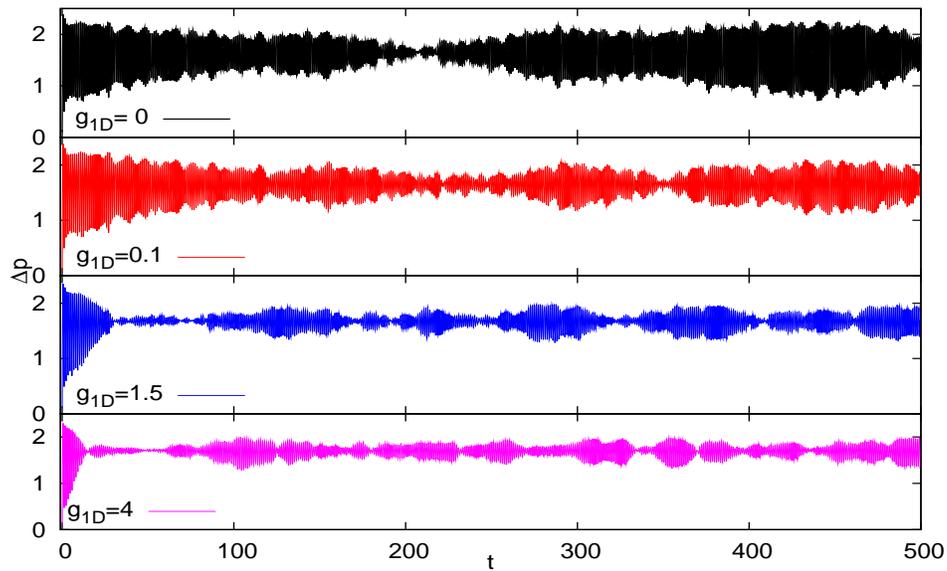

(b) Momentum dispersion vs time for lattice potential $V_0 = 8E_r$, non-linearity, $g_{1D} = 0$, $g_{1D} = 0.1$, $g_{1D} = 1.5$, $g_{1D} = 4$ and $\Delta p_0 = 0.1$.

Figure 3.11: Momentum dispersion vs time for deep lattice.

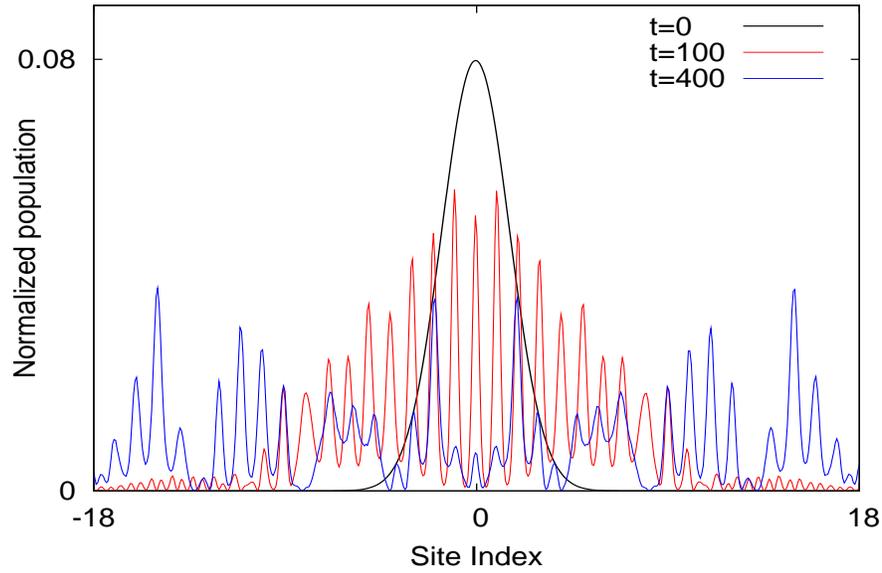

(a) Normalized population of repulsive condensed atoms at $t = 0$, $t = 100$ and $t = 400$. Other parameters are $V_0 = 2E_r$ and $g_{1D} = 0.7$.

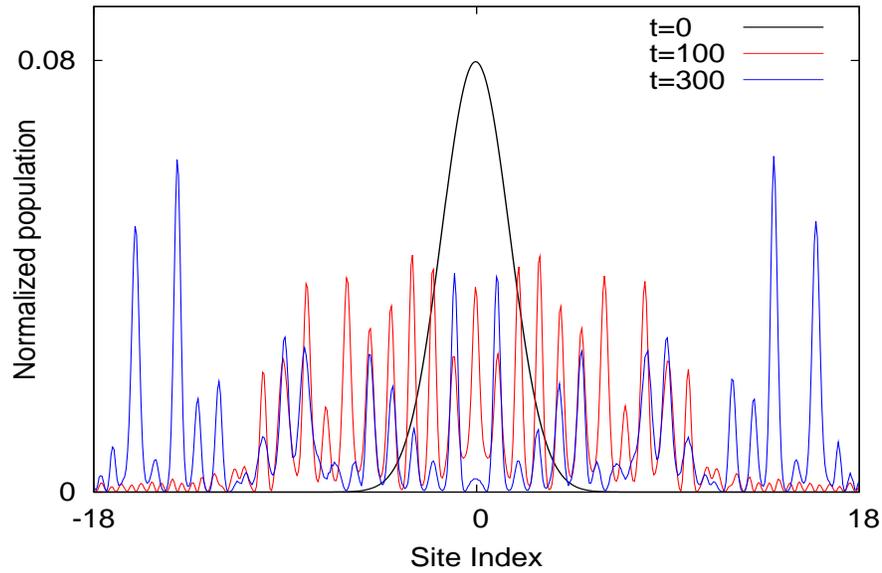

(b) Normalized population of repulsive condensed atoms at $t = 0$, $t = 100$ and $t = 350$. Other parameters are $V_0 = 2E_r$ and $g_{1D} = 1.5$.

Figure 3.12: Normalized population for repulsive condensate in optical lattice.

time which confirms that the self-trapping is responsible of suppression of the condensate. For the deep lattice case, the wave packet of finite width consists of a large number of atoms which have inhomogeneous unstable Bloch states which decay into trains of localized solitons [Dabrowska et al 2006].

### 3.3.3 Largest non-linear energy scale

In this regime, on-site interaction energy is largest energy scale, which implies that, it is larger than ground band width and band gap between lower and first excited band. Here, linear band concept is no longer applicable. Using perturbation theory, an effective potential concept is introduced [Choi 1999] which simplifies the dynamics in this regime. Analytical solutions exist in this regime [Bronski et al 2001] and energies obtained from these solutions are function of quasi-momentum and reveal loop structures instead of flat bands in linear case.

This approach explains the motion in an effective potential for each atom. The energy variations caused by non-linear term in the Gross-Pitaevskii equation and the external periodic potential contribute to the effective potential. Due to periodic potential the atomic density is maximum at the potential minima, the potential energy will be effectively increased (decreased) due to the attractive (repulsive) atom-atom interaction. The dynamics of atoms can be described in an effective potential $V_{eff}$ without non-linear term [Choi 1999], so that,

$$V_{eff} = \frac{V_0}{1 + 4g_{1D}} \cos(2z).$$ (3.40)

This effective potential approximation remains quite valid provided condensate density is nearly uniform which was experimentally confirmed for one dimensional potential [Morsch 2001]. This condition is fulfilled for weak external potential or strong atomic interactions. It was predicted that motion of homogeneous condensate with large non-linearity is naively changed in the presence of periodic potential.

Mostly, solutions in this regime can't be derived analytically. However, in the homogeneous case, solutions are obtained analytically for the potential form $V(z) = -V_0 sn^2(z, k_e)$ [Bronski et al 2001], where, the function $sn(z, k_e)$ is Jacobian elliptic sine functions and $k_e$ is elliptic modulus, such that, $0 \leq k_e \leq 1$. When $k_e = 0$, the potential describes an optical lattice. The stationary solutions exist with and without a non-trivial phase. The stability of the solutions depends on the background density of atoms but an addition of a constant background of atoms leads to a homogeneous non-linear energy and stabilize the solutions.

## Loops in Band Structures

When the non-linear on-site interaction energy is dominant, the concept of linear band structure is no longer applicable. However, the solutions, even in the presence of non-linearity still show some resemblance to the linear band spectrum with new features such as loop formation as shown in Fig-3.13. Investigation of loop structures show that instabilities appear in the band structure for large non-linearity near the boundary of the first Brillouin zone [Wu et al 2003; Machholm et al 2003; Seaman 2004]. When interaction energy per particle exceeds the lattice potential amplitude, the loop structure appear in the lowest Bloch bands near the Brillouin zone boundary and near the zone center in the higher bands. Fig-3.13 shows the loop structure in the energy bands. The swallow tail width increases with the interaction energy, which can extend into zone center. These loops seem as the circumstance of the periodic potential, for the Brillouin zone center, they are a general phenomenon which appear in the vanishing periodic potential [Machholm et al 2003]. In the vanishing potential limit, the loop formed between the second and third bands are degenerated with a special excited state of a condensate, termed as *dark soliton train*.

Spatio-temporal dynamics in this regime for condensate in deep lattice with non-linearity $g_{1D} = 4$ shows that condensate is mostly trapped in few lattice sites. A strong confining lattice potential suppresses Josephson tunneling and during the time evolution, non-linear Bloch states decay into gap soliton trains [Louis 2005]. Spatial dispersion in deep lattice case gains steady state in much shorter time as compared to the shallow lattice due to shorter band width. An introduction of non-linearity quickly drives the cloud to the band edge gap solitons appear in early evolution. For attractive case,

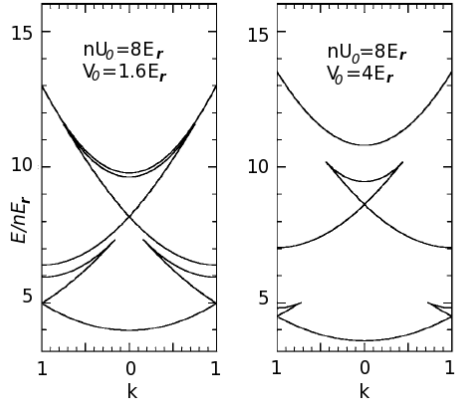

Figure 3.13: A modified band structure: Energy per particle is given as a function of quasi-momentum $k$ [Machholm et al 2003]. The figure shows that the loop structure become less dominant as the lattice potential is increased

enriched collapse and revival behavior in each occupant site is seen. Spatial dispersion of condensate for this regime attains almost steady state earlier than the other two regimes (non-linear energy is smallest and intermediate energy). Momentum dispersion for attractive condensate in this regime reveals collapse and revival in $\Delta p$ with time and revival period in much shorter than the other two regimes.

For repulsive condensate with non-linearity $g_{1D} = 4$, matter waves remain trapped in the potential wells where they were initially placed (Fig-3.9) dispersion in position space for this regime is smaller as compared to the repulsive intermediate non-linearity regime. The dispersion in momentum shows fluctuations with very small amplitude and revival behavior of $\Delta p$ is modified. Small fluctuations in $\Delta p$, and suppression in dispersion is due to suppression of Josephson tunneling and formation of gap soliton trains [Louis 2005].

For shallow lattice, spatio-temporal behavior for attractive condensate lattice in this regime for $g_{1D} = -1.5$ is shown in Fig-3.3. Condensate in this regime shows very interesting features at time $t = 200$, central lattice showing maximum density, neighboring lattice sites have negligible densities while, next to nearest neighboring lattice sites have non-negligible densities. After some time fragmented condensates constructively interference at central lattice site which shows the existence and role of next to nearest correlation in interference of matter wave for this case. Dispersion in position space (Fig-3.5b) is small compared to the other two regimes and settles to mean value with small fluctuations. The momentum dispersion in this regime also shows revival behavior with shorter revival time.

For repulsive condensate in shallow lattice in this regime, spatio-temporal dynamics is shown in Fig-3.4. The figure reveals that condensate diffuses to neighboring sites and spatial dispersion (Fig-3.5b) increases with time eventually settled to a steady state value. This steady state is attained in shorter time as compared to the other two non-linear regimes as strong repulsion derives cloud quickly to the band edges. Where, the condensate forms truncated non-linear Bloch waves (self-trapped states). This behavior is revealed in Fig-3.12b, where, normalized population is shown for different times. The formation of steep edges appear at $t = 300$ on both side of the condensate. These steep edges truncate the non-linear Bloch state [Alexander et al 2006]. While, for intermediate non-linear regime it appeared at $t = 400$ and edges are more steeper compared to the intermediate regime, indicating stronger self-trapping. Momentum dispersion (Fig-3.6b) shows that revival time does not change significantly but fluctuation of dispersion decreases in this case.

## 3.4 Stability of condensate in optical lattice

The discussion on the dynamics of condensate in optical lattice will be incomplete without considering the instabilities that arise due to spatial confinement of condensate in periodic potential and non-linearity. In the lattice potentials, two type of instabilities can exist: i) Landau instabilities, for which small perturbations lead to a lowering of the system energy; ii) dynamical instabilities, due to exponential growth of perturbations [Wu 2000; Wu 2001; Wu et al 2003; Machholm et al 2003].

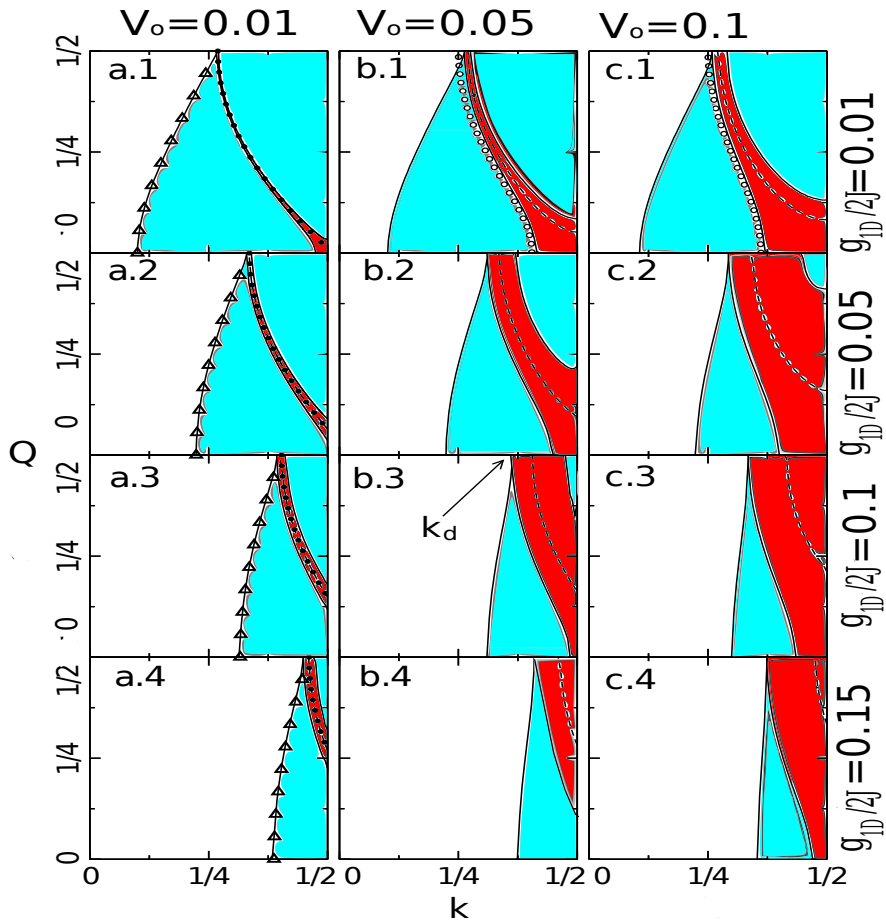

Figure 3.14: Stability diagram obtained by [Wu et al 2003]. The parameters describing the physical situation are the potential modulation $V_0$, the non-linearity $g_{1D}$ for the homogeneous case, and the quasi-momentum, $k$ of the homogeneous condensate. The parameter $Q$ is the corresponding wave vector of the perturbation. It is important to note that $k = 2\pi/d$ implies a modulation of the density with twice the period of the periodic potential. The regime in which the stationary states exhibit a Landau instability are indicated by the light shaded area. The results are symmetric in $k$ and $Q$, so only the parameter region, for $0 < k < 1/2$ and $0 < Q < 1/2$ are shown. The dark shaded area represents the dynamically unstable regime with associated critical quasi-momentum $k_d$.

### 3.4.1 Landau instability

Landau instability is generally discussed in the scenario of Bose liquids and their superfluidity, i.e., the Bose liquid can flow through tight spaces without facing friction provided that its speed is below a critical value. Landau argued that a quantum current faces friction only when the creation of excitations on the liquid lowers the energy of the system. The same is true for a BEC's confined in an optical lattice. In order to find out the influence of small perturbation on the energy of a given Bloch state $e^{i\,k\,z}\phi_k(z)$, the energy of a slightly perturbed Bloch state, that is,

$$\psi_k(z) = e^{i\,k\,z}\left\{\phi_k(z) + u_k(z,Q)e^{iQ\,z} + v_k^*(z,Q)e^{iQ\,z}\right\}, \qquad (3.41)$$

gives the signatures of instability. The functions $u_k(z,Q)$ and $v_k^*(z,Q)$ have the period of potential and $Q \in [2\pi/d, 2\pi/d]$. This perturbation can cause energy deviation. A detailed mathematical method is given in references: [Wu et al 2003; Machholm et al 2003]. If the perturbation increases the energy of the Bloch state, the condensate exhibits superflow as original Bloch state corresponds to a local energy minimum. If energy is negative, normal flow is expected. The numerical results are summarized in the Fig-3.14. A detailed analysis giving the quasi-momentum, $k_d$, for the existence of the both type of instabilities as a function of non-linearity, $g_{1D}$, and potential depth, $V_0$, is given [Machholm et al 2003]. In the stability phase diagrams for condensate, shown in the Fig-3.14, these results are symmetric in $k$ and $Q$, so only the parameter region, for $0 < k < 1/2$ and $0 < Q < 1/2$ are shown. The shaded area of each panel, corresponds to Bloch states have Landau instability. The dynamics depends on three parameters: i) The potential depth $V_0$; ii) The non-linearity $g_{1D}$; and iii) quasi-momentum $k$. The energy deviation is calculated as a function of the free parameter $Q = \pi/d$, which describes a perturbation with the spatial period $d$. In Fig-3.14, the light shaded area represents the Landau unstable region in which phonons emission can lower the systems energy.

### 3.4.2 Dynamical instability

Dynamical instability means that small deviations from the stationary solution in the system exponentially grow in time. Dynamical instability occurs in homogeneous condensates confined in an optical lattice only in the presence of attractive interactions but can be induced by the presence of a periodic potential even for repulsive interactions.

To analyze dynamical stability of the condensate, same procedure is adopted as in the case of Landau instability but now the time-dependent Gross-Pitaevskii equation is used. Taking only the linear term in the perturbation, linear differential equation is obtained which describes the time evolution of the small perturbation [Machholm et al 2003; Wu et al 2003]. The Bloch states are stable if the corresponding eigenvalues are real. However, complex eigenvalues are an indication that the perturbation may grow exponentially. It is important to know that energetically unstable Bloch states are responsible of dynamical instability. The mode that is unstable for the quasi-momentum of the condensate $k = k_d$ is specified by $k = 2\pi/d$. This implies that the corresponding unstable (exponentially growing) mode shows a period doubling [Machholm et al 2004], since the functions $v(z)$ and $u(z)$ in Eq. (3.41) have the same period as the periodic potential.

The Bloch states with $k$, lies outside the shaded area, represent superflow as they correspond to local energy minima. As $g_{1D}$ is increased these superflow regions expand and occupy the entire Brillouin zone for sufficiently large $g_{1D}$. Furthermore, the lattice potential $V_0$ does not significantly influence the super-flow regions as it can be seen in each row. The phase boundaries for $V_0 << 1$ are well reproduced from the analytical expression [Wu et al 2003] $k = \sqrt{\frac{Q^2}{4} + g_{1D}}$, for $V_0 = 0$ which is shown as triangles symbols in the first column.

Common feature of all the panels in Fig-3.14 is that there exists a critical Bloch wave number $k_d$ beyond which the Bloch states show dynamical instabilities. The onset of instability at $k_d$ always corresponds to $Q = 1/2$. This implies that, if we drive the Bloch state from $k = 0$ to $k = 1/2$, at $Q = \pm 1/2$, the first unstable mode appears which represents period doubling. Only for longer wavelength i.e., $k > k_d$, instabilities can occur. These unstable modes grow, drives the system far away from the Bloch state and breaks the translational symmetry of the system spontaneously. The dynamical instability have also been investigated with the reference of an effective-mass approximation [Konotop and Salerno 2002].

## 3.5  Wave packet revivals

In chapter-2, we discussed the condensate revivals of cold atoms in optical lattices both for deep and shallow optical lattices. In this section we study the role of non-linear interaction term on the time scales present in the system for deep and shallow lattices.

**Deep optical lattices**

We consider a deep optical lattice with potential depth $V_0 = 8E_r$, and a Gaussian condensate comprising of lowest energy band with initial momentum spread $\Delta p = 0.1E_r$ around the mean $p_0 = 0$.

Autocorrelation function for attractive condensate is shown in Fig-3.15a for $g_{1D} = 0$, $g_{1D} = -0.1$, $g_{1D} = -1.5$ and $g_{1D} = -4$. and for repulsive interaction in Fig-3.15b for the same non-linearity values. In the case of repulsive interaction the auto-correlation falls quickly as tunneling probability to the nearest neighbor and next to nearest neighbor is non-vanishing and later, the formation of solitonic trains as discussed in sec-3.3 suppresses tunneling wave packet revivals. Stronger the repulsive interaction smaller the auto-correlation. For $g_{1D} = 0.1$, the revival structure doesn't change significantly except $|A(t)|^2$ collapses to the lower value compared to the single particle wave packet dynamics ($g_{1D} = 0$). When $g_{1D} = 1.5$ (interaction energy is in intermediate range) and $g_{1D} = 4$ (interaction energy is dominant), the revival structures changes significantly. For attractive interaction, a small non-linearity $g_{1D} = 0.1$, modifies the revival structures. For large attractive interaction ($g_{1D} = -1.5$, $g_{1D} = -4$) revival time decreases. We study the condensate distribution at collapse time and first revival time for $g_{1D} = -1.5$, $V_0 = 8E_r$ in Fig-3.16. In this case collapse time $t = 30$ and quantum revival

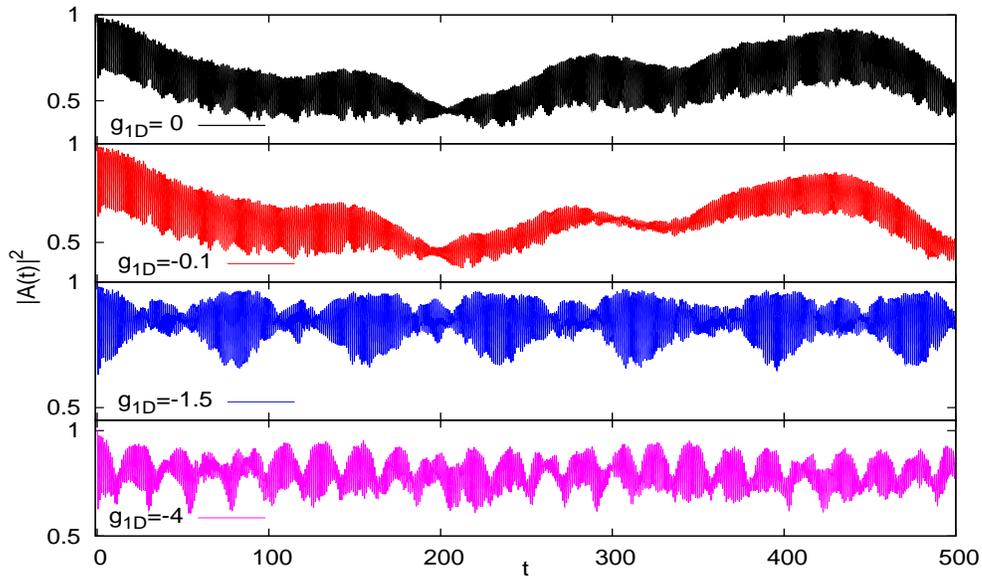

(a) Square of auto-correlation function of Gaussian condensate initially placed such that it spans lowest Bloch band of lattice with $V_0 = 8E_r$, for $g_{1D} = 0$, $g_{1D} = -0.1$, $g_{1D} = -1.5$ and $g_{1D} = -4$. Other parameters are $\Delta p = 0.1E_r$ and mean initial momentum $p_0 = 0$

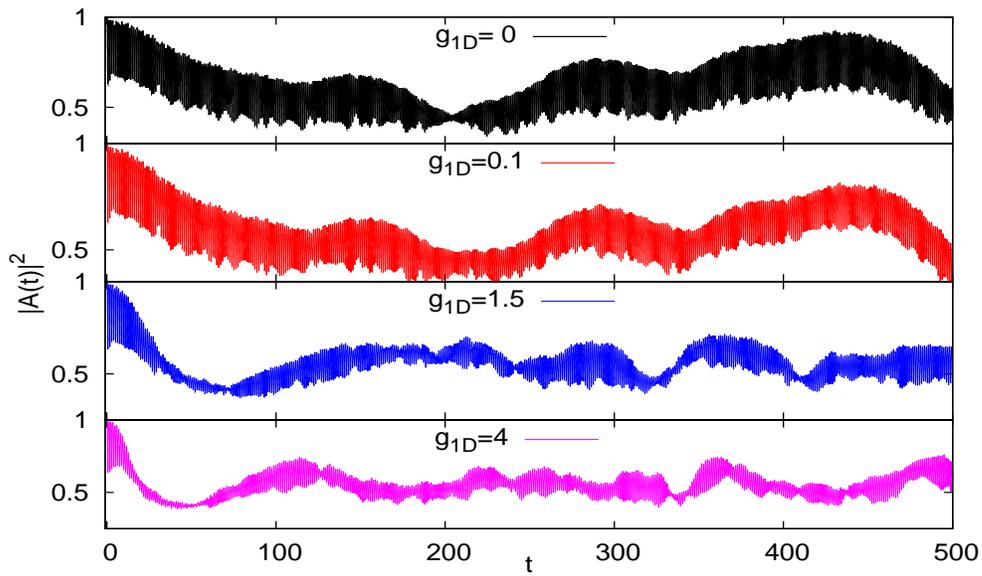

(b) Square of auto-correlation function of Gaussian condensate initially placed such that it spans lowest Bloch band of lattice with $V_0 = 8E_r$, for $g_{1D} = 0$, $g_{1D} = 0.3$, $g_{1D} = 0.7$ and $g_{1D} = 1.5$. Other parameters are $\Delta p = 0.1E_r$ and mean initial momentum $p_0 = 0$.

Figure 3.15: Square of auto-correlation function vs time for Gaussian condensate placed in deep lattice.

time is $t = 90$. At $t = 30$ condensate is localized in lattice wells and overlap with initial condensate is minimum while at $t = 90$ overlap with initial condensate is maximum.

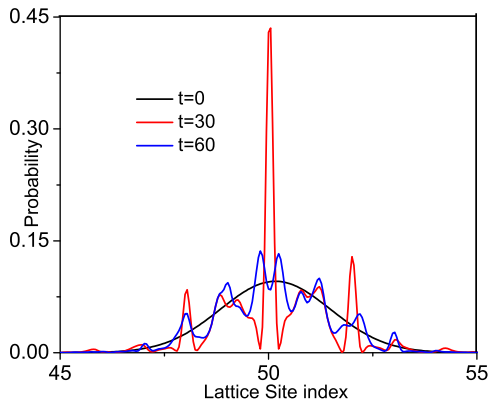

Figure 3.16: Distribution function for attractive BEC with $g = -1.5$ at time $t = 0$, which defines initial distribution, $t = 30$, at which the condensate displays first collapse and $t = 90$, for first revival. The lattice potential is $V_0 = 8E_r$.

### Shallow optical lattices

We consider a shallow optical lattice with potential depth $V_0 = 2E_r$, and plot auto-correlation function for a condensate with momentum dispersion $\Delta p = 0.1E_r$.

For strong homogeneous interaction (repulsive or attractive), an effective potential approximation given in Eq. (3.40) is valid in this regime and expressions for classical periods, quantum revivals and super revivals given in Sec-2.7 are valid. For attractive condensate with $g_{1D} = 0.3$, revival structure changes and auto-correlation falls to a low value which is due to the initially Josephson tunneling of cloud and later suppression of tunneling due to formation of solitons near band edges. As interaction increases auto-correlation stays near the unity and for large non-linearity ($g_{1D} = -0.7$ and $g_{1D} = -1.5$) revivals are more visible with indication that increase in non-linearity causes

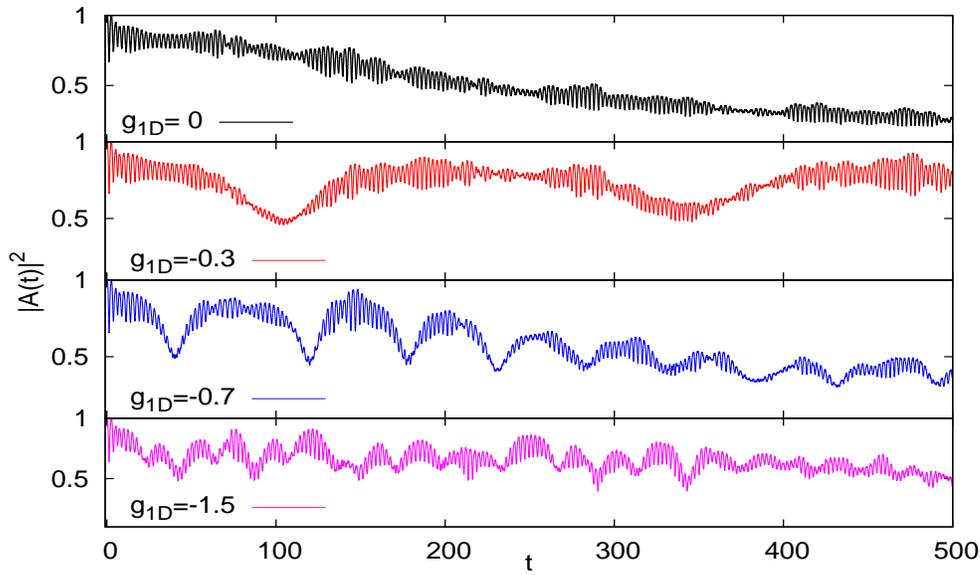

(a) Square of auto-correlation function of Gaussian condensate initially placed such that it spans lowest Bloch band of lattice with $V_0 = 2E_r$, for $g_{1D} = 0$, $g_{1D} = -0.3$, $g_{1D} = -0.7$ and $g_{1D} = -1.5$. Other parameters are $\Delta p = 0.1E_r$ and mean initial momentum $p_0 = 0$

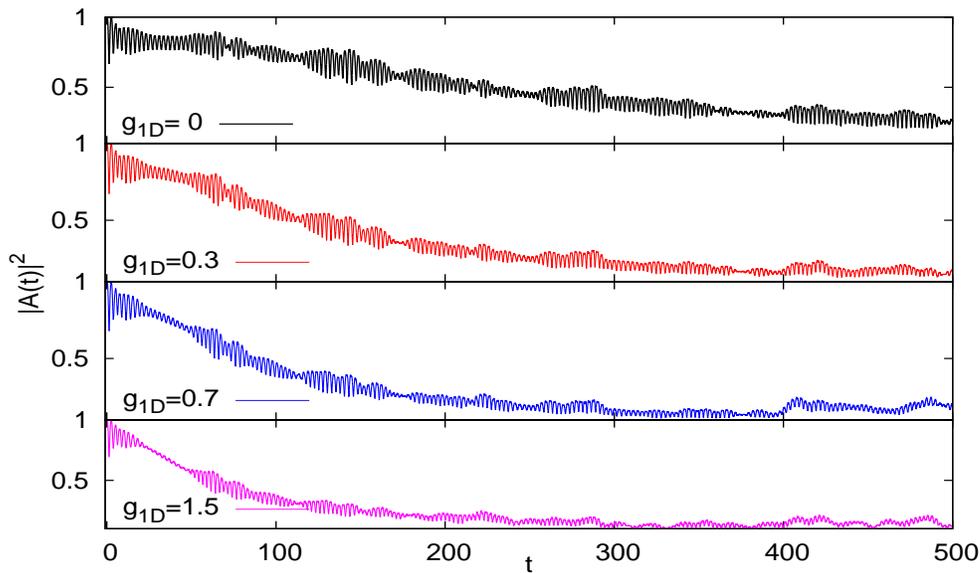

(b) Square of auto-correlation function of Gaussian condensate initially placed such that it spans lowest Bloch band of lattice with $V_0 = 2E_r$, for $g_{1D} = 0$, $g_{1D} = 0.3$, $g_{1D} = 0.7$ and $g_{1D} = 1.5$. Other parameters are $\Delta p = 0.1E_r$ and mean initial momentum $p_0 = 0$ a decrease in revival time Fig-3.17a. For repulsive interaction, increase in non-linearity causes a decrease in auto-correlation due to the fact that initially cloud has faster tunneling and latter due to steep edges tunnel is suppressed and auto-correlation falls with an increase in non-linearity due to self-trapping Fig-3.17b.

Figure 3.17: Square of auto-correlation function vs time for Gaussian condensate placed in shallow lattice.

## 3.6 Discussion

In this chapter, we summarize the dynamics of the condensate in optical lattice. It explains the influence of interaction term on the center of mass dynamic of the condensate. The dynamics are investigated by studying spatio-temporal behavior, auto-correlation function, momentum and spatial dispersion of the condensate. The inherent non-linearity of a condensate due to Bragg scattering of a matter-wave from an optical lattice and repulsive atomic interactions play its role in the dynamics. Under some circumstances, localization is possible as condensate shows anomalous dispersion at the edges of a Brillouin zone of the lattice and magnitude of this dispersion can be managed by tuning the lattice depth and interaction strength. Therefore, the condensate spreading either can be controlled by actively controlling the dispersion by varying the intensity of confining field or by utilizing the inter-atomic interaction. The later approach leads to non-linear localization of condensate. In shallow optical lattice, this non-linear localization is due to the formation of solitons or steep edges on both side of one-dimensional lattice which truncates Bloch states (self-trapping). While, for deep lattice case, localization is due to the trains of localizes gap solitons which are formed due to decay of inhomogeneous unstable Bloch states.

# Chapter 4

# Cold Atoms in Driven Optical Lattices

## 4.1 Introduction

In this chapter, we explain the dynamics of cold atoms in driven optical lattice and focuses our attentions on classically integrable regions i.e., nonlinear resonances in phase space. In our analysis we employ Poincaré surface of sections, momentum dispersion and their parametric dependencies. Poincaré surface of section reveals the stroboscopic map of regular and chaotic structures of phase space and its parametric dependencies. Classical dynamics of our system displays an intricate dominant regular and dominant stochastic dynamics, one after the other, as a function of increasing modulation amplitude in the limit where, lattice potential is small and modulation amplitude is large enough [Moore 1994; Raizen 1999].

For quantum wave packet dynamics, momentum dispersion, auto-correlation and spatio-temporal behavior are studied in the vicinity of nonlinear resonances. Due to spatial and temporal periodicity in driven optical lattices, the corresponding Schrödinger equation for driven lattice can be mapped on Mathieu equation which make the system a paradigm to understand the effects of periodic modulation on wave packet evolution.

Dynamical recurrences in quantum systems, which display chaotic dynamics in their classical counterpart, are different due to their parametric dependence on modulation effects and as they are limited to nonlinear resonances. Recurrences in such systems originate from the simultaneous excitation of discrete quasi-energy states. Our mathematical framework is developed in two classes: delicate recurrences which take place when lattice is weakly perturbed; and robust recurrences which manifest for sufficiently strong external driving force. These dynamical recurrences can also be applied as a probe to study quantum chaos [Saif 2000; Saif 2005a].

## 4.2   Cold Atoms in Driven Optical Lattices

External drive in optical lattice introduces a phase modulation, hence, the electric field in (2.8) is modified as

$$\hat{E}(x,t) = \hat{e}_y[\varepsilon_o \cos(k_L x - \Delta L k_L \sin \omega t)e^{-i\omega_L t} + c.c.],\qquad(4.1)$$

where, $\Delta L$ is amplitude and $\omega_L$ is frequency of external modulation. The dynamics of a cold atom of mass $M$ in phase modulated optical lattice is governed by the Hamiltonian,

$$H = \frac{p^2}{2M} + \frac{V_0}{2}\cos[2k_L\{x - \Delta L \sin(\omega_m t)\}],\qquad(4.2)$$

where, $k_L$ is wave number and $V_0$ define the potential depth of an optical lattice. Furthermore, $\Delta L$ and $\omega_m$ are amplitude and frequency. Introducing the dimensionless parameters, we scale time by phase modulation frequency and position $x$ by natural length scale i.e., standing wave period,

$$\tau = \omega_m t,$$

$$z = k_L x,$$

$$\text{and } \lambda = 2\Delta L k_L.$$

The momentum $\tilde{p}$ can be related to the original momentum $p$ as

$$\tilde{p} = \frac{k_L}{M\omega_m}p,$$

with commutation relation

$$[z,\tilde{p}] = \frac{k_L^2}{M\omega_m}i\hbar = i\frac{2\omega_r}{\omega_m} \equiv i\Bbbk,\qquad(4.3)$$

defining scaled Plank's constant

$$\bar{k} = \frac{2\omega_r}{\omega_m}.$$

The dimensionless Hamiltonian is written as

$$H = \frac{p^2}{2} + \frac{\tilde{V}_0}{2}\cos\{2z - \lambda\sin(\tau)\}, \qquad (4.4)$$

where, $\tilde{V}_0 = V_0\bar{k}/\hbar\omega_m$ is effective potential depth. Here and in later discussion, for simplicity, the tilde sign on $p$ have been ignored.

## 4.3  Classical Dynamics

In this section, we discuss the classical dynamics of particle in driven optical lattice. The classical dynamics are governed by the following Hamilton equations,

$$\dot{z} = \frac{\partial H}{\partial p} = p,$$

and

$$\dot{p} = -\frac{\partial H}{\partial z} = -\frac{\tilde{V}_0}{2}\sin\{2z - \lambda\sin(\tau)\},$$

or equivalently

$$\ddot{z} = -\frac{\tilde{V}_0}{2}\sin\{2z - \lambda\sin(\tau)\} = -\frac{\tilde{V}_0}{2}\sum_{m=-\infty}^{\infty}J_m(\lambda)\sin(2z - m\tau), \qquad (4.5)$$

where, $J_m(\lambda)$ is $m^{th}$ order Bessel function. The expression at the right side in last equation is written by invoking the *Jocobi-Anger* identity

$$e^{i\lambda\sin(\omega_m\tau)} = \sum_{m=-\infty}^{\infty}J_m(\lambda)e^{im\omega_m\tau}.$$

The effect of potential on the atomic motion depends on the velocity of the atoms. The phase modulation velocity of lattice field is $-\lambda\cos(\tau)$. The maximum velocity of potential is given by the phase modulation amplitude $\lambda$, which gives expectations that a particle can't be accelerated to the velocities larger than $\lambda$. We confirm this argument by stationary phase analysis. When the velocity of atoms and velocity of phase modulation coincide, the effect of motion is significant. The condition for phase $z(\tau) - \lambda \sin(\tau)$ be stationary is

$$\dot{\phi}(\tau_0) = \dot{z}_{\tau_0} - \lambda \cos(\tau_0) = 0.$$

For a particle with momentum $p$, the above stationary phase condition is only fulfilled if $\lambda \cos(\tau_0) = p$ or $|p| < \lambda$

The classical Hamiltonian in Eq. (4.4), can be written as

$$H^{(m)} = \frac{p^2}{2} + \frac{\tilde{V}_0}{2} J_0(\lambda) \cos(2z) - \frac{\tilde{V}_0}{2} \sum_{m=-\infty}^{\infty} J_m(\lambda) [\cos(2z - m\tau) + (-1)^m \cos(2z + m\tau)].$$
$$(4.6)$$

We can neglect the summation terms in Eq. (4.6) under the condition [Moore 1994],

$$\frac{\tilde{V}_0}{\sqrt{\lambda}} << 1, \tag{4.7}$$

and for small initial momentum, atoms follow the equation of motion,

$$\ddot{z} = -\frac{\tilde{V}_0}{2} J_0(\lambda) \sin(2z). \tag{4.8}$$

Which is equation of motion of an un-driven pendulum oscillating with frequency $\sqrt{\frac{\tilde{V}_0}{2} |J_0(\lambda)|}$ and hence the expression for classical period is $T_{cl} = \frac{2\pi}{\sqrt{\frac{\tilde{V}_0}{2} |J_0(\lambda)|}}$. Since the trapped atoms oscillate with large amplitude around lattice minima, the above harmonic approximation for classical period $T_{cl}$, is not quite valid. The energy conservation law gives,

$$E = \frac{p^2}{2} - \frac{\tilde{V}_0}{2} J_0(\lambda) \cos(2z) = -\frac{\tilde{V}_0}{2} J_0(\lambda) \cos(2z_0), \tag{4.9}$$

where, $z_0$ represents the position of lattice well and it is considered that momentum changes very slowly, which yields

$$p^2 = 2\tilde{V}_0 J_0(\lambda) [\sin^2(z_0) - \sin^2(z)].$$

We find that

$$d\tau = \frac{dz}{\sqrt{2\tilde{V}_0 J_0(\lambda) [\sin^2(z_0) - \sin^2(z)]}},$$

and oscillation period is given as

$$T = \frac{2}{\sqrt{2\tilde{V}_0|J_0(\lambda)|}} \int_0^{z_0} \frac{dx}{\sqrt{\sin^2(z_0) - \sin^2(z)}},$$

$$T = \frac{4}{\sqrt{2\tilde{V}_0|J_0(\lambda)|}} F(\frac{\pi}{2}, z_0),$$

where, $F(\frac{\pi}{2}, z_0)$ is complete elliptical integral of first kind. For $z_0 = 1$, $F(\frac{\pi}{2}, 0.5) = 1.8541$. When $\lambda = 10$ and $\tilde{V}_0 = 0.5$, then $T = 14.955$.

### 4.3.1 Poincaré Surface of Section

Numerically, the dynamics of classical particle are explored by studying Poincaré section (which is a stroboscopic map in $zp$−plane over one period of external modulation) for fixed value of modulation amplitude. To plot Poincaré surface section, 200 initial conditions randomly distributed in phase space are taken and each initial point is evolved for the time, $\tau = 1000\pi$. After each period of phase modulation, the snap shot of position and momentum is taken. Taking advantage of periodicity of the potential $V(z) = V(z + \pi)$, the position of the atom is also folded back to the interval $[-\pi/2, \pi/2]$.

From Poincaré sections in Figs-4.1-4.3, it is revealed that for finite and small $\lambda$, the hyperbolic fixed points survive, while there stable and unstable manifolds don't coincide. Rather, they intersect each other many times generally in homoclinic points shaping as a *homoclinic tangle* and chaos is in a narrow chaotic layer around the separatrix which separate bounded and unbounded motion in the un-driven case. Moreover, all orbits with periods, a rational multiples of the period of external modulation are destroyed and replaced by a chain of hyperbolic (unstable) and elliptic (stable) orbits. The unstable and stable manifolds of the hyperbolic orbits intersect again and form a *heteroclinic tangle* in secondary separatrix layers around the *elliptic islands* i.e., in the domain of stability of the elliptic orbits.

This structure is repeated again and again within each elliptic island on rapidly decreasing scale. As modulation strength increases, stochastic region around the separatrix increases on the expanse of regular region. While, large regular left-running ($p < 0$) and right-running ($p > 0$) solutions still exist but all oscillatory orbits, even those whose periods are the "most irrational" multiples of the driving period, are destroyed and the chaotic domain spread gradually in the whole domain between the right-running ($p > 0$) and left-running ($p < 0$) regular solution. This is the domain of dominant chaos, with a threshold at some value $\lambda = \lambda_c$. Beyond the $\lambda_c$, almost all available phase space in chaotic and only few islands of regular motion survive around the elliptic fixed point.

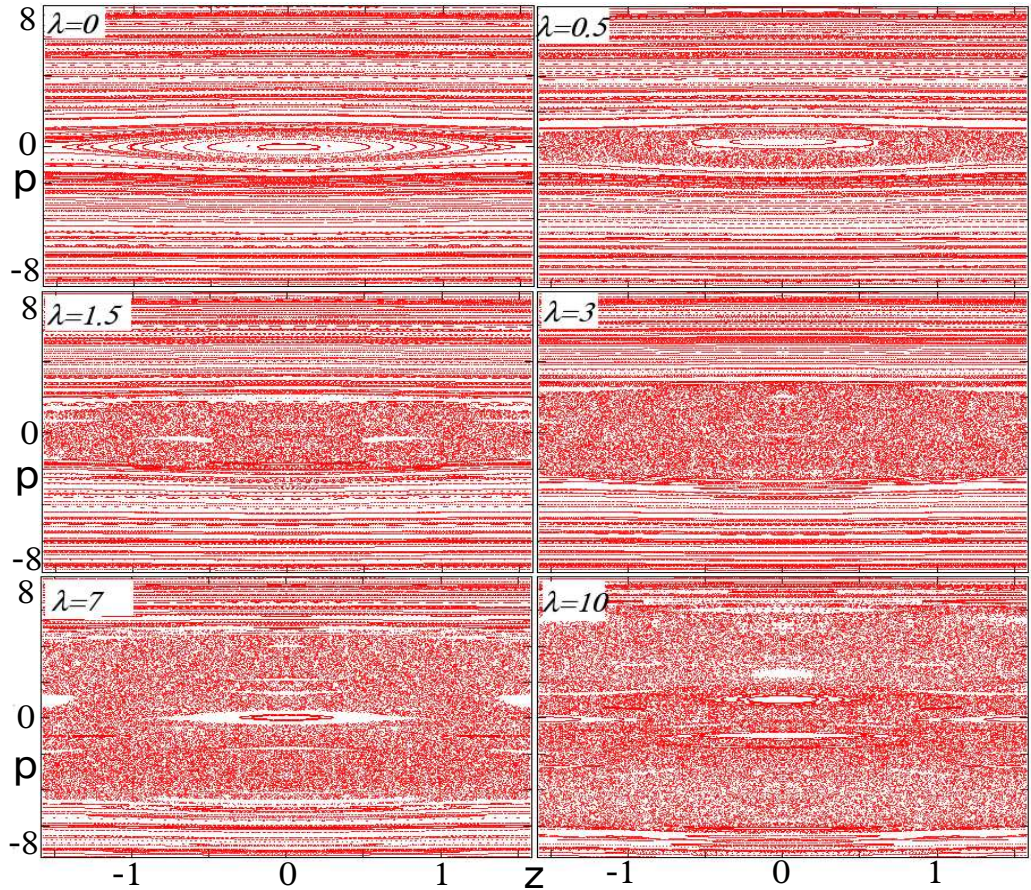

Figure 4.1: Poincaré surface of section for different modulation amplitudes with $\tilde{V}_0 = 0.36$.

However, in the case, where, the condition, $\frac{\tilde{V}_0}{\sqrt{\lambda}} << 1$, is satisfied, potential can be expressed by the approximation Eq. (4.7). The resonances reappear and their positions can be recognized by the sign of zero order Bessel function. In Fig-4.1 for modulation amplitudes $\lambda = 7$ and $\lambda = 10$, it is seen that whenever, the Bessel function, $J_0(\lambda)$ is positive, the resonances are positioned at $z = 0$ and for negative $J_0(\lambda)$, these elliptic fixed point are located at $z = \pi$.

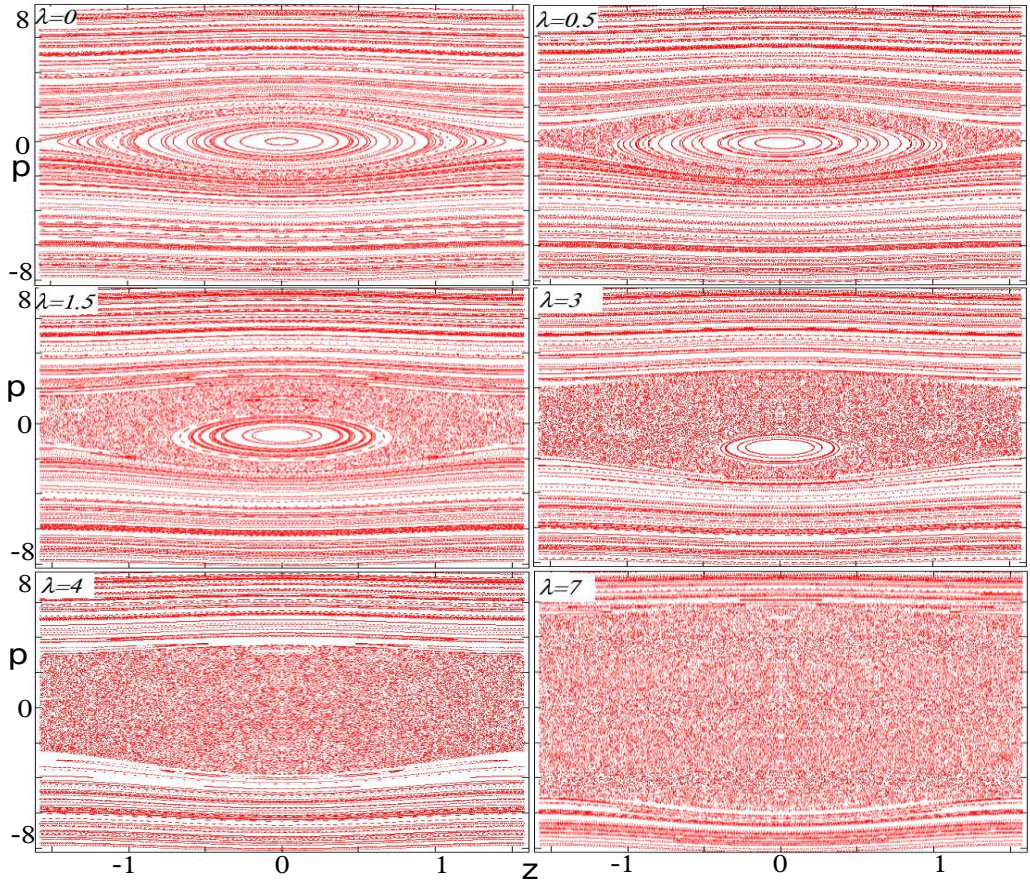

Figure 4.2: Poincaré surface of section for different modulation amplitudes with $\tilde{V}_0 = 2$.

In the previous literature [Moore 1994; Raizen 1999], it is stated that reappearance of classical resonances is only dependent on condition Eq. (4.7). While, extensive numerical study presented here, reveals that resonances reappear only when $\tilde{V}_0 < 1$ along with the condition Eq. (4.7). Otherwise, a stochastic sea appears beyond the critical value of modulation amplitude, and extends its boundaries with the increase in modulation amplitude.

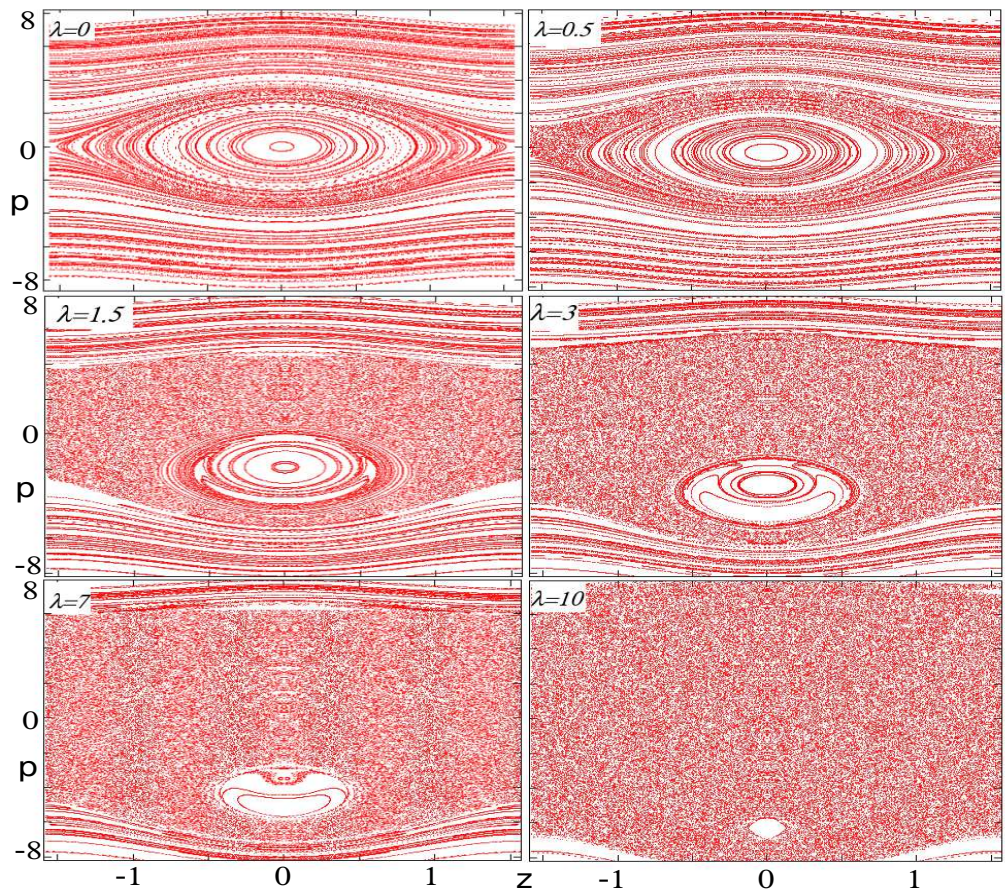

Figure 4.3: Poincaré surface of section for different modulation amplitudes with $\tilde{V}_0 = 5.7$.

We can also enter in the chaotic domain by keeping $\lambda$ constant and increasing $\tilde{V}_0$ from a very small value. Increasing the effective potential $\tilde{V}_0$ from zero, orbits with period rationally related to the period of driving force are destroyed and replaced by resonances. The global chaos condition is now attained at some critical value of $\tilde{V}_0$.

## 4.3.2 Momentum Dispersion

We consider expansion Eq. (4.6), the terms in summation in this expansion describe resonance [Chirikov et al 1982] when

$$p = \dot{z} = \frac{m}{2},$$ (4.10)

with width

$$\Delta p_m = 4\sqrt{\frac{\tilde{V}_0 |J_m(\lambda)|}{2}}.$$ (4.11)

The resonance width exponentially decreases for $|m| > \lambda$, and approximate value for $|J_m(\lambda)| \lesssim \sqrt{\pi \lambda}$ can be used when $|m| < \lambda$ and $\lambda >> 1$. Thus among infinitely many resonances, only those have considerable size with $m \lesssim \lambda$. The resonance overlap criteria [Chirikov et al 1982] give the condition of onset of chaos when

$$\tilde{V}_0 > \frac{\sqrt{\pi \lambda}}{20}.$$ (4.12)

The effect of driving force on undriven resonance near $\dot{z} = p = 0$ can also be understand by finding moving center of driven resonance as $\dot{z} = \lambda \cos t$. The resonance is now a moving region around $p = 0$ and oscillate with modulation frequency and amplitude. The width of moving resonance remain time independent [Chirikov et al 1982] i.e., $\Delta p = \sqrt{8\tilde{V}_0}$. Thus every point in the region $|p| < \lambda$ is crossed by the resonance twice in each driving period. For the case when resonance crosses much faster than the system points, $p$ changes by

$$\Delta p \simeq -\frac{\sqrt{\pi}\tilde{V}_0}{2\lambda} \sin(z_r \pm \pi/4).$$ (4.13)

This gives the diffusion constant

$$D \simeq \frac{<\Delta p>}{T/2} = \frac{\tilde{V}_0}{4\lambda},$$ (4.14)

and it is predicted that due to diffusive spreading of $p$ over the domain $|p| < \lambda$, saturated momentum dispersion is

$$<\Delta p^2> \simeq \frac{4}{\sqrt{3}}\lambda.$$ (4.15)

The time evolution of classical and quantum mechanical dispersion is shown in Figs-4.6-4.9 for modulation amplitudes $\lambda = 0.5$, 1.5, 3. In the classical case a Gaussian distribution of 10000 non-interacting particles is evolved. In both cases (classical and quantum mechanical) plots are averaged values computed for different initial positions such that $z_0 = -\frac{\pi}{2}$, $-\frac{\pi}{3}$, $-\frac{\pi}{4}$, 0, $\frac{\pi}{4}$, $\frac{\pi}{3}$, $\frac{\pi}{2}$, initial momentum $p_0 = 0$, initial momentum dispersion $\Delta p = 0.5$ and position dispersion is $\Delta z = \frac{1}{2\Delta p}$. From Figs-4.6-4.9, it is seen that classical momentum dispersion increases quickly for small modulation and fluctuates around a mean value showing saturation behavior as for small $\lambda$, the diffusion rate, $D = \frac{\tilde{V}_0}{4\lambda}$, is larger and velocity of atoms match with the velocity of driving field and momentum transfer is more likely. As modulation is increased, the saturation time increases as diffusion decreases inversely with modulation. While saturation value increases as classical resonance boundary limits the momentum space $\Delta p_{max} \approx \frac{4\lambda}{\sqrt{3}}$. On the other hand, for fixed value of modulation, classical momentum dispersion, $\Delta p_{cl}$, saturates at higher value for deep lattice potential. The fluctuation in the classical curves is due to the presence of resonance (which can be seen in corresponding Poincaré section) as it trap a part of the ensemble which oscillates inside the resonance with classical period.

Classical evolution of non-interacting Gaussian distribution is shown in Fig-4.4. In this figure, density plots, position and momentum distributions of 10000 particles with initial momentum and position dispersion $\Delta p = 0.5$ and $\Delta z = 1$ respectively are shown for time $\tau = 0$ (left column) and $\tau = 200$ for two different initial conditions: i) When initially distribution is placed near the center of resonance (middle column); ii) When it is located initially at stochastic region (right column). Classical dynamics show that when particles are placed inside the resonance region they remain their and only those particles diffuse which are part of distribution tail and initially find themselves near the separatrix (Fig-4.4-b). Dispersion in position and momentum space is almost negligible as seen from Figs-4.4-(e),(h). On the other hand, when particles are initially evolved in chaotic region, they quickly diffuses to all available phase space and only a small fraction of particles remains trapped in the resonance. Again this trapped fraction is a tail of distribution which find itself, initially, trapped in the resonance and remains there.

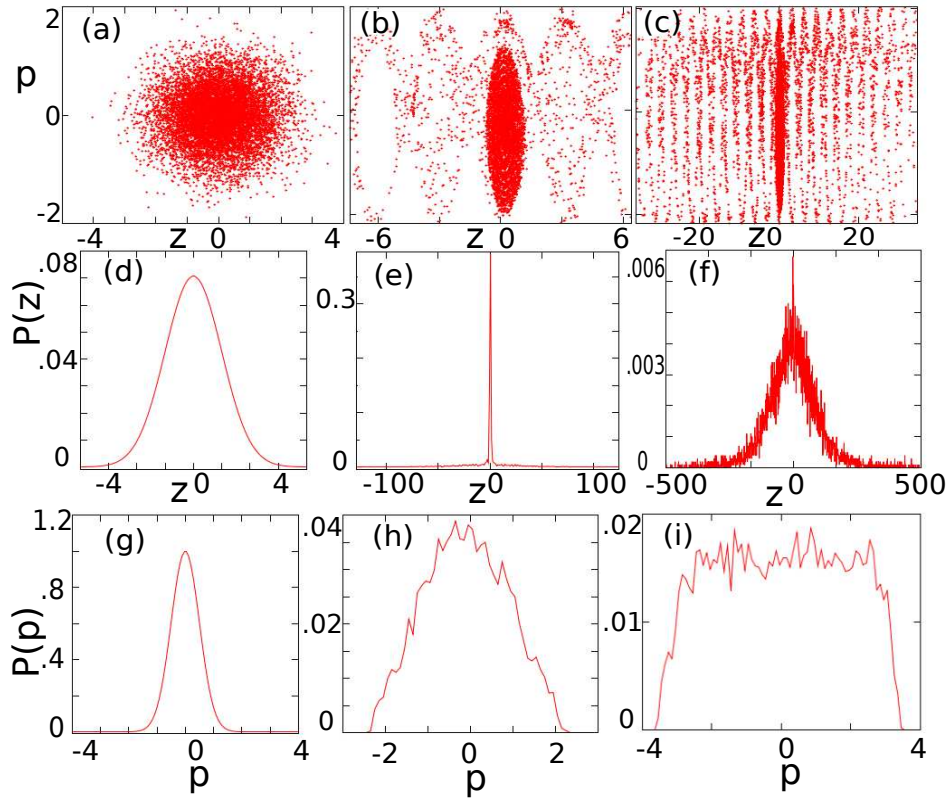

Figure 4.4: Density plots, position and momentum distributions of 10000 non-interacting particles with initial momentum $\Delta p = 0.5$ and position dispersion $\Delta z = 1$. Initial particle density (a), position distribution (d) and momentum distribution (g) are shown in left column. For modulation $\lambda = 3$, when particle is initially placed near the center of resonance, particle density, position and momentum distribution are shown in (b), (e) and (h) respectively at $\tau = 200$ in the middle column. While, (c), (f) and (i) show density, position and momentum distribution respectively, when particles are initially placed in chaotic region at time, $\tau = 200$. $\tilde{V}_0 = 2$ in this case.

## 4.4   Quantum Dynamics

In this section quantum dynamics of Gaussian wave packet are studied in driven optical lattices. Parametric dependencies of momentum dispersion and dependencies on initial excitation condition are studied.

### 4.4.1   Atomic Momentum Distribution

We solve the Schrödinger equation using following ansatz for wave function

$$\psi(z, \tau) = \exp[\frac{i}{k} pz - \frac{i}{k} \frac{p^2}{2} \tau] \phi(z, \tau). \tag{4.16}$$

with the assumption that amplitude $\phi(z, \tau)$ is slowly varying. Using our ansatz we find

$$i\hbar \frac{\partial \psi}{\partial \tau} = [\frac{p^2}{2} \phi - i\hbar \frac{\partial \phi}{\partial \tau}] \exp[\frac{i}{k} pz - \frac{i}{k} \frac{p^2}{2} \tau], \tag{4.17}$$

and

$$-\frac{\hbar^2}{2} \frac{\partial^2 \psi}{\partial z^2} = [\frac{p^2}{2} \phi - i\hbar p \frac{\partial \phi}{\partial z} - \frac{\hbar^2}{2} \frac{\partial^2 \phi}{\partial z^2}] \exp[\frac{i}{k} pz - \frac{i}{k} \frac{p^2}{2} \tau]. \tag{4.18}$$

As $\phi(z, \tau)$ is slowly varying amplitude, we can neglect its second order derivative with respect to position space and get first order differential equation.

$$\frac{\partial \phi(z, \tau)}{\partial \tau} + \frac{\partial \phi(z, \tau)}{\partial z} = \frac{i}{k} \frac{\tilde{V}_0}{2} \cos[2z - \lambda \sin(\tau)] \phi(z, \tau), \tag{4.19}$$

The solution of this differential equation with initial condition $\phi(z_0, t_0)$ is given as

$$\phi(z, \tau) = \int_{m=-\infty}^{\infty} dz_0 Q(z, \tau, z_0, \tau_0) \phi(z_0, \tau_0), \tag{4.20}$$

where,

$$Q(z, \tau, z_0, \tau_0) = \delta[z - \bar{z}(z_0, \tau)] \exp[\frac{i}{k} \frac{\tilde{V}_0}{2} \int_0^{\tau_0} d\tau \cos[2\bar{z}(\tau) - \lambda \sin \tau]]. \tag{4.21}$$

here, $\bar{z}(\tau) = z_0 + p(\tau - \tau_0)$ is given by the solution of characteristic equation,

$$\frac{d\bar{z}}{d\tau} = p. \qquad (4.22)$$

The exact solution of Eq. (4.19) is

$$\phi(z, \tau) = \exp[\frac{i}{\hbar}\frac{\tilde{V}_0}{2}\int_0^\tau d\tau \cos(z + p(\bar{t} - \tau)\lambda\sin\tau)]\phi(z - p\tau, \tau = 0). \qquad (4.23)$$

In last equation, we set $\tau_0 = 0$ and eliminated $z_0$ using equations $\bar{z}(\tau) = z_0 + p(\tau - \tau_0)$ and $z = z_0 + p(\tau - \tau_0)$.

We consider a plane wave initially with momentum $p_0$ and $\phi(z, \tau = 0) = (2\pi\hbar)^{-1/2}$. We find the solution

$$\psi_{p_0}(z, \tau) = \frac{1}{\sqrt{2\pi\hbar}}\exp[\frac{i}{\hbar}(p_0 z - \frac{p_0^2}{2}\tau + \frac{\tilde{V}_0}{2}\int_0^\tau d\tau \cos(z - p_0(\tau - \bar{t})\lambda\sin\tau)]. \qquad (4.24)$$

By taking Fourier transformation, wave function $\psi_{p_0}(z, \tau)$ can be written in momentum space $\psi_{p_0}(p, \tau)$ (see Appendix-6). The momentum distribution $|\psi_{p_0}(p, \tau)|^2$ for an integer number $N$ of modulation periods can be written as

$$P(p = p_0 + m\hbar, \tau = 2\pi N) = J_m^2(\frac{\pi\tilde{V}_0}{\hbar}\frac{\sin(N\pi p_0)}{\sin(\pi p_0)}\mathbf{J}_{p_0}(-\lambda)) \qquad (4.25)$$

where, $\mathbf{J}_{p_0}$ is *Anger function* and momentum distribution is normalized as

$$\sum_{m=-\infty}^{\infty} P(p_0 + m\hbar) = \sum_{n=-\infty}^{\infty} J_n^2 = 1 \qquad (4.26)$$

For the case, when $p_0 = 0$, $\mathbf{J}_0(\lambda) = J_0(-\lambda) = J_0(\lambda)$ and $J_m(0) = \delta_{n,0}$. In this case momentum distribution modifies as

$$P(p = p_0 + m\hbar, \tau = 2\pi N) = \begin{cases} J_m^2(\frac{\tilde{V}_0\tau}{2\hbar}) & \text{if } \lambda = 0, \\ J_m^2(\frac{\tilde{V}_0 J_0(\lambda)\tau}{2\hbar}) & \text{if } \lambda \neq 0. \end{cases}$$

Hence in the absence of modulation ($\lambda = 0$), Bessel function envelope for momentum distribution is obtained. But in the presence of phase modulation, the interaction time is modified by the factor $J_0(\lambda)$, which is depending on modulation strength. In an other limiting case when $p_0$ is an integer, $\mathbf{J}_{p_0}(-\lambda) = J_{p_0}(-\lambda) = (-1)^{p_0} J_{p_0}(\lambda)$ where, momentum distribution is defined as

$$P(p = p_0 + m\bar{k}, \tau = 2\pi N) = \begin{cases} \delta_{m,0} & \text{if } \lambda = 0, \\ J_m^2(\dfrac{\tilde{V}_0 J_{p_0}(\lambda)\tau}{2\bar{k}}) & \text{if } \lambda \neq 0. \end{cases}$$

Now we shall calculate the width of the momentum distribution $\Delta p = (\bar{p^2} - \bar{p}^2)^{1/2}$

$$\Delta p^2 = \sum_{m=-\infty}^{\infty} (p_0 + m\bar{k})^2 P(p_0 + m\bar{k}) - [\sum_{m=-\infty}^{\infty} (p_0 + m\bar{k}) P(p_0 + m\bar{k})]^2,$$

$$= \sum_{m=-\infty}^{\infty} (p_0^2 + m^2\bar{k}^2) J_m^2(\eta) - p_0^2 [\sum_{m=-\infty}^{\infty} J_m^2(\eta)]^2, \tag{4.27}$$

here, we have used abbreviated Bessel argument $\eta$ instead of the argument in Eq. (4.25). As the summation rule gives

$$\sum_{m=-\infty}^{\infty} m^2 J_m^2(\eta) = \frac{\eta^2}{2}, \tag{4.28}$$

the expression for momentum dispersion is

$$\Delta p^2 = (p_0^2 + \bar{k}^2 \eta^2/2) - p_0^2. \tag{4.29}$$

Replacing the $\eta$ by original expression, we get

$$\Delta p(\tilde{V}_0, \lambda, \tau = 2N\pi) = \frac{\pi \tilde{V}_0}{\sqrt{2}} |\frac{\sin(N\pi p_0)}{\sin(\pi p_0)} \mathbf{J}_{p_0}(-\lambda)|. \tag{4.30}$$

The right hand side of the above expression is vanished for the zeros of $\mathbf{J}_{p_0}(-\lambda)$. Actually the dynamics of an atom with momentum $p_0$ are governed by a potential scaled by Anger function. For the zeros of Anger function, atoms undergo free evolution and $\Delta p$ remains constant.

In most experiments, the center of mass motion is approximated by Gaussian distribution. By considering Gaussian distribution as initial condition with $\Delta z$ and $\Delta p$ as initial dispersion in position and momentum respectively.

A normalized minimum uncertainty wave packet initially located at position $z_0$ with a mean momentum $p_0$ in position space is given as

$$\psi(z,0) = \frac{1}{(2\pi\Delta z^2)^{\frac{1}{4}}} \exp[-\frac{(z-z_0)^2}{4\Delta z^2} + \frac{i}{k}p_0 z], \qquad (4.31)$$

and in momentum representation

$$\psi(p,0) = \frac{1}{(2\pi\Delta p^2)^{\frac{1}{4}}} \exp[-\frac{(p-p_0)^2}{4\Delta p^2} - \frac{i}{k}pz_0]. \qquad (4.32)$$

The wave function, evaluated by performing minor algebra is

$$\psi(z,\tau) = \frac{1}{[2\pi(\Delta z(\tau))^2]^{1/4}}[\frac{2\Delta z^2 + ik\tau}{2\delta z^2 - ik\tau}]^{1/4}\exp[\frac{\frac{(z-z_0)^2}{4\Delta z^2} - \frac{i}{k}(z-z_0)p_0 + \frac{i}{k}\frac{p_0^2}{2}\tau}{1 + \frac{2i\tau}{k}\Delta p^2}]$$

$$\times \sum_{m=-\infty}^{\infty} b_m(\tau)e^{imz}, \qquad (4.33)$$

and position distribution

$$P(z,\tau) = \frac{1}{\sqrt{2\pi}\Delta\bar{z}} \exp[\frac{(z-z_0-p_0\tau)^2}{2(\Delta\bar{z})^2}], \qquad (4.34)$$

here,

$$\Delta\bar{z} = \Delta z[1 + (\frac{\Delta p}{\Delta z})^2]^{1/2}, \qquad (4.35)$$

gives the dispersion in position space with time.

## 4.4.2 Momentum distribution versus modulation

The momentum distribution of a minimum uncertainty Gaussian wave packet with initial width $\Delta p = 0.5$, mean momentum $p_0 = 0$ is numerically explored. The wave packet is evolved for fixed time, $\tau = 200$, and the results are shown in Fig-4.5 for lattice potentials $\tilde{V}_0 = 0.36,\ 2,\ 5.7,\ 8$. In Figs-(4.5)-(4.9), results are averaged values computed for different initial positions such that $x_0 = -\frac{\pi}{2},\ -\frac{\pi}{3},\ -\frac{\pi}{4},\ 0,\ \frac{\pi}{4},\ \frac{\pi}{3},\ \frac{\pi}{2}$, such that wave packet is always adiabatically loaded in the lowest energy band of respective lattice.

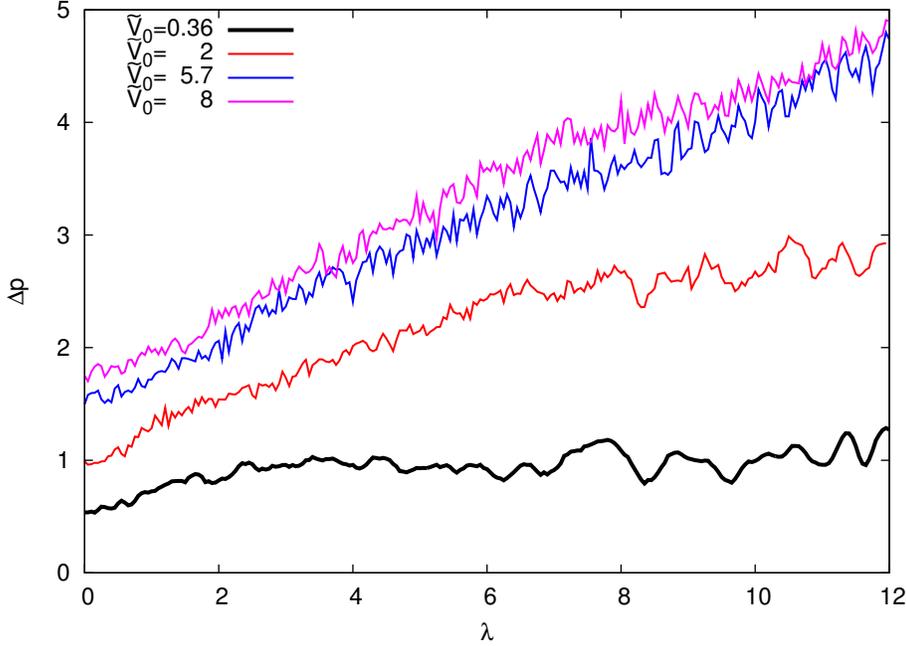

Figure 4.5: Momentum distribution vs modulation of wave packet adiabatically loaded in the lattice with mean initial momentum $p_0 = 0$. Other parameters are $\Delta p = 0.5$ and $\bar{k} = 1$

The momentum width for shallow optical lattice remain small and saturates earlier as compared to deep lattice. In deep lattice, stochastic region grow quickly near the separatrix, resonance islands become smaller and smaller and disappear. While, in the case of shallow lattice, chaotic regions develop slowly around the separatrix, resonance islands reappear as seen in Fig-4.6 for modulation amplitudes $\lambda = 7$, 10. For lattice potential where quantum mechanical localization length $\frac{\pi \tilde{V}_0^2}{2\lambda \bar{k}}$ is smaller than classical resonance boundary (see Eq. (4.15)), diffusion in momentum space is limited by quantum mechanical localization i.e., quantum mechanical suppression of classical diffusion. The $\Delta p$ curve in Fig-4.5 for $\tilde{V}_0 = 0.36$ saturates almost for the value $\lambda = 2.5$, in the case of $\tilde{V}_0 = 2$, it saturates at $\lambda = 6.5$ while, in the case of deep lattice it keeps on growing.

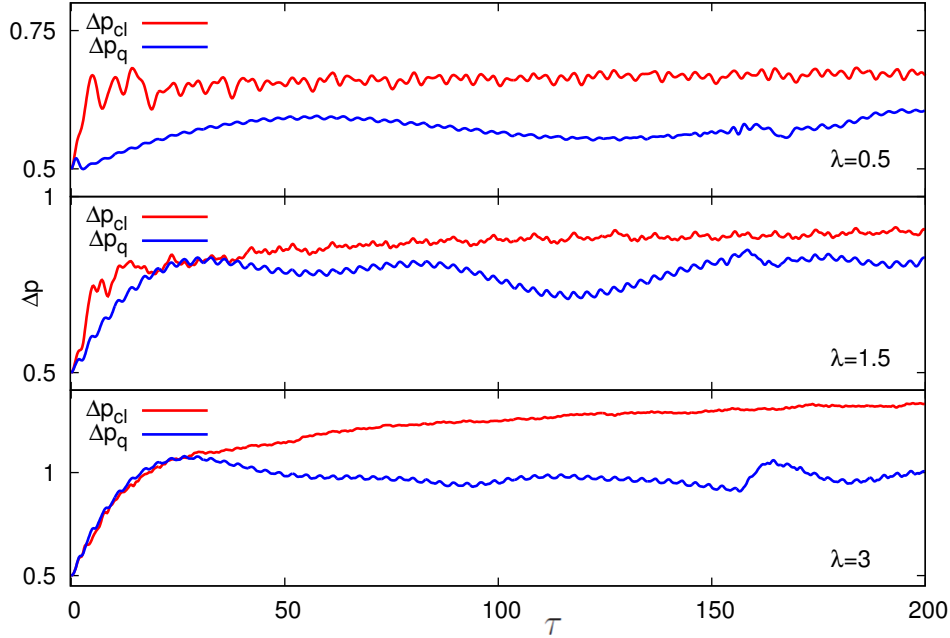

Figure 4.6: Classical and quantum momentum dispersion verses time for different modulation amplitudes for shallow lattice $\tilde{V}_0 = 0.36$. Other parameters are same as in Fig-4.5

### 4.4.3   Momentum distribution versus time

The momentum dispersion is studied for fixed value of modulation amplitudes versus time both for shallow and deep lattice potentials. In Figs-4.6 to 4.9, time evolution of momentum dispersion versus time is shown for effective lattice potentials $\tilde{V}_0 = 0.36$, 2, 5.7, 8, respectively for fixed modulation amplitudes $\lambda = 0.5$, 1.5, 3 for each case.

From time evolution of momentum dispersion in Figs-4.6 to 4.9, it is noted that, the quantum mechanical momentum dispersion $\Delta p_q$ is smaller than classical momentum dispersion $\Delta p_{cl}$ as diffusion is suppressed due to localization effect of modulation [Bardroff 1995] and as modulation increases, saturation value of classical dispersion also increases. This classical effect can be understand by studying the respective Poincaré surface of section Fig-4.1, where, it is seen that global KAM-boundary surrounding stochastic sea and

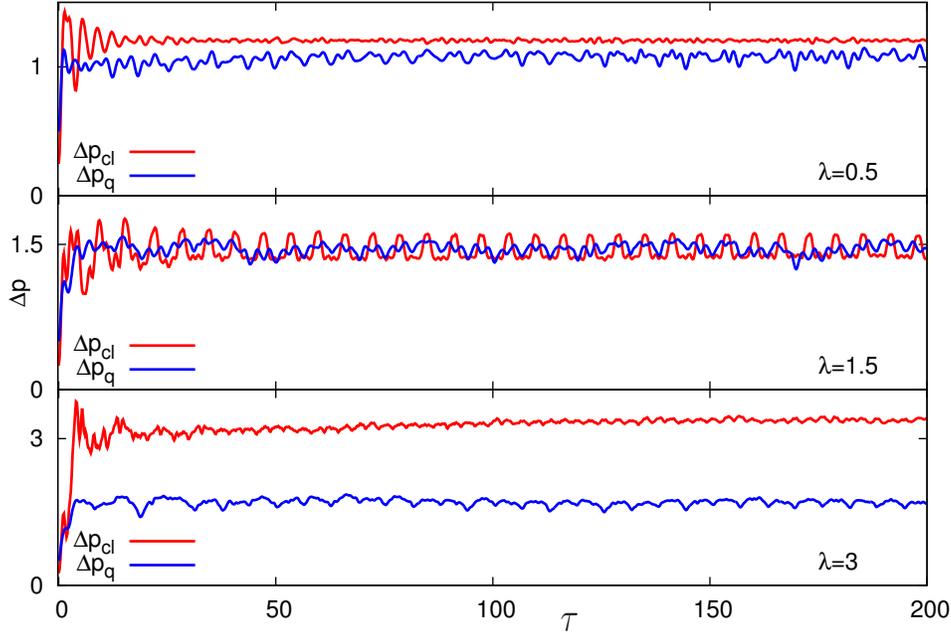

Figure 4.7: Classical and quantum momentum dispersion verses time for different modulation amplitudes for shallow lattice $\tilde{V}_0 = 2$. Other parameters are same as in Fig-4.5

regular regions is expanding with modulation and particles from the classical ensemble initially placed in stochastic region explore the the entire stochastic region but unable to escape from KAM-surfaces. In Fig-4.7, as modulation amplitude increases saturation value of $\Delta p_q$ increases but at a value $\lambda = 3$, diffusion is suppressed earlier and at the lower value than $\lambda = 1.5$. It is analytically shown that the quantum break time increases and localization length $\frac{\pi \tilde{V}_0^2}{2\lambda \hbar k}$ decreases as $\lambda$ increases. A similar behavior of classical and quantum dispersion in Fig-4.7 for $\lambda = 1.5$ can be explained by respective Poincaré surface of section Fig-4.2. It is seen for stated initial condition a major part of wave packet falls in classical resonance and classical dispersion is suppressed. As lattice potential is increased the saturation value for $\Delta p_q$ for fixed value of modulation also increases. When $\tilde{V}_0 < 1$ (see Fig-4.6), there are two mechanism of suppression of diffusion: i) The classical resonance boundary is larger than quantum localization length $\frac{\pi \tilde{V}_0^2}{2\lambda \hbar k}$ and diffusion is limited by quantum

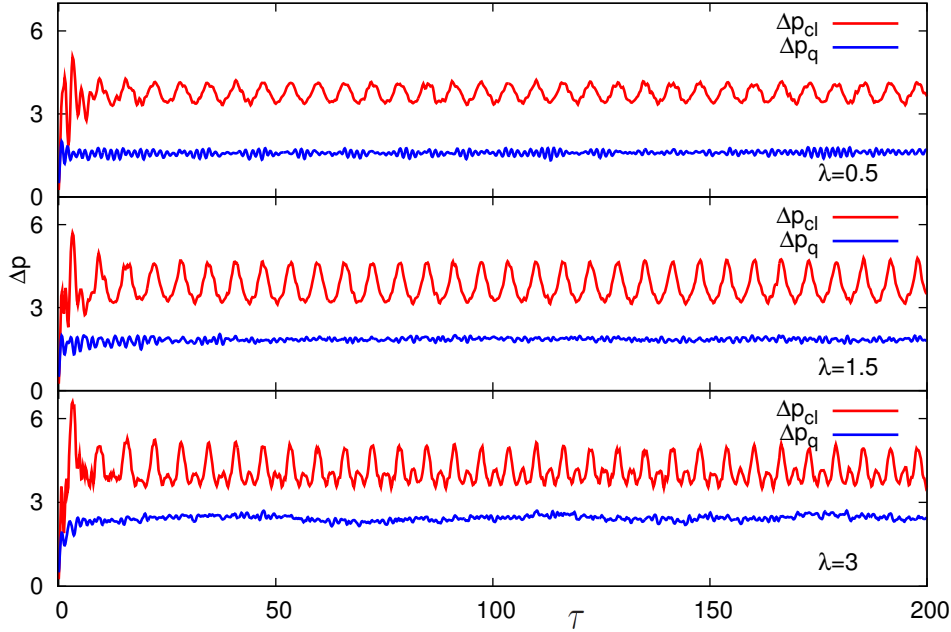

Figure 4.8: Classical and quantum momentum dispersion verses time for different modulation amplitudes for the lattice $\tilde{V}_0 = 5.7$. Other parameters are same as in Fig-4.5

localization in this case and $\Delta p_q$ increases as $\lambda$ increases; ii) The classical resonance boundary is smaller than quantum localization length, dispersion in momentum space increases with modulation. Quantum localization effect dictates $\Delta p$ to increase following localization length for small values of lattice potential but as soon as the classical boundary becomes smaller than quantum localization length, dispersion grows only due to resonance broadening with the growth in $\tilde{V}_0$.

## 4.5  Floquet Theory of Nonlinear Resonances

In this section, we extend the Floquet formalism discussed in chapter-1, to nonlinear resonances and derive expressions for energy spectrum. In the periodically driven potentials energy is no more a constant of motion. For the reason we solve the time dependent Schrödinger equation by using secular

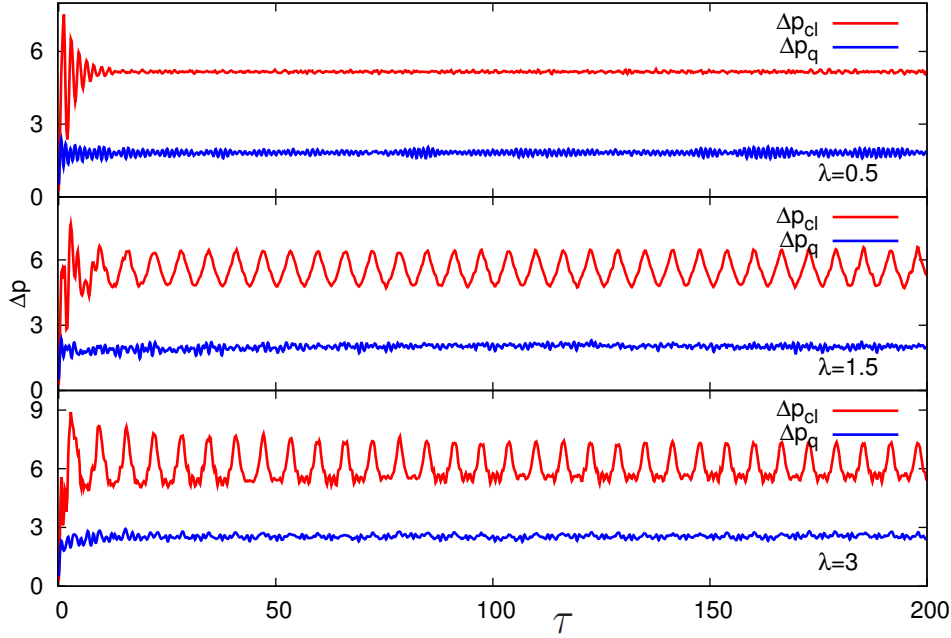

Figure 4.9: Classical and quantum momentum dispersion verses time for different modulation amplitudes for deep lattice $\tilde{V}_0 = 8$. Other parameters are same as in Fig-4.5

perturbation approximation as suggested by Max Born [Born 1960]. Therefore, the solution is obtained by averaging over rapidly changing variables. This leads us to find out a partial solution of the periodically driven systems. As a result, we find quasi energy eigen-states and the quasi eigen energies of the dynamical system for non-linear resonances.

In order to study the quantum nonlinear resonances of the TDS, we consider the scaled Hamiltonian,

$$H = H_0 + \lambda V(x) \sin(\tau). \tag{4.36}$$

Classically, this Hamiltonian is characterize by a single dimensionless parameter $\lambda$. In quantum mechanical solutions, another scaled parameter, scaling Plank's constant, $\hbar$ appears. The parameter $\lambda$ controls the degree of non-integrability while, scaled Plank's constant controls the scale at which its quantum mechanical counterpart can resolve phase space structures.

The solution of the Schrödinger equation corresponding to the Hamiltonian (4.36) in the vicinity of $N^{th}$ resonance can be written in the form [Saif 2001; Flatté 1996; Berman 1977],

$$|\psi(\tau)\rangle = \sum_n C_n(\tau)|n\rangle \exp\{-i[E_{\bar{n}} + (n - \bar{n})\frac{\hbar}{N}]\frac{\tau}{\hbar}\}, \qquad (4.37)$$

here, $E_{\bar{n}}$ is the mean energy, $C_n(\tau)$ is time dependent probability amplitude, $\bar{n}$ is mean quantum number, $\hbar$ is scaled Plank's constant and $|n\rangle$ are eigen states of undriven system. On substituting (4.37) in the time dependent Schrödinger equation, we find that the probability amplitude $C_n(\tau)$, changes with time following the equation, $i\hbar\dot{C}_n(\tau) = [E_n - E_{\bar{n}} - (n - \bar{n})\frac{\hbar}{N}]C_n(\tau) + \frac{\lambda V}{2i}(C_{n+N} - C_{n-N})$. Where, $V = V_{n-N} \approx V_{n+N}$ are off-diagonal matrix elements and are approximately constant near the potential minima for tight-binding approximations.

We take the initial excitation such that it is narrowly peaked around the mean value, $\bar{n}$. For the reason, we take slow variations in the energy, $E_n$, around the $\bar{n}$ in a nonlinear resonance, and expand it up to second order in Taylor expansion. Thus, the Schrödinger equation for the probability amplitudes, $C_n(\tau)$, is

$$i\hbar\dot{C}_n = \hbar(n - \bar{n})(\omega - \frac{1}{N})C_n(\tau) + \frac{1}{2}\hbar(n - \bar{n})^2\zeta C_n(\tau)$$
$$+ \frac{\lambda V}{2i}(C_{n+N} - C_{n-N}). \qquad (4.38)$$

In (4.38), fast oscillating terms are averaged out and only the resonant ones are kept. Parameters $\omega = \frac{\partial E_n}{\hbar \partial n}|_{n=\bar{n}}$, and $\zeta = \frac{\partial^2 E_n}{\hbar^2 \partial n^2}|_{n=\bar{n}}$ are the frequency and non-linearity of the time independent system respectively.

We introduce the Fourier representation for $C_n(\tau)$ as,

$$C_n(\tau) = \frac{1}{2N\pi}\int_0^{2N\pi} g(\theta, \tau)e^{-i(n-\bar{n})\theta/N}d\theta, \qquad (4.39)$$

which helps us to express (4.38) as the Schrödinger equation for $g(\theta, \tau)$, such that $i\hbar\dot{g}(\theta, \tau) = H(\theta)g(\theta, \tau)$. Here, the Hamiltonian $H(\theta)$ is given as, $H(\theta) = -\frac{N^2\hbar^2\zeta}{2}\frac{\partial^2}{\partial\theta^2} - iN\hbar(\omega - \frac{1}{N})\frac{\partial}{\partial\theta} - \lambda V\sin\theta$. In order to obtain this equation, we consider the function $g(\theta, \tau)$ as $2N\pi$ periodic, in $\theta$ coordinate.

Due to the time-independent behavior of the Hamiltonian, we write the time evolution of $g(\theta, \tau)$, as $g(\theta, \tau) = \tilde{g}(\theta) e^{\frac{-i\mathcal{E}\tau}{\hbar}}$. Therefore, Schrödinger equation for $g(\theta, t)$ reduces to the standard Mathieu equation [Saif 2006],

$$[\frac{\partial^2}{\partial \theta^2} + a - 2q \cos 2\theta] \tilde{g}(\theta) = 0, \qquad (4.40)$$

here, $\tilde{g} = \chi(z) \exp\left(-i2(N\omega - 1)z/N^2\zeta\hbar\right)$ and $\theta = 2z + \pi/2$. The Mathieu characteristic parameters [Abramowitz 1970; McLachlan 1947] are

$$a = \frac{8}{N^2\hbar^2\zeta}[\frac{(N\omega - 1)^2}{2N\zeta} + \mathcal{E}_{\mu, \nu}], \qquad (4.41)$$

$$\text{and} \qquad q = \frac{4\lambda V}{N^2\hbar^2\zeta}. \qquad (4.42)$$

The $\pi$-periodic solutions of Eq. (4.40) correspond to even functions of the Mathieu equation whose corresponding eigenvalues are real [Abramowitz 1970]. These solutions are defined by Floquet states, *i.e.,* $\tilde{g}(\theta) = e^{i\mu\theta} P_\mu(\theta)$, where, $P_\mu(\theta) = P_\mu(\theta + \pi)$, and $\mu$ is the characteristic exponent. In order to have $\pi$-periodic solutions in $\tilde{g}(\theta)$, we require $\mu$ to be defined as $\mu = \mu(j) = 2j/N$, where, $j = 0, 1, 2, ...., N - 1$.

The allowed values of $\mu(j)$ can exist as a characteristic exponent of solution to the Mathieu equation for discrete $\nu$ (which takes integer values) only for certain value $a_\nu(\mu(j), q)$, when $q$ is fixed. Hence, with the help of Eq. (4.41), we obtain the values of unknown $\mathcal{E}$. Therefore, we may express the quasi energy of the system as [Breuer and Holthaus 1991]

$$\mathcal{E}_{\mu, \nu} = \left[\frac{N^2\hbar^2\zeta}{8} a_\nu(\mu(j), q) + \hbar \tilde{\alpha} j\right] \mod \hbar\omega, \qquad (4.43)$$

where, the index $\nu$ takes the definition $\nu = \frac{2(n - \bar{n})}{N}$, and $\tilde{\alpha}$, defines the winding number.

## 4.6  Quantum Recurrence Times Based on Quasi Energy Spectrum

In this section we discuss two cases correspond to nonlinear resonances. We consider weakly coupled $q < 1$ or strongly coupled $q \gg 1$ potentials. In $q < 1$ situation, for large $\nu$ [Saif 2001] and in $q \gg 1$ situation for small $\nu$, near the center of resonance matrix elements are constant. In the following discussion, we analyze the wave packet dynamics in these regimes and show their parametric dependencies.

The time scales, $T^{(j)}$ at which recurrences of an initially well-localized wave packet occur depend on the quasi-energy of the respective system, The recurrence times are obtained as, $T^{(j)} = 2\pi/\Omega^{(j)}$, hence, the values of $j$ as $j = 1, 2, 3...$, correspond to, respectively, classical, quantum, super, and higher order revival times. With the help of Eq. (4.41) and Eq. (4.43), we obtain the frequencies $\Omega^{(j)}$ as

$$
\begin{aligned}
\Omega^{(1)} &= \frac{1}{k} \left\{ \frac{\partial \mathcal{E}_{\mu,\nu}}{\partial \mu} + \frac{\partial \mathcal{E}_{\mu,\nu}}{\partial \nu} \right\}, \\
\Omega^{(2)} &= \frac{1}{2! k^2} \left\{ \frac{\partial^2 \mathcal{E}_{\mu,\nu}}{\partial \mu^2} + \frac{2\partial^2 \mathcal{E}_{\mu,\nu}}{\partial \mu \partial \nu} + \frac{\partial^2 \mathcal{E}_{\mu,\nu}}{\partial \nu^2} \right\}, \\
\Omega^{(3)} &= \frac{1}{3! k^3} \left\{ \frac{\partial^3 \mathcal{E}_{\mu,\nu}}{\partial \mu^3} + 3\frac{\partial^3 \mathcal{E}_{\mu,\nu}}{\partial \mu^2 \partial \nu} + 3\frac{\partial^3 \mathcal{E}_{\mu,\nu}}{\partial \mu \partial \nu^2} + \frac{\partial^3 \mathcal{E}_{\mu,\nu}}{\partial \nu^3} \right\}.
\end{aligned}
\tag{4.44}
$$

After a few mathematical steps Eq. (4.44) reduce to

$$
\begin{aligned}
\Omega^{(1)} &= \frac{1}{k} \left\{ \frac{\partial \mathcal{E}_{\mu,\nu}}{\partial \nu} + \alpha k \right\}, \\
\Omega^{(2)} &= \frac{1}{2! k^2} \left\{ \frac{\partial^2 \mathcal{E}_{\mu,\nu}}{\partial \nu^2} + \alpha k \frac{2\partial \mathcal{E}_{\mu,\nu}}{\partial \nu} \right\}, \\
\Omega^{(3)} &= \frac{1}{3! k^3} \left\{ \frac{\partial^3 \mathcal{E}_{\mu,\nu}}{\partial \nu^3} + 3\alpha k \frac{\partial^2 \mathcal{E}_{\mu,\nu}}{\partial \nu^2} \right\}.
\end{aligned}
\tag{4.45}
$$

which lead to classical period, $T^{(1)} = T_\lambda^{(cl)}$, quantum revival time, $T^{(2)} = T_\lambda^{(rev)}$, and super revival time, $T^{(3)} = T_\lambda^{(spr)}$, when calculated at the mean values.

## Delicate Dynamical Recurrences

The condition, $q < 1$, may be satisfied in the presence of weak perturbation due to external periodic force (and/or), for large nonlinearity and/or for large effective Plank's constant [Saif 2001; Saif 2005a; Saif 2005; Iqbal 2006]. The Mathieu characteristic parameters, $a_\nu$ and $b_\nu$ are given [Abramowitz 1970], as

$$a_\nu \simeq b_\nu = \nu^2 + \frac{q^2}{2(\nu^2 - 1)} + \dots \text{ for } \nu \geqslant 5 \qquad (4.46)$$

The above expressions are not limited to integral value of $\nu$ and are very good approximations when $\nu$ is of the form, $m + \frac{1}{2}$. In case of integral value of $\nu$, i.e., $\nu = m$, the series holds only up to the terms not involving $\nu^2 - m^2$ in the denominator.

The energy spectrum for weakly modulated periodic potentials [Saif 2006] can be defined using Eq. (4.43) and Eq. (4.46). The relations obtained for classical period, quantum revival time and super revival time, in the presence of small perturbation for primary resonance $N = 1$, index $j$ takes the value, $j = 0$, and time scale are simplified as

$$T_\lambda^{(cl)} = (1 - M^{(cl)})T_0^{(cl)}\Delta, , \qquad (4.47)$$

$$T_\lambda^{(rev)} = (1 - M^{(rev)})T_0^{(rev)}, , \qquad (4.48)$$

$$T_\lambda^{(spr)} = \frac{\pi\omega^2}{2\lambda V\zeta\Delta^2}\frac{(1 - \mu_1^2)^4}{\mu_1}, \qquad (4.49)$$

here, $T_0^{(cl)} = \frac{2\pi}{\omega}$ is classical period and $T_0^{(rev)} = 2\pi/(\frac{1}{2!}\hbar\zeta)$ is quantum revival time in the absence of external modulation. Furthermore, $\Delta = (1 - \frac{\omega_N}{\omega})$. The modification factors are given as

$$M^{(cl)} = -\frac{1}{2}(\frac{\lambda V\zeta\Delta^2}{\omega^2})^2\frac{1}{(1 - \mu_1^2)^2},$$

and

$$M^{(rev)} = \frac{1}{2}(\frac{\lambda V\zeta\Delta^2}{\omega^2})^2\frac{3 + \mu_1^2}{(1 - \mu_1^2)^3},$$

here, $\mu_1 = \frac{\hbar\zeta\Delta}{2\omega}$ is re-scaled non-linearity.

**Robust Dynamical Recurrences**

The condition, $q \gg 1$, may be satisfied in the presence of large amplitude, $\lambda$, of external modulation by periodic force in a dynamical system. In addition, we may get the regime by considering a system with very small linearity i.e., $\zeta \approx 0$ and/or by taking effective Plank's constant $\hbar$ be very small. Quasi energies for nonlinear resonance are defined in terms of Mathieu characteristic parameters, which are

$$a_\nu \approx b_{\nu+1} \approx -2q + 2s\sqrt{q} - \frac{s^2+1}{2^3} - \frac{s^3+3s}{2^7\sqrt{q}} - ..., \tag{4.50}$$

where, $s = 2\nu + 1$.

The energy spectrum of nonlinear resonances for strongly modulated periodic potentials can be defined using Eq. (4.43) and Eq. (4.50). Here, in the deep potential limit, the band width is,

$$b_{\nu+1} - a_\nu \simeq \frac{2^{4\nu+5}\sqrt{\frac{2}{\pi}}q^{\frac{\nu}{2}+\frac{3}{4}}\exp(-4\sqrt{q})}{\nu!}. \tag{4.51}$$

Keeping lower order terms in $s$, in Eq. (4.50), we get quasi energy spectrum for nonlinear resonances as $(\nu + \frac{1}{2})\hbar\omega_h$, which resembles to the harmonic oscillator energy spectrum for $\omega_h = 2\sqrt{V_0}$.

Similarly the relations for classical period, quantum revival time and super revival time for strongly driven case are obtained as,

$$T_\lambda^{(cl)} = \frac{4\pi}{N^2k\zeta\sqrt{q}\left\{1 - \frac{2v+1}{8\sqrt{q}} - \frac{3(2v+1)^2+3}{2^8q}\right\} + 2\alpha}, \tag{4.52}$$

$$T_\lambda^{(rev)} = \frac{32\pi}{N^2\zeta\left\{1 + \frac{3(2v+1)}{16\sqrt{q}} + 8\alpha k\sqrt{q} - (2v+1)\ \alpha k\right\}}, \tag{4.53}$$

$$T_\lambda^{(spr)} = \frac{32\pi}{N^2\zeta\alpha}[1 - \frac{1}{8\alpha k\sqrt{q}}\{1 + \frac{3\alpha k(2v+1)}{2}\}]. \tag{4.54}$$

For primary resonance $N = 1$, index $j$ takes the value, $j = 0$, and time scale are simplified as

$$T_\lambda^{(cl)} = T_0^{(cl)}\frac{\Delta}{8\mu_1}\left[1 - \frac{(4+\mu_1)\sqrt{\zeta}}{8\mu_1\sqrt{q}} - \frac{(4+\mu_1)^2\zeta}{(8\mu_1)^2q}\right], \tag{4.55}$$

$$T_\lambda^{(rev)} = T_0^{(rev)}2\left[1 - \frac{3(\mu_1+4)}{16\mu_1\sqrt{q}} + \frac{9(4+\mu_1)^2}{(16\mu_1)^2q}\right], \tag{4.56}$$

and

$$T_\lambda^{(spr)} = \frac{2^5\pi\sqrt{q}}{k\zeta}. \tag{4.57}$$

# 4.7 Revival Times of Cold Atoms in Driven Optical Lattices

We consider ultracold atoms in standing wave field with phase modulation due to acousto-optic modulator. The atoms in the presence of phase modulation experience an external force, thus exhibit dispersion both in classical and quantum domain. In quantum dynamics atoms experience an additional control due to effective Plank's constant. Here, we report in long time evolution, material wave packet display quantum recurrence phenomena.

As time scales of driven optical lattice are expressed in terms of undriven system parameters, for the reason, we investigate, classical period, quantum revival time and super revival time scales of undriven optical lattice [Ayub et al 2009; Drese et al 1997; Dyrting 1993].

In the case of undriven shallow optical potential classical period is

$$T_0^{(cl)} = \{1 + \frac{q_0^2}{2(\bar{n}^2 - 1)^2}\}\frac{\pi}{\bar{n}}, \qquad (4.58)$$

where, $q_0 = \frac{V_0}{4\omega_r}$ is rescaled potential depth in units of recoil energy. The quantum revival time is

$$T_0^{(rev)} = 2\pi\{1 - \frac{q_0^2}{2}\frac{(3\bar{n}^2 + 1)}{(\bar{n}^2 - 1)^3}\}, \qquad (4.59)$$

and super revival time is

$$T_0^{(spr)} = \frac{\pi(\bar{n}^2 - 1)^4}{q_0^2\bar{n}(\bar{n}^2 + 1)}, \qquad (4.60)$$

In shallow lattice potential limit, i.e., $q_0 \lesssim 1$, neglecting the higher order terms in $q_0$ the classical frequency, $\omega = 2\bar{n}\{1 - \frac{q_0^2}{2(\bar{n}^2 - 1)^2}\}$ and non-linearity, $\zeta = 2 + \frac{q_0^2}{2}\frac{3\bar{n}^2 + 1}{(\bar{n}^2 - 1)^3}$. On the other hand for deep optical lattice, the classical time period is

$$T_0^{(cl)} = \frac{\pi}{2\sqrt{q_0}}\{1 + \frac{s}{8\sqrt{q_0}} + \frac{3(s^2 + 1)}{2^8 q_0}\}, \qquad (4.61)$$

where, $s = 2\bar{n} + 1$. The quantum revival time

$$T_0^{(rev)} = 4\pi(1 - \frac{3s}{16q_0}), \qquad (4.62)$$

and super revival time is

$$T_0^{(spr)} = 32\pi\sqrt{q_0}. \tag{4.63}$$

In deep optical potential limiting case, i.e., $q_0 >> 1$, the classical frequency is $\omega = 4(\sqrt{q_0} - \frac{2\bar{n}+1}{8})$ and non-linearity is $\zeta = |-1 - \frac{3(2\bar{n}+1)}{2^4\sqrt{q_0}}|$.

### 4.7.1 Delicate dynamical recurrences

Now for weakly driven shallow or deep lattice $q \lesssim 1$. Time scales for primary resonance $N = 1$ are [Ayub et al 2011]

$$T_\lambda^{(cl)} = T_0^{(cl)} \left[ 1 + \frac{q^2}{2} \frac{1}{\{4(l+\beta)^2 - 1\}^2} \right] \Delta, \tag{4.64}$$

$$T_\lambda^{(rev)} = T_0^{(rev)} \left[ 1 - \frac{q^2}{2} \frac{12(l+\beta)^2 + 1}{\{4(l+\beta)^2 - 1\}^3} \right], \tag{4.65}$$

$$\text{and} \quad T_\lambda^{(spr)} = \frac{\pi\{4(l+\beta)^2 - 1\}^4}{2\zeta k q^2(l+\beta)\{4(l+\beta)^2 + 1\}}. \tag{4.66}$$

Where, $T_0^{(cl)} = \frac{2\pi}{\omega(l+\beta)}$ is the classical time period, $T_0^{(rev)} = \frac{4\pi}{k\zeta}$ is quantum revival time for unmodulated system and $\beta = \frac{2(N\omega-1)}{N^2 k\zeta}$. In this case $\omega$ and $\zeta$ are frequency and non-linearity of respective undriven lattice. Here, time scales for weakly driven shallow lattice and weakly driven deep lattice look similar but there behavior is not similar as parameters $\omega$ and $\zeta$ are different due different energy spectrum corresponding to undriven lattices.

Behavior of classical periods, quantum revivals and super revivals of matter waves in modulated optical crystal in nonlinear resonances versus modulation is shown in Fig-4.10. In each plot of this figure, left vertical axis shows the time scales when shallow optical lattice is modulated and right axis shows the time scales when deep optical lattice is modulated. The upper row of Fig-4.10 represents the time scales for small $q$ values i.e., delicate dynamical recurrences as a function of modulation $\lambda$, while, lower row represents the time scales when $q \gg 1$, as a function of modulation $\lambda$, i.e., robust dynamical recurrences. Here, left column shows the results related to the classical periods. Quantum revival times are plotted in middle column, while, right column shows super revival times.

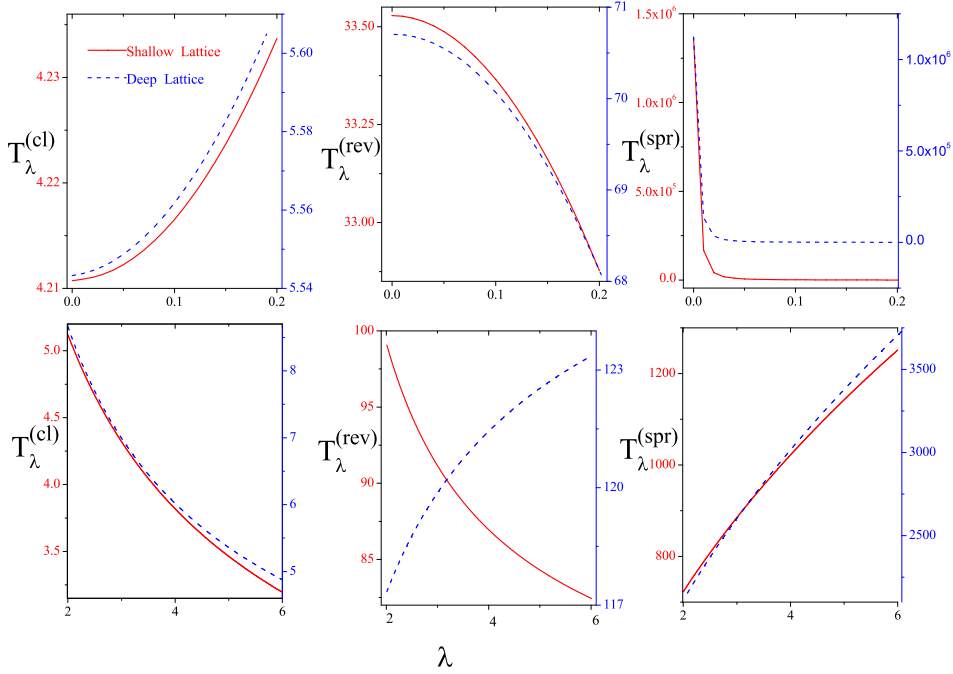

Figure 4.10: Left panel: Classical time period versus $\lambda$ for weak external modulation (a) and for strong external modulation (b). Middle panel: Quantum revival time versus $\lambda$ for weak external modulation (c) and for strong external modulation (d). Right panel: Super revival time versus $\lambda$ for weak modulation (e) and for strong modulation (f). In this figure, for deep lattice $V_0 = \tilde{1}6E_r$, for shallow lattice $\tilde{V}_0 = 2E_r$ and $\tilde{k} = 0.5$.

We note that when optical lattice is perturbed by weak periodic force, the classical period increases with modulation, as given in Eq. (4.64). Classical period for weakly driven shallow lattice potential changes slowly as compared to weakly driven deep lattice potential as shown in Fig-4.10(a). Quantum revival time in nonlinear resonances versus modulation is shown in middle column of Fig-4.10. For delicate dynamical recurrences (Fig-4.10(c)), the quantum revival time decreases as modulation increases. The behavior of quantum revival time is given by Eq. (4.65) for delicate dynamical recurrence. For weakly driven shallow optical lattice or weakly driven deep lattice, qual-itative and quantitative behavior of revival time is almost similar. Classical period and quantum revival time for delicate dynamical recurrences show good numerical and analytical resemblance for the system with our previous work [Saif 2005a; Iqbal 2006; Saif 2006].

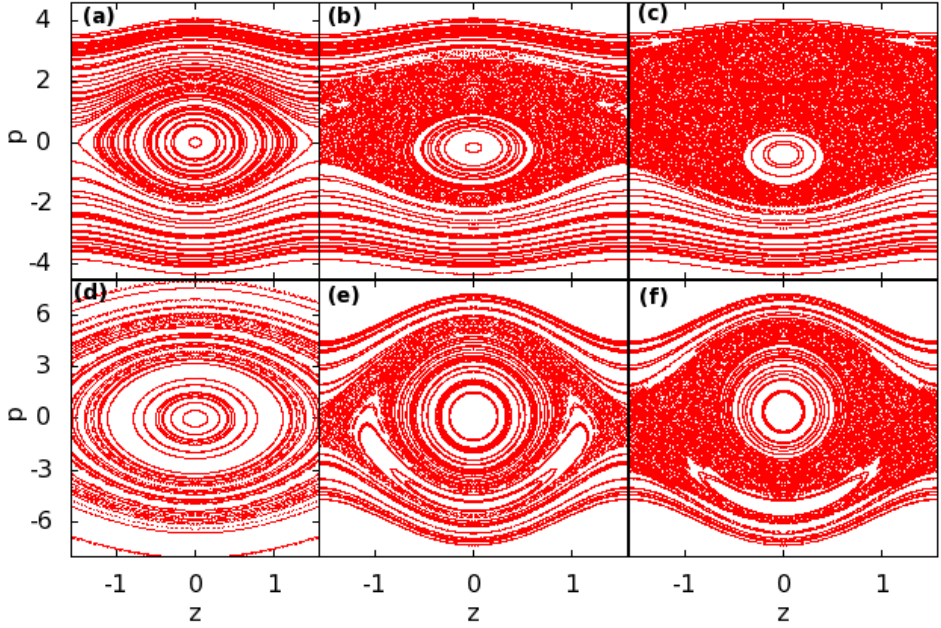

Figure 4.11: Poincaré phase space for different modulation strengths and lattice potentials. Upper row: (a) $\lambda = 0$, (b) , 0.5, (c) 1.5 and $\tilde{V}_0 = 2E_r$; Lower row: (d) $\lambda = 0$, (e) 0.5, (f) 1.5 and $\tilde{V}_0 = 16E_r$.

To have an idea about the classical dynamics of the system, we plot the Poincaré surface of section for shallow optical lattice ($\tilde{V}_0 = 2E_r$) with modulation strengths $\lambda = 0, 0.5, 1$ and for deep lattice ($\tilde{V}_0 = 16E_r$) with modulation strengths $\lambda = 0, 0.5, 1.5$ as shown in Fig-4.11. From phase space plot, we see the appearance of 1:1 resonance for $\lambda > 0$. This resonance emerges when the time period of external force matches with period of unperturbed system. One effect of an external modulation is the development of stochastic region near the separatrix. As modulation is increased, while the frequency is fixed, the size of stochastic region increases at the cost of regular region.

In order to observe the numerical dynamics of a quantum particle inside a resonance, we evolve a well localized Gaussian wave packet in the driven optical lattice. Numerical results are obtained by placing a wave packet in a primary resonance with $N = 1$. This can be realized in experimental set up of Mark Raizen at Austin, Texas. The authors [Moore 1994] worked with sodium atoms to observe the quantum mechanical suppression of classical diffusive motion and employed the $(3S_{\frac{1}{2}}, F = 2) \rightarrow (3P_{\frac{3}{2}}, F = 3)$ transition at $589nm$ , with $\omega_0/2\pi = 5.09 \times 10^{14} Hz$. The detuning was $\delta_L/2\pi = 5.4 \times 10^9 Hz$. The recoil frequency of sodium atoms was $\omega_r/2\pi = 25kHz$ for selected laser frequency and modulation frequency was chosen $\omega_m/2\pi = 1.3MHz$, whereas, the other parameters were $\bar{k} = 0.038$ and $q = 55$ (or $\tilde{V}_0 = 0.16$). In the other experiment [Dahan et al 1996], Bloch oscillations of ultracold atoms were observed with the $6S_{\frac{1}{2}} \rightarrow 6P_{\frac{3}{2}}$ transition in cesium atoms setting $\lambda_L = 852nm$ and $\omega_0/2\pi = 3.52 \times 10^{14} Hz$. The detuning was $\delta_L/2\pi = 3 \times 10^{10} Hz$, with $q$ up to 1.5. The recoil frequency was $\omega_r/2\pi = 2.07kHz$, so that a driving frequency $\omega_m/2\pi = 10^3 Hz$, three orders of magnitude lower than in the former experiment, gives $\bar{k} \approx 4$, taking the dynamics to the deep quantum regime. We numerically investigate the validity of our results with known theoretical and experimental [Moore 1994] results with dimensionless rescaled Plank's constant $\bar{k} = 0.16$, $\tilde{V}_0 = 0.36$, which is a case of strong modulation to a deep optical lattice of potential depth $V_0 = 28.13E_r$. The wave packet is initially well localized in such a way that localization length is less than or order of lattice spacing.

Fig-4.12 shows spatio-temporal evolution of an initially well localized wave packet in a lattice potential well. Fig-4.12-(a) is spatio-temporal dynamics of atomic wave packet for $\lambda = 0.05$, while Fig-4.12-(b) presents the case, for external modulation $\lambda = 0.1$. Spatiotemporal evolution of wave packet in optical lattice shows that wave packet diffuses to the neighboring lattice sites by tunneling and splits into small wavelets. Later, these wavelets constructively interfere and wave packet revival takes place. It is clear that as the modulation increases, revival time changes as shown in Fig-4.10 which is due to a change in interference pattern and confirms analytical results as discussed above.

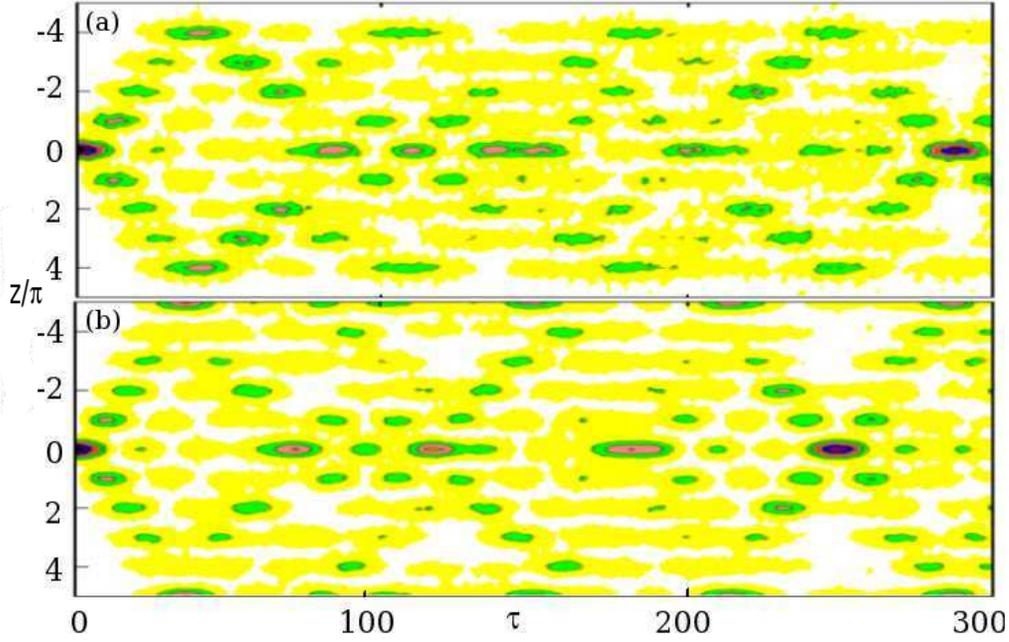

Figure 4.12: Spatio-temporal behavior of atomic wave packet for $\lambda = 0.05$, (a) and $\lambda = 0.1$ (b). Other parameters are $\tilde{V}_0 = 2$ $\Delta p = 0.5$ and $k = 0.5$. Dark regions show represent the maximum probability to find the particle.

### 4.7.2 Robust dynamical recurrences

On the other hand, for strongly driven optical lattice $q \gg 1$. Time scales for the atomic wave packet for primary resonance with $N = 1$ are given as [Ayub et al 2011]

$$T_\lambda^{(cl)} = \frac{2\pi}{k\zeta\{\sqrt{q} - \dfrac{4(l+\beta)+1}{8}\}}, \tag{4.67}$$

$$T_\lambda^{(rev)} = \frac{8\pi}{k\zeta}\left[1 - \frac{3\{4(l+\beta)+1\}}{16\sqrt{q}}\right], \tag{4.68}$$

and

$$T_\lambda^{(spr)} = \frac{32\pi \sqrt{q}}{k\zeta}. \tag{4.69}$$

In case of strongly driven lattice, when external modulation frequency is close to the harmonic frequency, matrix elements, $V$ can be approximated by those of harmonic oscillator and Mathieu parameter $q$ can be approximated as $q \approx \frac{4\sqrt{n+1}\lambda}{q_0^{\frac{1}{4}}k^2\zeta}$ [Drese et al 1997] . Under this approximation time scales are

$$T_\lambda^{(cl)} = \frac{16\pi q_0^{\frac{1}{8}}/\sqrt{\zeta}}{16(\bar{n}+1)^{\frac{1}{4}}\sqrt{\lambda} - \{4(l+\beta)+1\}q_0^{\frac{1}{8}}k\sqrt{\zeta}}, \qquad (4.70)$$

$$T_\lambda^{(rev)} = \frac{8\pi}{k\zeta}\left[1 - \frac{3\{4(l+\beta)+1\}q_0^{\frac{1}{8}}k\sqrt{\zeta}}{32(\bar{n}+1)^{\frac{1}{4}}\sqrt{\lambda}}\right],$$

$$\text{and } T_\lambda^{(spr)} = \frac{64\pi(\bar{n}+1)^{\frac{1}{4}}\sqrt{\lambda}}{k^2\zeta^{\frac{3}{2}}q_0^{\frac{1}{8}}}. \qquad (4.71)$$

When lattice is strongly modulated by an external periodic force, the classical period decreases as modulation increases. Classical period for strongly driven optical lattice is given by Eq. (4.67). The behavior of classical period for strongly driven lattice versus modulation is qualitatively of the same order for both strongly driven shallow lattice and strongly driven deep lattice as shown in Fig-4.10(b). The behavior of classical period in strongly driven lattice case is understandable as strong modulation influence more energy bands of undriven lattice to follow the external frequency and near the center of nonlinear resonance the energy spectrum is almost linear, as can be inferred from Eq. (4.43) and Eq. (4.50) with assumptions $q \gg 1$ and $l$ is small, i.e., wave packet is placed near the center of resonance.

Quantum revival time in nonlinear resonances versus modulation is shown in middle column of Fig-4.10. For robust dynamical recurrences, the behavior of quantum revival time is given by Eq. (4.68). The qualitative behavior of revival time for strongly driven shallow lattice is different from that of strongly driven deep lattice Fig-4.10(d), as in the later case, change in revival time is almost one order of magnitude larger than the former case, for equal changes in modulation. Here, revival time differently depends on $\lambda$ for shallow and deep lattice in the case of strong modulation. The energy spectrum of deep and shallow optical lattices are different i.e. frequencies associated with the lattice spectrum are different for the two cases. For deep lattice case energy spectrum (band structure) near the bottom is almost equally spaced and spacing decreases as band index increases (shown in Fig-2.2). In addition, band gaps are wider near the bottom of the lattice potential compared to shallow lattice and $\beta = \frac{2(N\omega-1)}{N^2 \bar{k} \zeta}$ is positive near the center of primary resonance ($N = 1$). Here, $\omega$ is transition frequency between the two lowest energy levels. The contribution from second term in Eq. (4.68) decreases as $\lambda$ or $q$ increases and revival time increases. This can also be explained as follow. When an external modulation of some fixed frequency $\omega_m$ is applied and modulation is strong enough then lattice frequencies which are slightly different from modulation frequency, follow modulation frequency and spectrum near the nonlinear resonances is linear as more and more bands follow the external modulation and energy spectrum near the resonance center have minimum non-linearity. Hence, stronger the modulation, smaller the non-linearity in the energy spectrum and longer the revival time as revival time is inversely proportional to the non-linearity in energy spectrum. On the other hand, the shallow lattice energy spectrum has wide energy bands and narrow band gaps as shown in Fig-2.3 and Table-2.1. Near the lattice potential minima, the constant $\beta$ for $N = 1$ is negative and sign before second term in Eq. (4.68)) becomes positive. Now as $\lambda$ or $q$ is increased, second term decreases and hence revival time decreases in this case.

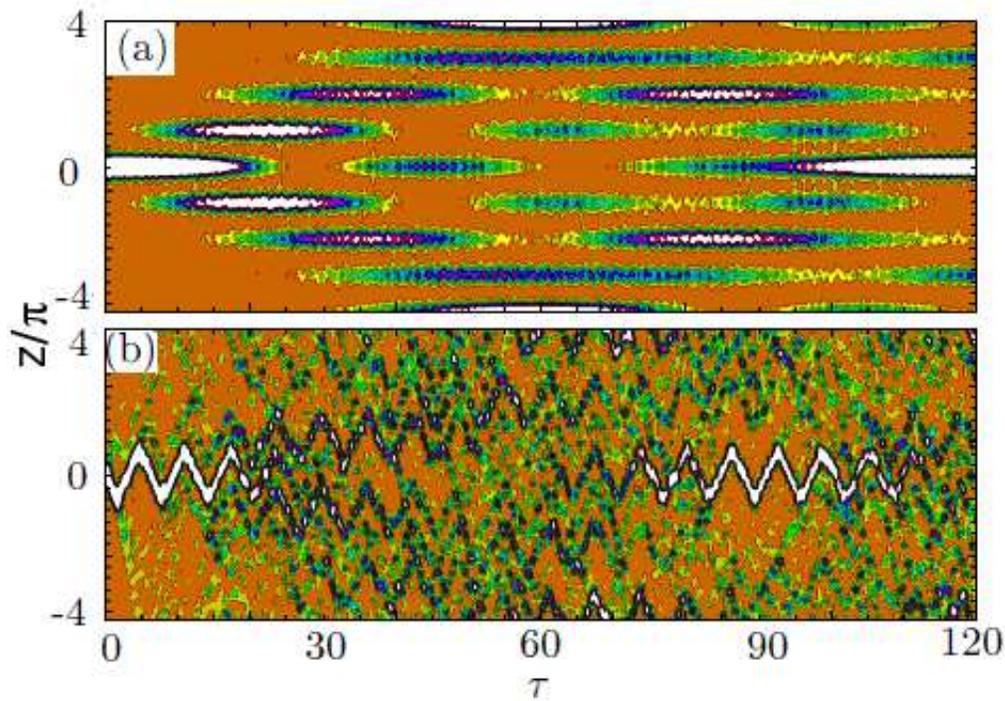

Figure 4.13: Spatiotemporal behavior of atomic wave packet. (a) Wave packet dynamics in the absence of modulation. (b) Dynamic when $\lambda = 3$. Other parameters are $\tilde{V}_0 = 0.36 \ \Delta p = 0.1$ and $\bar{k} = 0.16$ . Here, white regions show the maximum probability to find the particle and regions with dark color has higher probability than the lighter colors.

Fig-4.13 shows spatio-temporal evolution of an initially well localized wave packet in a lattice potential well. Fig-4.13-(a) is for the spatio-temporal dynamics of atomic wave packet in the absence of periodic modulation, while Fig-4.13-(b) presents the case when external modulation, $\lambda = 3$. Spatio-temporal dynamics of atomic wave packet shows that revival time changes as given analytical expression (4.68) plotted in Fig-4.10.

The super revival time behavior versus modulation, shown in right column of Fig-4.10. For robust dynamical recurrences Eq. (4.69) gives the time scale for super revivals. The super revival for robust dynamical recurrences increases with modulation as shown in the Fig-4.10(f). Here, qualitative behavior of super revival time is same for strongly driven shallow lattice and strongly driven deep lattice but quantitatively super revival time increases almost two times faster.

Square of auto-correlation function $(\mid A(t) \mid^2)$ for the minimum uncertainty wave packet is plotted as a function of time in Fig-4.14 and Fig-4.15 for $\lambda = 1.5$ and $\lambda = 0.5$ respectively. Other parameters are $q = 27.78$, $\Delta p = \Delta z = 0.5$, $\bar{k} = 0.5$ and $V_0 = 16E_r$. In these figures upper inset is an enlarged view which displays classical periods, while lower inset of the figures show quantum revivals. Lower panel shows the existence of super revivals. The classical period, quantum revival time and super revival time are indicated by arrows and showing there characteristics. Numerical results are in good agreement with analytical expressions. Figures-4.12 and 4.13 show spatiotemporal dynamics of Gaussian wave packet inside a resonance for $V_0 = 16E_r$ and $V_0 = 2E_r$ respectively. For other parameters, see figure caption. These numerical results are in agreement with analytical results.

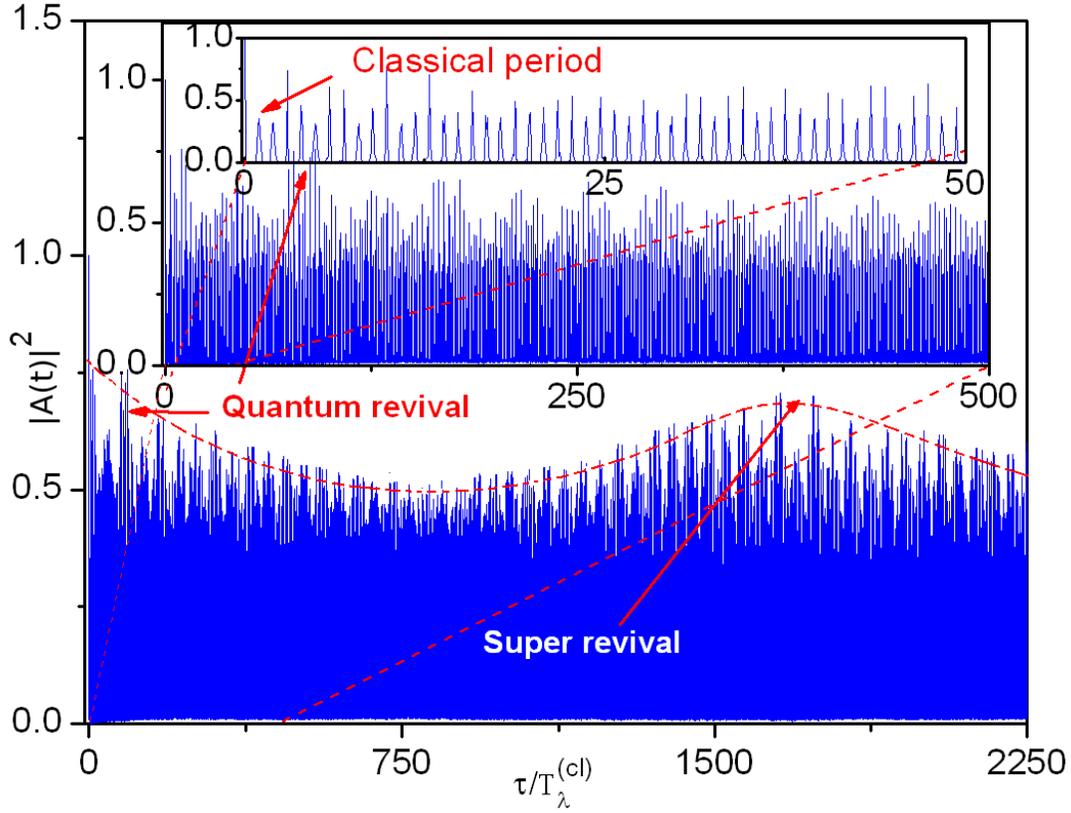

Figure 4.14: Square of auto-correlation function of Gaussian wave packet. Here, $\lambda = 0.5$, $\bar{k} = 0.5$, $\tilde{V}_0 = 16 E_r$, $\Delta z = \Delta p = 0.5$ and wave packet is initially, well localized at the central lattice well around the second band of undriven lattice. Classical period, revival time and super revival time are indicated by arrows.

## 4.8 Discussion

In this chapter, classical and quantum dynamics are studied. Classical dynamics are explored by studying Poincaré sections, classical momentum dispersion and time evolution of Gaussian distribution. It is noted that for $\tilde{V}_0 < 1$, the periodically driven lattice have classical dynamics, neither regular nor completely chaotic but displays an intricate dominant regular dynamics and dominant stochastic dynamics one after the other as a function of increasing modulation amplitude. While, for effective lattice potential, $\tilde{V}_0 \geq 1$, the intricate dominant regular and dominant stochastic dynamics one after the other is not seen. Beyond a critical value of modulation amplitude, stochastic region bounded by left running ($p < 1$) and right running ($p > 1$) regular solutions, increases with modulation. These regular solutions bounding stochastic region have an upper bound on momentum dispersion.

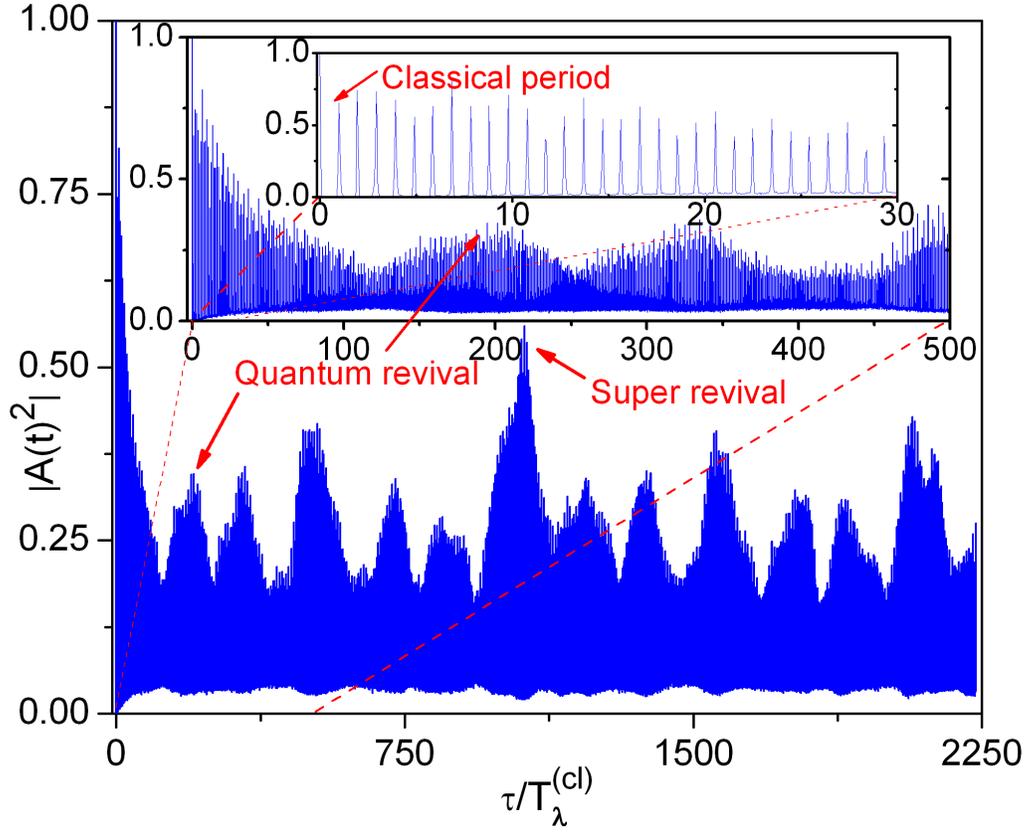

Figure 4.15: Square of auto-correlation function of Gaussian wave packet. Here, $\lambda = 1.5$, $\bar{k} = 0.5$, $\tilde{V}_0 = 16E_r$, $\Delta z = \Delta p = 0.5$ and wave packet is initially, well localized at the central lattice well around the second band of undriven lattice. Classical period, revival time and super revival time are indicated by arrows.

Quantum dynamics are explored by studying quantum momentum dispersion, temporal and spatio-temporal behavior of wave packet in inside nonlinear resonances that exist in the classical counterpart. Dynamics of wave packet inside nonlinear resonances show dynamical recurrences behavior. As a function of time the recurrence behavior of wave packet is analyzed close to nonlinear resonances. Our mathematical framework based on Floquet theory is developed in two classes: delicate recurrences which take place when lattice is weakly perturbed; and robust recurrences which manifest for sufficiently strong external driving force Classical periods of a wave packet initially localized near the center of resonance increases with modulation for delicate dynamics. While, quantum revivals and super revivals decreases with modulation. Similarly, condition $q \gg 1$ is satisfied when shallow or deep potential is strongly modulated. Classical periods in robust case decreases and super revival times increase with modulation. Here, quantum revival time for the case when deep lattice is strongly modulated is increases with modulation, while, it decreases when shallow lattice is strongly modulated. The difference in the behavior is due to the contrast in energy spectrum of undriven lattices. When modulation is increased in deep lattice case more and more energy levels are influenced by external modulation (Fig-4.11 (e), (f)), non-linearity near the center of resonance decreases and revival time increases. Parametric dependence of analytical results are confirmed by exact numerical solutions both for delicate and robust dynamical recurrences. Both temporal (Fig-4.14 and Fig-4.15) and spatio-temporal (Fig-4.12 and Fig-4.13) dynamics shows that the non-linearity of the undriven system, and the initial conditions on the excitation contribute to the classical period, quantum revival time and super revival time.

# Chapter 5

# Condensate in Driven Optical Lattices

In this chapter, we study the dynamics of condensate in driven optical lattices by analyzing dispersive behavior in position and momentum space. The dispersion behavior gives the parametric limits where, the condensate is stable in driven lattices. These stability limits are verified by studying the spatio-temporal dynamics of the condensate in driven lattice numerically. Latter, revival times are studies for different interaction regimes.

The dynamics of a condensate in driven one dimensional optical lattice, strongly confined by radial trap is governed by the Hamiltonian [Staliunas et al 2006; Eckardt 2010],

$$H = \frac{p^2}{2M} + \frac{V_o}{2} \cos[2k_L\{x - \Delta L \sin(\omega_m t)\}] + g_{1D}|\tilde{\psi}|^2,$$

where, $k_L$ is wave number, $V_o$ defines the potential depth of an optical lattice, $\Delta L$ and $\omega_m$ are amplitude and frequency of external derive, whereas, $M$ is the mass of an atom. Furthermore, $g_{1D} = \hbar k_L \omega_\perp a_s$ defines the effective two body interaction coefficient, $\omega_\perp$ is radial trap frequency and $a_s$ is inter-atomic s-wave scattering length.

The unitary transformation [1] (a special case of Kramers-Henneberger transformation) $\tilde{\psi} = \psi(x,t)(x,t) \exp[\frac{i}{\hbar}\{\omega_m M \Delta L \cos(\omega_m t)x + \beta(t)\}]$, where,

---

[1] The unitary transformation is time periodic and preserves the quasi-energy spectrum.

$\beta(t) = \frac{\omega_m^2 \Delta L^2 M}{4} \left[ \frac{\sin(2\omega_m t)}{2\omega_m} + t \right]$, to a frame co-moving with the lattice, modifies the Hamiltonian as

$$H = \frac{p^2}{2M} + \frac{V_o}{2} \cos 2k_L x - Fx \sin \omega_m t + g_{1D} |\psi|^2, \qquad (5.1)$$

where, $F = M \Delta L \omega_m^2$ is amplitude of inertial force emerging in the oscillating frame. To examine the dynamics of condensate in driven optical lattices numerically, Hamiltonian (5.1) is expressed in dimensionless quantities. We scale the quantities so that $z = k_L x$, $\tau = \omega_m t$, $\psi = \frac{\psi}{\sqrt{n_0}}$, where, $n_0$ is average density of a condensate. Multiplying the Schrödinger wave equation by $\frac{2\omega_r}{\hbar \omega_m^2}$, where, $\omega_r = \frac{\hbar k_L^2}{2M}$ is single photon recoil frequency, we get dimensionless Hamiltonian, viz.,

$$\tilde{H} = -\frac{\tilde{k}^2}{2} \frac{\partial^2}{\partial z^2} + \frac{\tilde{V}_o}{2} \cos 2z + \lambda z \sin \tau + G|\psi|^2. \qquad (5.2)$$

Here, $G = \frac{g_{1D} n_0 \tilde{k}}{\hbar \omega_m}$, and $\lambda = \frac{F d \tilde{k}}{\hbar \omega_m} = k_L \Delta L$, is scaled modulation amplitude. Also $\tilde{V}_o = \frac{V_o \tilde{k}}{2\hbar \omega_m}$, and $\tau$ is scaled time in the units of modulation frequency, $\omega_m$.

In the tight binding regime, wave function $\tilde{\psi}$ can be be expressed in Wannier states. Furthermore, as condensate is loaded in lowest vibrational state of optical lattice, $\tilde{\psi}$ can be expanded in terms of lower band Wannier functions,

$$\tilde{\psi} = \sum_j c_j(z, \tau) w_o(z - jd), \qquad (5.3)$$

where, $w_o(z - jd)$ is first band Wannier function centred at $jth$ lattice site and $c_j$ is complex amplitude of mini condensate associate with $jth$ lattice site.

Substituting Eq. (5.3) in Eq. (5.2), we get discrete GPE

$$i \frac{\partial c_j}{\partial \tau} = -J \left( c_{j+1} + c_{j-1} \right) + \lambda \sin \tau \; j \; c_j + G|c_j|^2, \qquad (5.4)$$

with $\tilde{k} = 1$,

$$J = -\int dz w_o^*(z) H_o w_o(z - d) dz, \qquad (5.5)$$

$$H_o = -\frac{1}{2} \frac{\partial^2}{\partial z^2} + \frac{\tilde{V}_o}{2} \sin(2z). \qquad (5.6)$$

where, $J$ is tunneling or the hoping matrix element between nearest neighboring lattice sites. In Eq. (5.4) the overall energy shift $\varepsilon_o$ is given by

$$\varepsilon_o = \int dz \; w_o^*(z) H_o w_o(z) dz, \tag{5.7}$$

which is set to zero.

In the case of dilute condensate, i.e., $G = 0$, Eq. (5.4) describe the dynamics of single-particle whose analytical solutions can explicitly be derived using the following gauge transformation [Kolovsky et al],

$$c_j(\tau) \rightarrow \exp\left[-i \; j \left(\lambda \sin(\tau)\right)\right] \tilde{c}_j(\tau). \tag{5.8}$$

Then the Eq. (5.4) becomes

$$i\frac{\partial \tilde{c}_j(\tau)}{\partial t} = -J(e^{-iD(\tau)}\tilde{c}_{j+1}(\tau) + e^{iD(\tau)}\tilde{c}_{j-1}(\tau)), \tag{5.9}$$

with, $D(\tau) = \lambda \; \sin(\tau)$. Sine this gauge transformation retrieves back the translation symmetry in the above equation, we can consider the spatially periodic boundary conditions, i.e., $\tilde{c}_j(\tau) = \tilde{c}_{L+j}(\tau)$. Now the corresponding semi-classical Hamiltonian is

$$H(\tau) = -J\sum_j \left(e^{-iD(\tau)}\tilde{c}_j^* \tilde{c}_{j+1}(t) + e^{iD(\tau)}\tilde{c}_{j+1}^* \tilde{c}_j(\tau)\right). \tag{5.10}$$

We use the Bloch-wave representation to obtain the solutions for $\tilde{c}_j(\tau)$,

$$\tilde{c}_j(\tau) = L^{-1/2} \sum_k e^{ikj} b_k, \tag{5.11}$$

where, the quasi-momentum is defined as $k = 2\pi n/L$ and $L$ represents lattice length or number of lattice sites in discrete version of GPE such that $-\pi \le k < \pi$, $n = 0, \pm 1, \ldots, \pm(L-1)/2$ when $L$ is odd, and $n = 0, \pm 1, \ldots, \pm L/2$ when $L$ is even. Now equations showing the time evolution of $b_k$ take the simple form

$$i\frac{\partial b_k}{\partial \tau} = -2J\cos\left[k - \lambda \; \sin(\tau)\right] b_k, \tag{5.12}$$

with solutions expressed as

$$b_k(\tau) = b_k(0)\exp\left\{i2J\left[\cos(k)S(\tau) - \sin(k)C(\tau)\right]\right\}. \tag{5.13}$$

Here, $b_k(0)$ is the integral constant, and $S(\tau)$ and $C(\tau)$ are defined as follow

$$S(\tau) = \sum_{m=-\infty}^{+\infty} \frac{\sin(m\tau)}{m} \mathcal{J}_m(\lambda), \tag{5.14}$$

$$C(\tau) = \sum_{m=-\infty}^{+\infty} \frac{\cos(m\tau) - 1}{m} \mathcal{J}_m(\lambda), \tag{5.15}$$

where, $\mathcal{J}_m$ is the $m$th order Bessel function of the first kind.

The the inverse Fourier transformation can also give analytical solutions $c_j$ for $G = 0$. These solution can be the starting points to derive the trial solutions for $G \neq 0$. Our concern is about $b_k(0) = \delta_{k,p}$ case, which gives the final expressions of Eq. (5.4) as

$$\begin{aligned} c_j(\tau) &= \exp\{i\, j[p - \lambda \sin(\tau)]\} \\ &\quad \times \exp\{i2J[\cos(p)S(\tau) - \sin(p)C(\tau)] - ig\tau\}. \end{aligned} \tag{5.16}$$

These solutions are applicable for $L \to \infty$ in which the boundary conditions make no difference.

## 5.1 Stability of condensate in driven lattice

The stability of BEC in driven optical lattice is studied taking into account the mean field dynamics of condensate by [Creffield 2009] for large modulation frequency with the condition that $\hbar\omega_m$ is less than the energy gap between two lowest bands. For the undriven case with repulsive interaction, the dynamical instability does not occur [Wu et al 2003]. As $\lambda \to 0$, the critical value of interaction $G_{cr}$ at which instability occur approaches to infinity. With the increase of modulation strength $\lambda$, the critical value for interaction, $G_{cr}$, abruptly drops and touches a wide local minimum value. Then it attains a maximum value at first root ($\lambda = 2.4048$) of Bessel function of order zero $\mathcal{J}_0$, as in tight binding regime tunneling of condensate is suppressed and stability is regained. Beyond the first root of zeroth-order Bessel function, the condensate becomes unstable for any positive value of interaction. Near the second root of Bessel function another maximum of $G_{cr}$ value is attained

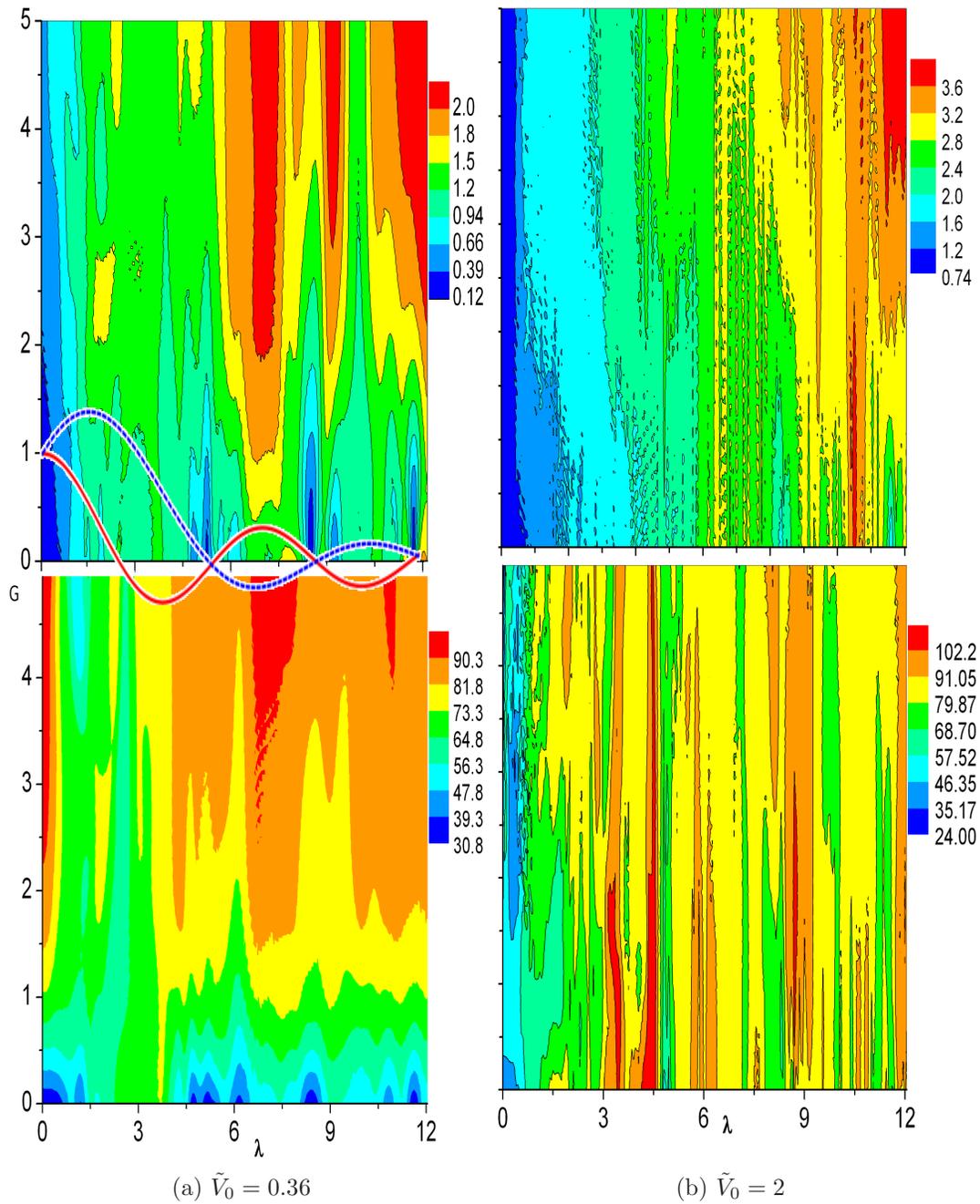

(a) $\tilde{V}_0 = 0.36$                    (b) $\tilde{V}_0 = 2$

Figure 5.1: Momentum and position dispersion of condensate vs modulation.
The dark blue region corresponds to minimum dispersion and thus maximum
density, whereas, bright red colour displays maximum dispersion leading to
minimum stability. A systematic colour drops from dark red to light red
indicates the variation of dispersion from minimum to maximum value in
momentum space (upper row) and position space (lower row). The dispersion
is measured at $\tau = 100\pi$ (after 50 periods of modulation) for $\bar{k} = 1$, $\Delta p = 0.1$.
In this graph, solid line shows zero order Bessel $\mathcal{J}_0(\lambda)$ and dotted line shows
function given in Eq. (5.17).

and this pattern repeats. These stability zones expand for larger external frequency.

The behavior of instability for the modulation strength lying between the two zeros of Bessel function can be explained by introducing an effective tunneling, $J_{eff} = J\mathcal{J}_0(\lambda)$. When $\lambda$ is larger than 2.4048, then $J_{eff}$ is negative and $J_{eff}G_{cr}$ becomes negative and dynamical instability can occur. The physical significance of this change in sign is experimentally realized [Lignier et al 2007], as it caused the momentum distribution function to be discretely shifted by $\pi$. Hence, the tunneling not only renormalized in amplitude, but also acquires a phase-factor of $e^{i\pi}$ [Creffield and Sols 2008]. A similar feature was explored in the analysis of a condensate in accelerated lattice [Zheng et al 2004a].

In the next section, we explore dispersion of condensate in driven lattice as a function of modulation and interaction strength for a fixed time both in position and momentum space. Suppression of dispersion in position and momentum displays the localization and hence the stability of the condensate.

## 5.2 Momentum and position dispersion of condensate

the momentum and position dispersion of repulsive condensate in driven optical lattice as a function of interaction and modulation amplitude. We provide density plots to explain dispersion as function of modulation and interaction. In these density plots, the dark blue regions corresponds to minimum dispersion and thus maximum density, whereas, bright red colour displays maximum dispersion leading to minimum stability. A systematic colour drops from dark red to light red indicates the variation of dispersion from minimum to maximum value in momentum space (upper row) and position space (lower row). In Fig-5.1, we plot momentum and position dispersion for effective potential $\tilde{V}_0 = 0.36$ and $\tilde{V}_0 = 2$. In Fig-5.1a, we consider $\tilde{V}_0 = 0.36$ and in Fig-5.1b, we take $\tilde{V}_o = 2$. In this figure the dark line is a plot of zeroth-order Bessel function while dotted line represents the function $f(\lambda)$, which is sum of first three order Bessel functions of first kind and defined as

$$f(\lambda) = \mathcal{J}_0(\lambda) + \mathcal{J}_1(\lambda) + \mathcal{J}_2(\lambda) + \mathcal{J}_3(\lambda). \tag{5.17}$$

For small value of $\tilde{V}_0$, where, the condition, $\frac{\tilde{V}_0}{\sqrt{\lambda}}$, is satisfied, dispersion is minimum at the zeroes of zeroth-order Bessel. For the values of modulation and effective potential where condition $\frac{\tilde{V}_0}{\sqrt{\lambda}}$ is not satisfied, zeros of the function expressed by the Eq. (5.17) define dispersion minima. Near the zeros of the Bessel function, momentum and position dispersion is very small which shows that instabilities in this parameter regime are suppressed due to dynamical localization (DL) [Creffield 2009]. In DL and coherent destruction of tunneling (CDT), the initial localized quantum state never diffuses under a periodic external field. In this aspect, these phenomena are similar. However, the dissimilarities are: i) CDT is derived approximately in an extreme case of a small value of the transfer matrix element while DL is an exact result obtained in an infinite driven system and is valid irrespective of the magnitudes of the transfer matrix element. On the other hand, in DL the initial distribution oscillates around the initial value whereas, in the CDT, the initial distribution is frozen [Kayanuma and Saito 2008].

Now as the interaction is increased, the dispersion stays constant for some critical value of $G_{cr}$ beyond which dispersion increases with interaction. Beside stable regions, there exists regions which display maximum dispersion. These are unstable regions for condensate. Our findings are in good agreement with earlier obtained results, for example in Ref: [Creffield 2009], where, it is shown that for small external modulation, $G_{cr}$ falls quickly to zero while in our case dispersion remain small for small modulation strength and small interaction $G_{cr}$. Moreover, the position dispersion has large values for large $G$ and dispersion minima disappear for small critical value of interaction $G_{cr}$.

In Fig-5.4, momentum and position dispersion versus modulation and interaction is shown for $\tilde{V}_0 = 5.7$ and same colour scheme is used as in Fig-5.1. Again dispersion can be understood in the same manner as in Fig-5.1 and coherent destruction of tunneling limits the dispersion.

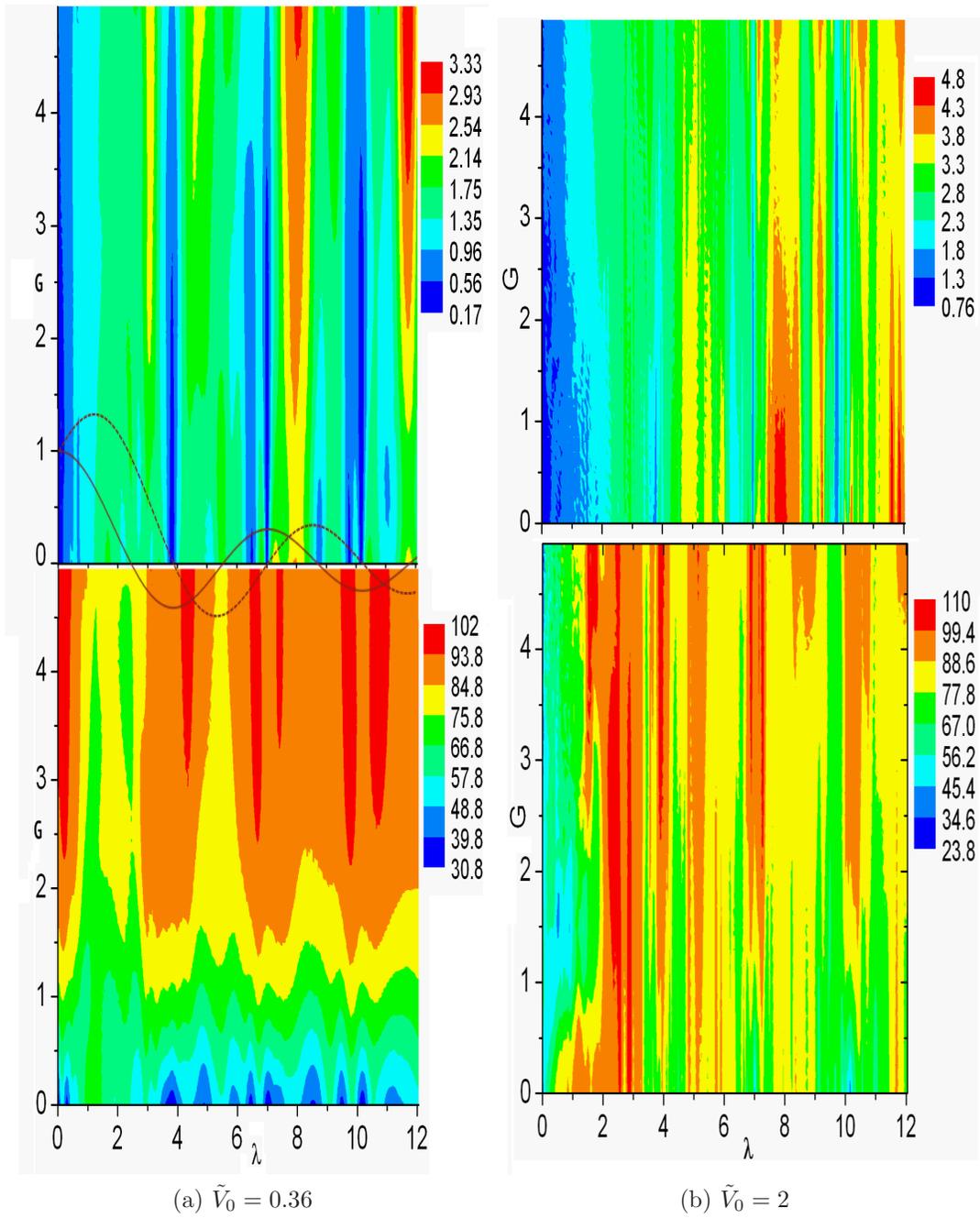

(a) $\tilde{V}_0 = 0.36$                    (b) $\tilde{V}_0 = 2$

Figure 5.2: Momentum and position dispersion of condensate vs modulation for $\omega_m = 2$. The dispersion is measured at $\tau = 100\pi$ (after 50 periods of modulation) for $\bar{k} = 1$, $\Delta p = 0.1$. In this graph, solid line shows zero order Bessel function of first kind $\mathcal{J}_0(\lambda)$ and dotted line shows behavior of first three terms in function given by Eq. (5.17).

From Eq. (5.2), we see that effective potential, $\tilde{V}_0$, is scaled by modulation frequency, $\omega_m$, and for large modulation frequency, $\omega_m$, effective potential is small. If we consider the large modulation frequency, near the roots of Bessel function the value of critical interaction $G_{cr}$, increases and windows of minimum dispersion is widen up as shown in Fig-5.2a and Fig-5.2b provided that $\hbar\omega_m$ is not larger than the energy gap between the lowest energy bands. In Fig-5.2a, the number of dispersion minima both in momentum and position are large compared to Fig-5.1a. It is noted that these minima exit at the zeros of $\mathcal{J}_0(\lambda) + \mathcal{J}_1(\lambda) + \mathcal{J}_2(\lambda)$. With increase in modulation frequency $\omega_m$, the role of higher order terms are decreased and for sufficient large modulation, potential minima exist at the zeros of first order Bessel function as dicussed in [Creffield 2009].

To understand the dispersion behavior fully, we excite minimum uncertainty condensate for different initial conditions i.e., the points in Fig-5.1a, where, the dispersion is minimum and a point where, dispersion is maximum. Fig-5.3 shows time evolution of momentum and position dispersion for modulation and interaction, $(\lambda, G) = (7.7, 0), (7.7, 2), (8.4, 2)$ where, dispersion is maximum in momentum and position space. Whereas, for $(\lambda, G) = (8.4, 0), (8.4, 0.5)$, dispersion is minimum in momentum space and in position space it increases slowly compare to the previous case as shown in Fig-5.3. From Fig-5.1a, we see that when condensate is excited with modulation and interaction $(\lambda, G) = (7.7, 0)$, or $(\lambda, G) = (7.7, 2)$, it is excited in the region with maximum dispersion both in momentum and position space and time evolution of condensate in Fig-5.3 shows increasing behavior both in position and momentum space. While, dispersion in momentum remains unchanged, when condensate is excited with $(\lambda, G) = (8.4, 0)$ or $(\lambda, G) = (8.4, 0.5)$ as these points lie in minimum momentum dispersion regions. The dispersion in position increases slowly with time for $(\lambda, G) = (8.4, 0.5)$ as this point doesn't lie in the minimum dispersion region. The small wiggling behavior seen in momentum dispersion is due to collapse and revival of mini-condensates trapped in each effective potential minima due to tunneling suppression. This suppression in tunneling is due to coherent destruction of tunneling around the zeros of Bessel function.

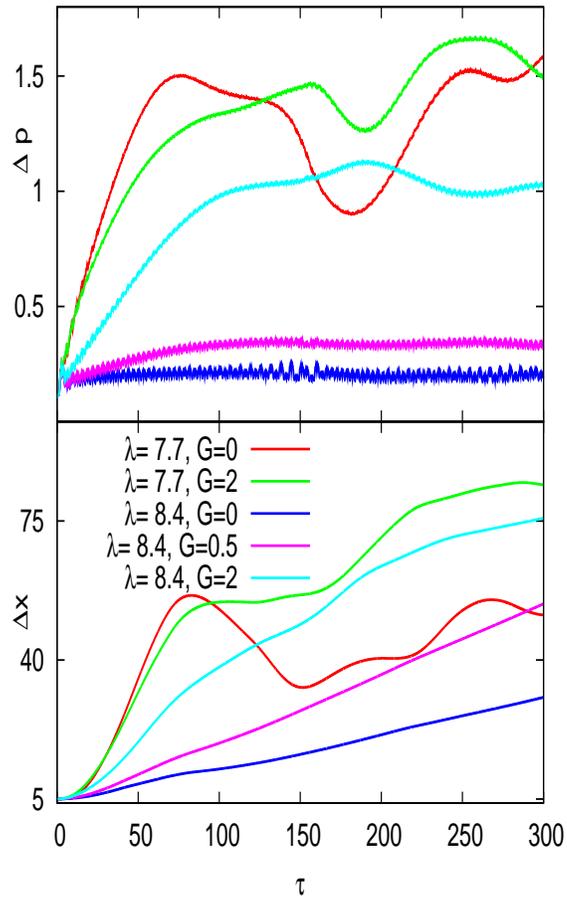

Figure 5.3: Momentum and position dispersion of condensate vs time for modulation and interaction, $(\lambda, G) = (7.7, 0)$, $(7.7, 2)$, $(8.4, 2)$ where, dispersion is maximum in position and momentum space. Whereas, for $(\lambda, G) = (8.4, 0)$, $(8.4, 0.5)$ dispersion is negligible in momentum space while it is slower in position space compare to the previous case as shown in Fig-5.1. Other parameters are same as in Fig-5.1a.

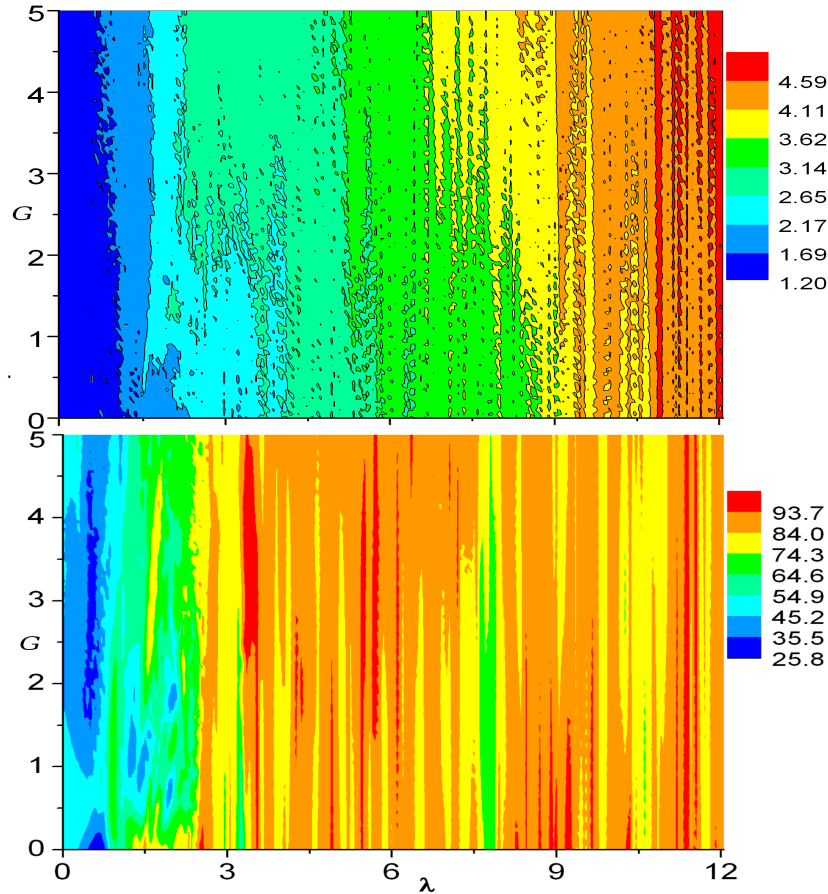

Figure 5.4: Momentum and position dispersion of condensate vs modulation. The dispersion is measured at $\tau = 100\pi$ (after 50 periods of modulation) for $\tilde{V}_0 = 5.7$. Other parameters are same as in Fig-5.1

## 5.3 Spatio-temporal dynamics of condensate in driven optical lattice

In this section, spatio-temporal dynamics of condensate are studied by initializing a Gaussian condensate in driven optical lattice. Fig-5.5 shows the spatio-temporal dynamics of condensate initially excited in localized windows for $(G, \lambda) = (0, 1.5)$, $(G, \lambda) = (2, 1.5)$, left column and delocalization windows $(G, \lambda) = (0, 9)$, $(G, \lambda) = (2, 9)$, right column. Other parameters are same as in Fig-5.4. When condensate is excited in localization region where coherent destruction in tunneling causes the matter wave to be localized in the respective lattice site, the condensate mimic signatures of localization in position and momentum space as shown in Fig-5.5a. While, for the case when condensate is initially excite in the region of maximum dispersion, spatio-temporal dynamics of condensate are diffusive as shown in right column of Fig-5.5a. In Fig-5.5b, probability to find the condensate at position $z$ is plotted at time, $\tau = 0,\ 50,\ 100$ for the two contrasting regions with $\lambda = 1.5$ and $\lambda = 9$. When condensate is excited with parameters belong to the region where relevant density plot show minimum dispersion for $\lambda = 1.5$, no significant change in spatio-temporal profile and probability is noted with time. On the other hand, when condensate is excited with parameters belong to the region where relevant density plot show maximum dispersion for $\lambda = 9$, it shows dispersive behavior both in spatio-temporal profile and density plots.

## 5.4 Recurrence behavior of condensate in driven optical lattice

In this section, we study dynamical recurrences of a Bose-Einstein condensate in optical crystal subject to periodic external driving force. The recurrence behavior of the condensate is analyzed as a function of time close to nonlinear resonances occurring in the classical counterpart. Our mathematical formalism for the recurrence time scales is presented as: delicate recurrences which take place for instance when lattice is weakly perturbed; and, robust recurrences which may manifest themself for sufficiently strong external driving force. The analysis is not only valid for dilute condensate but also applicable for strongly interacting homogeneous condensate provided, the external modulation causes no significant change in density profile of the condensate. We explain parametric dependence of the dynamical recurrence times which can easily be realized in laboratory experiments. In addition, we find a good agreement between the obtained analytical results and numerical calculations.

For the dynamics of weakly interacting BECs in optical lattice the interaction term can safely be neglected [Eckardt 2010], which is a situation

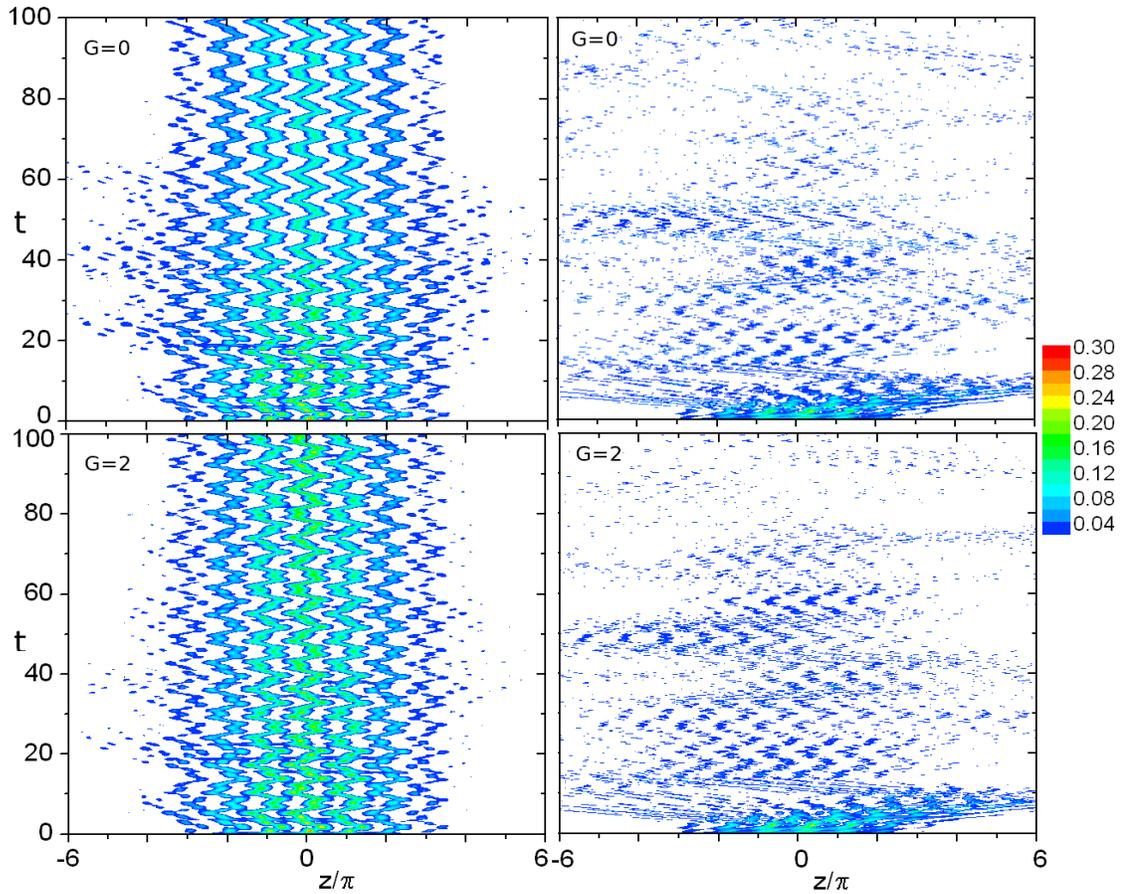

(a) Spatio-temporal dynamics of condensate for $\lambda = 1.5$, interaction $G = 0, \ 2$ left column and $\lambda = 9$, interaction $G = 0, \ 2$. Other parameters are same as in Fig-5.4

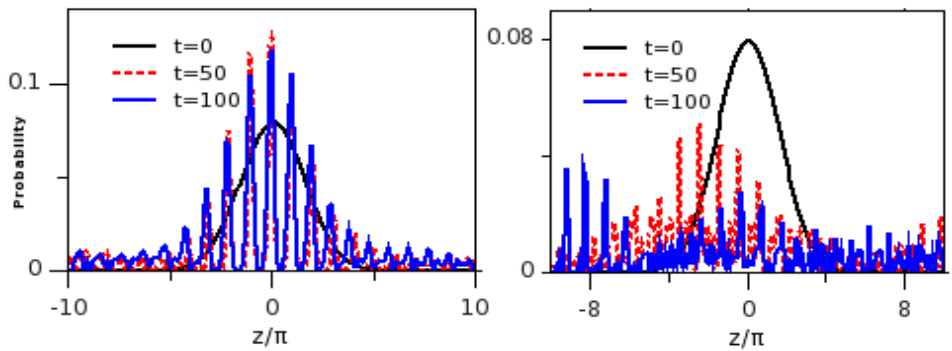

(b) Wave function snapshot for different times $t = 0, \ 50, \ 100$ for the two cases: left side $\lambda = 1.5$ and right side $\lambda = 9$.

Figure 5.5: Spatio-temporal dynamics of condensate (upper row) and stroboscopic plots of probability at different times (lower row).

achievable in the present day experiments with the advent of Feshbach scattering resonances. In addition, the effective potential $\frac{\tilde{V}_o}{2}\cos 2z + G|\psi|^2$ seen by each atom may as well be written [Choi 1999] as,

$$V_{eff} = \frac{\acute{V}}{2}\cos(2z) + const,$$

where, $\acute{V} = \frac{\tilde{V}_o}{1+4G}$.

The analytical result, calculated using perturbation theory [Choi 1999], is valid as long as the condensate density is nearly uniform, i.e., $\acute{V} << 1$ which describes either a weak effective potential $\tilde{V}_o$ or a strong atomic interaction $G$. The condition is experimentally confirmed for one dimensional potential [Morsch 2001].

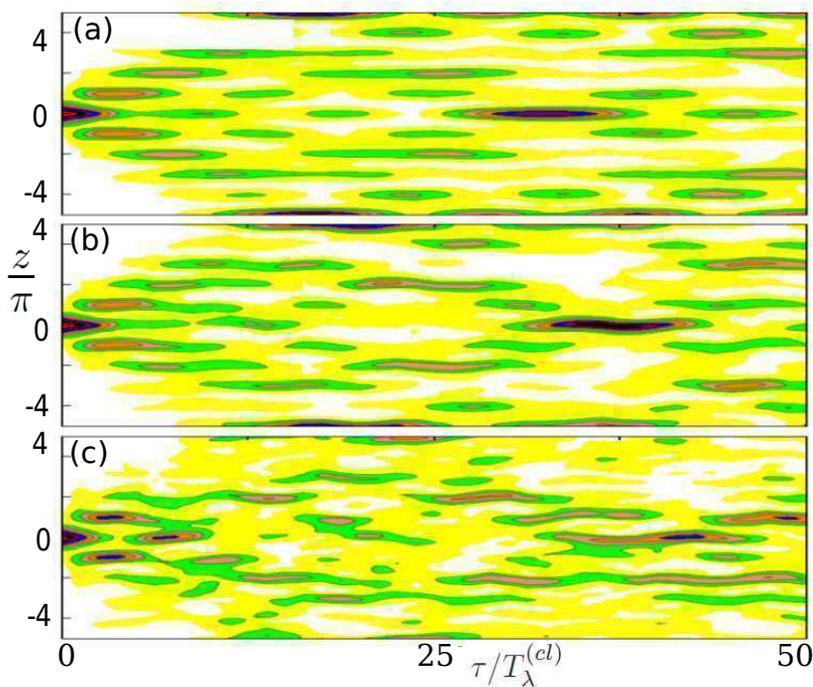

Figure 5.6: Spatio-temporal behavior of atomic condensate for $q = 0(\lambda = 0)$, $\acute{V} = 2$ (a) $q = 0.2\beta_0$, $\acute{V} = 2$(b) and $q = 0.2\beta_0$, $\acute{V} = 0.3$ (c). Other parameters are $\Delta p = 0.5$, $\beta_0 = \frac{\hbar^2 \zeta}{4V}$ and $\hbar = 1$. Dark regions in this figure show higher density.

Fig-5.6 shows spatio-temporal evolution of an initially well localized condensate in a crystal potential well. Fig-5.6a is spatio-temporal dynamics in undriven crystal, while Fig-5.6b and Fig-5.6c present the case, for external modulation $q = 0.2\beta_0$ and different value of $q_o$. Spatio-temporal evolution of the condensate in optical crystal shows that condensate diffuses to the neighboring lattice sites by tunneling and splits into smaller wavelets. Later, these wavelets constructively interfere and condensate revival takes place. Recurrence time calculated numerically is the same as obtained from analytical results. Keeping modulation constant as $q = 0.2\beta_0$ but for different $\acute{V}$, which may be a consequence of varied atom-atom interaction, recurrence time is modified. We note that in the absence of interaction term, $G$, revival time changes with $\acute{V}$ and interference pattern is similar. But with the introduction of interaction term not only revival time is modified due to change in $\acute{V}$, interference pattern is also modified as seen in Fig-5.6c.

Fig. 5.7 shows spatio-temporal evolution of an initially well localized wave packet in a lattice potential well inside a resonance for $\acute{V} = 16$. Fig-5.7a is for the spatio-temporal dynamics of atomic condensate in the absence of periodic modulation, Fig-5.7b presents the case for external modulation, $q = 3\beta_0$, while, Fig-5.7c shows the behavior of condensate for $q = 9\beta_0$. In the Fig-5.7c only classical periods are seen as in deep effective potential higher order time scales are very large. The quantum revival times in Fig-5.7 seen numerically are the same as obtained analytically in Eq. (4.68).

The suggested theoretical results may be realized in experimental set up of recently performed experiments at Pisa [Lignier et al 2007], where, dynamical control of matter wave tunneling is studied in strongly shaken optical crystals. A BECs of about $5 \times 10^4$ $^{87}$Rb atoms was evolved in a dipole trap which was realized using two intersecting Gaussian laser beams at 1030 nm wavelength and a power of around 1 W per beam focused to waists of 50 $\mu$m. After obtaining pure condensate, trap beams were readjusted to obtain elongated condensates with the trap frequencies (80 Hz in radial and $\approx 20$ Hz in the longitudinal direction). Along the axis of one of the dipole trap beams a one-dimensional optical crystal potential was introduced and the power of the lattice beams ramped up in 50 ms in order to avoid excitations of the BEC to the non-condensed atoms. The optical crystals used was created using two counter-propagating Gaussian laser beams ($\lambda_L = 852$ nm) with $120\,\mu m$ waist and a resulting optical crystal spacing $d_L = \lambda_L/2 = 0.426\,\mu$m. The depth $V_o$ of the resulting periodic potential is measured in units of $E_r = \hbar^2\pi^2/(2Md_L^2)$. In laboratory, accessible quantity like scaled optical lattice depth $\acute{V}$ ranges from 1 to 20.

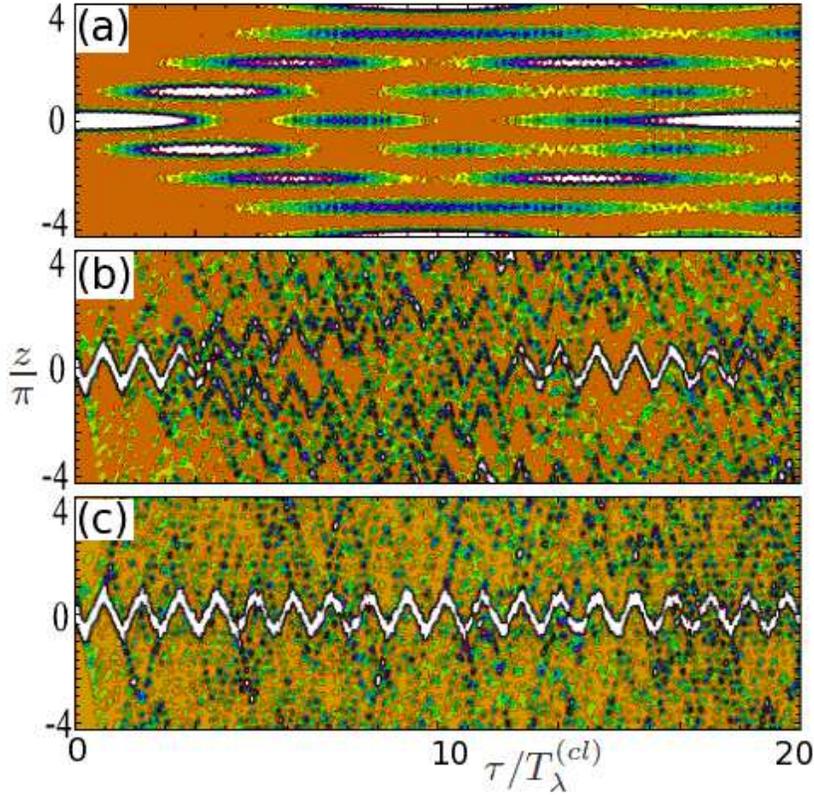

Figure 5.7: Spatio-temporal behavior of atomic condensate. Wave packet dynamics in the absence of modulation (a), when modulation $q = 3\beta_0$ (b) and when modulation $q = 9\beta_0$ (c). Other parameters are $\acute{V} = 0.36$, $\Delta p = 0.1$ and $\acute{k} = 0.16$. Here, white regions show the maximum density and regions with dark color has higher density than the lighter colors.

For optical crystal with potential depth $\acute{V} = 2$, and $\acute{V} = 16$, the mean separation of the two lowest bands is $\approx 3.15 kHz$ and $20.784 kHz$, respectively. For the driving frequency, $\omega_m/2\pi$, ranging from $3kHz - 9kHz$, the rescaled Plank's constant $\acute{k}$ ranges from 0.668 to 2.066.

# Chapter 6

# Discussion

In this thesis, we considered the center of mass dynamics of ultra-cold atoms and Bose-Einstein condensate in one dimensional optical lattice, both in the absence and presence of external modulation.

In chapter-2, we discuss the center of mass dynamics of the single particle wave packet in the absence of external modulation. We provide analytical study in two regimes, that is deep optical lattice and shallow lattice potential. In this chapter, we first time discuss the wave packet behavior and their parametric dependence both analytically and numerically. In deep lattice case, the energy bands collapse to a single level as tunneling is suppressed near the lattice potential minima. In this regime, we observe collapse and revival behavior in wave packet dynamics. As for sufficiently deep lattice, energy spectrum is linear near the bottom of lattice potential, a single particle wave packet revives after each classical period. Beyond linear regime, energy dependence is quadratic and wave packet dynamics in this region shows complete quantum revivals enveloping many classical periods, however, beyond this region quantum revival time is no longer constant but decreases as $\bar{n}$ increases, where, $\bar{n}$, is mean band index. This dependence of revival time on mean quantum number can be associated with the non-linearity in the energy spectrum. For sufficiently deep lattice potential, near the lattice depth, non-linearity is almost zero as we move up from lowest energy band to the higher one, non-linearity emerges and dictates the revival time. Since, revival time is inversely proportional to non-linearity, the quantum revival time decreases as $\bar{n}$ increases. This behavior of wave packet revival can only be seen in sufficiently deep lattice if we are away from the top of the lattice potential where, continuum in energy is seen and semi-classical limit in reached at which according to Correspondence Principle quantum revival time excepted to diverge. The higher order time scale exist in the region of changing non-linearity. super revival time in this region can be seen. Again this region expands as potential height is increased but super revival time in this region is directly proportional to square root of potential height. Above this region other time scales also exist where, super revival time decreases as $\bar{n}$ increases but these time scales are too long to consider them finite.

In shallow optical lattices, quantum tunneling plays its role and we encounter with wide energy bands instead of energy levels. Two energy scales, band gap and band width play active role in the dynamics of the system. If the band gap is larger than band width, tight binding approximations can be applied. But in situations where, lattice is very shallow, next to nearest tunneling can't be ignored and tight binding is no longer valid. The wave packet in this regime shows dispersive behavior in position space with time causing a decrease in revival amplitude.

In chapter-3, we discuss center of mass dynamics of the condensate in an optical lattice for the sake of completeness of the discussion. It is first time shown that now nonlinear phenomena like solitonic behavior and self-trapping of condensate can be explained by studying momentum and spatial dispersion, spatio-temporal behavior and wave packet revivals of the condensate. The inherent non-linearity of a condensate due to interatomic interactions and Bragg scattering of a matter-wave by an optical lattice play its role in the dynamics. Localization is possible as condensate shows anomalous dispersion at the edges of a Brillouin zone of the lattice and magnitude of this dispersion can be managed. Therefore, the condensate spreading either can be controlled by actively controlling the lattice parameters or by utilizing the atom-atom interaction. The second approach leads to non-linearly localized states.

For periodically driven optical lattice, classical and quantum dynamics are explored. Classical dynamics are explained by studying Poincaré surface sections, classical momentum dispersion and time evolution of Gaussian distribution. The Poincaré surface of sections plotted for different modulation and effective potentials. These plots show that the the dynamics of classical particle in driven lattice is neither regular nor completely chaotic but displays an intricate dominant regular dynamics and dominant stochastic dynamics one after the other as a function of increasing modulation amplitude as long as the ratio between effective lattice potential, $\tilde{V}_0$, and square root of modulation amplitude, $\lambda$ is much smaller than unity and can effectively be explained by zeroth-order Bessel function of first kind.

Quantum dynamics of a wave packet inside non-linear resonances shows dynamical recurrences. As a function of time the recurrence behavior of wave packet is analyzed close to nonlinear resonances. Our mathematical framework based on Floquet theory is developed in two classes: delicate recurrences which take place when lattice is weakly perturbed; and robust recurrences which manifest for sufficiently strong external driving force. Analytical expression for the classical period, revival time and super revival time are developed. Classical periods of a wave packet initially localized near the center of resonance increases with modulation for delicate dynamics. While, quantum revivals and super revivals decreases with modulation. Similarly, robust dynamics condition is satisfied when shallow or deep potential is strongly modulated. Classical periods in robust case decreases and super revival times increase with modulation. Here, quantum revival time for the case when deep lattice is strongly modulated increases with modulation, while, it decreases when shallow lattice is strongly modulated. The difference in the behavior is due to the contrast in energy spectrum of undriven lattices. When modulation is increased in deep lattice case more and more energy levels are influenced by external modulation non-linearity in the energy spectrum near the center of resonance decreases and revival time increases. Parametric dependence of analytical results are confirmed by exact numerical solutions both for delicate and robust dynamical recurrences. Temporal and spatio-temporal dynamics shows that the non-linearity in the energy spectrum of the undriven system, and the initial conditions on the excitation contribute to the classical period, quantum revival time and super revival time.

Dynamics of the condensate in driven optical lattice crystal is analyzed by studying position dispersion, momentum dispersion, dynamical recurrences and spatio-temporal dynamics. The dispersion behavior gives the parametric limits where, the condensate is stable in driven lattices. These stability limits are verified by studying the spatio-temporal dynamics of the condensate in driven lattice numerically. Later, revival times are studies for different interaction regimes. The dynamical stability of condensate in driven optical lattice is measured by the dispersion behavior of condensate excited in driven optical lattice for a variable range of modulation and interaction parameters. The dispersion is minimum near the zeros of Bessel function and hence condensate dynamics are stable provided interaction is smaller than the critical value $G_{cr}$. Now as the interaction is increased, the dispersion stays constant for some critical value of $G_{cr}$ beyond which dispersion increases with interaction. This critical value of interaction increases as the modulation frequency increases. Beside stable regions, there exists regions which display maximum dispersion. These are the regions where condensate is dynamically unstable.

Our analysis for classical period, quantum revival time and super revival time represented for single particle wave packet is also valid for sufficiently dilute BECs or the condensate for which the inter-particle interaction can be tuned to zero by exploiting Feshbach resonance. In addition, we suggest that the analysis is valid for strongly interacting homogeneous condensates as well where the nonlinear term can be replaced by an effective potential provided the external modulation causes slight changes in the density profile of the condensate.

# Appendices

## Appendix A: Solution of an Arbitrary Potential

An arbitrary potential $U(r)$, around its minima can be solved by taking its Taylor's expansion [Liboff 2002], that is

$$U(r) = U(r_m) + G^{(1)}(r - r_m) + G^{(2)}(r - r_m)^2 + G^{(3)}(r - r_m)^3 + .. \quad (6.1)$$

where, $G^{(j)} = (j!)^{-1}\partial^j U(r = r_m)/\partial r^j$, and $j$ is an integer. The value of $G^{(j)}$ for odd $j$ is zero as potential is $\cos(x)$ and it is calculated at the potential minima $r = r_m$. Thus in the presence of weak non-linearity the term $G^{(6)} << G^{(4)} << G^{(2)}$.

In our analysis we consider the expansion up to second order term as unperturbed Hamiltonian, $H_0$. The eigen functions and eigen energies of this Hamiltonian are those of harmonic oscillator. The effect of the higher order terms in Taylor's expansion is discussed as perturbation to eigen energies and eigen functions of the harmonic oscillator. We express the effective Hamiltonian governing the dynamics of atom around the potential minima as

$$H_0 \cong \frac{\hat{p}^2}{2} + U(r_m) + G^{(2)}(r - r_m)^2. \quad (6.2)$$

The eigen functions and eigen energies of the harmonic potential are $\phi_n(x) = \sqrt{\frac{\beta}{2^n n! \sqrt{\pi}}} H_n(\beta x) \exp(\frac{-\beta^2 x^2}{2})$, and $E_n^{(0)} = \hbar\sqrt{V_0}(2n+1) + U(r_m)$, where, $H_n(\beta x)$ are Hermite polynomials.

The first order correction and second order correction to energy is quite well known. The first order correction to the energy of quadratic potential is $E_n^{(p,a)} = \langle \phi_n | H^{(p)} | \phi_n \rangle$, and the second order correction is obtained from the relation $E_n^{(p,b)} = \sum_{m \neq n} \frac{|\langle \phi_n | H^{(p)} | \phi_m \rangle|^2}{E_n^{(0)} - E_m^{(0)}}$. The third order correction is determined by

$$E_n^{(p,c)} = \sum_{j \neq n} \sum_{l \neq n} \frac{\langle n | \hat{H}^{(p)} | j \rangle \langle j | \hat{H}^{(p)} | l \rangle \langle l | \hat{H}^{(p)} | n \rangle}{(E_n^{(0)} - E_j^{(0)})(E_n^{(0)} - E_l^{(0)})} - \langle n | \hat{H}^{(p)} | n \rangle \sum_{j \neq n} \frac{\langle n | \hat{H}^{(p)} | j \rangle \langle j | \hat{H}^{(p)} | n \rangle}{((E_n^{(0)})^2 - (E_j^{(0)})^2)},$$

where, $p = 3, 4, 5, 6, \ldots\ldots$.

The presence of the perturbation term $H^{(4)}$, leads to the Hamiltonian

$$H = H_0 + H^{(4)}. \tag{6.3}$$

Here, the first order correction to eigen functions, due to the $H^{(4)}$ term, is obtained as

$$|\phi_n^{(1a)}\rangle = D_1(\eta_1 \phi_{n-4} + \eta_2 \phi_{n-2} - \eta_3 \phi_{n+2} - \eta_4 \phi_{n+4}), \tag{6.4}$$

where, $D_1 = G^{(4)}(\frac{1}{4\sqrt{q_0}})^2 \frac{1}{\hbar \omega_h}$

$$\eta_1 = \sqrt{n(n-1)(n-2)(n-3)}/4,$$
$$\eta_2 = (2n-1)\sqrt{n(n-1)},$$
$$\eta_3 = (2n+3)\sqrt{(n+1)(n+2)},$$
$$\text{and } \eta_4 = \sqrt{(n+1)(n+2)(n+3)(n+4)}/4.$$

Now the eigen functions in the presence of first order perturbation, due to the correction $H^{(4)}$, is given as

$$\phi_n^q(x) = \phi_n(x) + \phi_n^{(1a)}(x),$$

and second order correction due to $H^{(4)}$ term is

$$\phi_n^{(1,b)} = D_2[\delta_1 \phi_{n-8} + \delta_2 \phi_{n-6} + \delta_3 \phi_{n-4} + \delta_4 \phi_{n-2} + \delta_5 \phi_{n+2} + \delta_6 \phi_{n+4} + \delta_7 \phi_{n+6} + \delta_8 \phi_{n+8}. \tag{6.5}$$

Here, $D_2 = (G^{(4)})^2(\frac{1}{4\sqrt{q_0}})^4(\frac{1}{\hbar\omega_h})^2$,

$$\delta_1 = \frac{\sqrt{n\,(n-1)\,(n-2)\,(n-3)\,(n-4)\,(n-5)\,(n-6)\,(n-7)}}{32},$$

$$\delta_2 = \frac{(6n-11)}{12}\sqrt{n\,(n-1)\,(n-2)\,(n-3)\,(n-4)\,(n-5)},$$

$$\delta_3 = \left(2n^2 - 9n + 7\right)\sqrt{n\,(n-1)\,(n-2)\,(n-3)},$$

$$\delta_4 = \frac{(56n^3 - 228n^2 + 214n - 146)}{8}\sqrt{n\,(n-1)},$$

$$\delta_5 = \frac{(56n^3 + 396n^2 + 838n + 645)}{8}\sqrt{(n+1)\,(n+2)},$$

$$\delta_6 = \frac{(31n^2 + 197n + 258)}{16}\sqrt{(n+1)\,(n+2)\,(n+3)\,(n+4)},$$

$$\delta_7 = \frac{(11n+27)}{24}\sqrt{(n+1)\,(n+2)\,(n+3)\,(n+4)\,(n+5)\,(n+6)},$$

and $\delta_8 = \frac{\sqrt{(n+1)(n+2)(n+3)(n+4)(n+5)(n+6)(n+7)(n+8)}}{32}.$

Hence, following the same procedure, the first order correction due to $H^{(6)}$ term changes the Hamiltonian of the system as

$$H = H_0 + H^{(4)} + H^{(6)}. \tag{6.6}$$

Hence the corrected eigen function in presence of correction due to $H^{(4)}$ and $H^{(6)}$ terms appear as

$$\phi_n^{(s)} = \phi_n + \phi_n^{(1,a)} + \phi_n^{(1,b)} + \phi_n^{(2,a)},$$

where,

$$\phi_n^{(2,a)} = D_6[\chi_1\phi_{n-6} + \chi_2\phi_{n-4} + \chi_3\phi_{n-2} + \chi_4\phi_{n+2} + \chi_5\phi_{n+4} + \chi_6\phi_{n+6}.$$

Here, $D_6 = G^{(6)}(\frac{1}{4\sqrt{q_0}})^3\frac{1}{\hbar\omega_h}$,

$$\chi_1 = 6\sqrt{n(n-1)(n-2)(n-3)(n-4)(n-5)},$$

$$\chi_2 = \frac{3}{4}(2n-3)\sqrt{n(n-1)(n-2)(n-3)},$$

$$\chi_3 = \frac{15}{2}(n^2 - n + 1)\sqrt{n(n-1)},$$

$$\chi_4 = \frac{15}{2}(n^2 + 3n + 3)\sqrt{(n+1)(n+2)},$$

$$\chi_5 = \frac{3}{4}(2n+5)\sqrt{(n+1)\,(n+2)\,(n+3)\,(n+4)},$$

$$\chi_6 = 2\sqrt{(n+1)\,(n+2)\,(n+3)\,(n+4)\,(n+5)\,(n+6)}.$$

Close to the minima of potential, we can find the eigen energies up to a considerable accuracy by using perturbation theory. The leading correction comes from $H^{(4)}$ using first order and second order perturbation theory respectively. The result is as under:

$$E_n^{(4)} = (\alpha_2 n^2 + \alpha_1 n + \alpha_0)\hbar\omega_h,$$

where, $\alpha_0 = 3D_a$, $\alpha_1 = 6D_a$, $\alpha_2 = 6D_a$, $C_{b=}(G^{(3)})^2(\frac{1}{4\sqrt{q_0}})^3\frac{1}{\hbar\omega_h}$ and $D_a = G^{(4)}(\frac{1}{4\sqrt{q_0}})^2$. At next order, the first order perturbation of $H^{(6)}$ and second order perturbation of $H^{(4)}$ contribute. The result can be written as:

$$E_n^{(6)} = (\beta_3 n^3 + \beta_2 n^2 + \beta_1 n + \beta_0)\hbar\omega_h,$$

where, $\beta_0 = 3I_a - 21J_b$, $\beta_1 = 8I_a - 59J_b$, $\beta_2 = 6I_a - 51J_b$, $\beta_3 = 4I_a - 34J_b$, and $I_a = 5G^{(6)}(\frac{1}{4\sqrt{q_0}})^3\frac{1}{\hbar\omega_h}$, $J_b = 2(G^{(4)})^2(\frac{1}{4\sqrt{q_0}})^2(\frac{1}{\hbar\omega_h})^2$.

At the next higher order, we need to evaluate three contributions; $H^{(8)}$ in first order, $H^{(6)}$ and $H^{(4)}$ in second order and $H^{(4)}$ in third order. The energy expression is

$$E_n^{(8)} = (\gamma_4 n^4 + \gamma_3 n^3 + \gamma_2 n^2 + \gamma_1 n + \gamma_0)\hbar\omega_h,$$

where, $\gamma_0 = 3X - 12Y - 111Z$, $\gamma_1 = 8X - 35Y - 347Z$, $\gamma_2 = 10X - 46Y - 472Z$, $\gamma_3 = 4X - 22Y - 250Z$, $\gamma_4 = 2X - 11Y - 125Z$, and $X = 35G^{(8)}(\frac{1}{4\sqrt{q_0}})^4\frac{1}{\hbar\omega_h}$, $Y = 30G^{(6)}G^{(4)}(\frac{1}{4\sqrt{q_0}})^5(\frac{1}{\hbar\omega_h})^2$, $Z = 48(G^{(4)})^3(\frac{1}{4\sqrt{q_0}})^6(\frac{1}{\hbar\omega_h})^2$.

Now the energy of the system is

$$\begin{aligned}
E_n &= E_n^{(0)} + E_n^{(4)} + E_n^{(6)} + E_n^{(8)}, \text{ or} \\
E_n &= (\kappa_4 n^4 + \kappa_3 n^3 + \kappa_2 n^2 + \kappa_1 n + \kappa_0)\hbar\omega_h + U(r_m),
\end{aligned}$$
(6.7)

where, $\kappa_0 = \alpha_0 + \beta_0 + \gamma_0 + \frac{1}{2}$, $\kappa_1 = \alpha_1 + \beta_1 + \gamma_1 + 1$, $\kappa_2 = \alpha_2 + \beta_2 + \gamma_2$, $\kappa_3 = \beta_3 + \gamma_3$, and $\kappa_4 = \gamma_4$.

# Appendix B: Anger function and distribution functions

In this appendix, we solve the equation,

$$\psi_{p_0}(z,t) = \frac{1}{\sqrt{2\pi k}}\exp[\frac{i}{k}(p_0 z - \frac{p_0^2}{2}t)]\exp[\frac{i\tilde{V}_0}{2k}\varphi(z,t)], \qquad (6.8)$$

where, phase $\varphi(z,t)$ is defined as

$$\varphi(z,t) = \int_0^t d\tau \cos(z - p_0(t-\tau)\lambda\sin\tau). \qquad (6.9)$$

We solve the above equation to find momentum distribution by considering interaction time as a integer multiple of $2\pi$. We decompose the integral in to integrals of interval $2\pi$ and find

$$\varphi(z,t=2N\pi) = Re[\sum_{\nu=0}^{N-1} e^{-ip_0 2\pi N}\int_{2\pi\nu}^{2\pi(\nu+1)} d\tau \exp[i(p_0(\tau) - \lambda\sin\tau]e^{iz}, \quad (6.10)$$

which is modified when we replace $\tau' = \tau - 2\pi\nu$

$$\varphi(z,t=2N\pi) = Re[\sum_{\nu=0}^{N-1} e^{ip_0 2\pi\nu}e^{-ip_0 2\pi N}\int_0^{2\pi} d\tau' \exp[i(p_0(\tau') - \lambda\sin\tau']e^{iz}. \qquad (6.11)$$

As

$$\sum_{\nu=0}^{N-1} e^{ip_0 2\pi\nu} = \frac{e^{ip_0 2\pi N} - 1}{e^{ip_0 2\pi\nu} - 1} = e^{ip_0\pi(N-1)}\frac{\sin(N\pi p_0)}{\sin(N\pi p_0)}, \qquad (6.12)$$

$$\varphi(z,t=2N\pi) = \frac{\sin(N\pi p_0)}{\sin(\pi p_0)}Re[\int_0^{2\pi} d\tau' \exp[i\{p_0(\tau'-\pi) - \lambda\sin\tau'\}]e^{i(z-p_0\pi N)}]. \qquad (6.13)$$

Substituting $\tau = \tau' - \pi$, in above equation yields *Anger function*,

$$\mathbf{J_{p_0}}(\xi) = \frac{1}{2\pi}\int_{-\pi}^{\pi} d\tau \exp[i\{p_0\tau - \xi\sin\tau\} = \frac{1}{\pi}\int_0^{\pi} d\tau \cos[p_0\tau - \xi\sin\tau], \quad (6.14)$$

and hence the phase

$$\varphi(z,t=2N\pi) = \frac{\sin(N\pi p_0)}{\sin(\pi p_0)}\mathbf{J_{p_0}}(-\lambda)\cos(z - p_0\pi N). \qquad (6.15)$$

Now we can explicitly write the wave function in position space for $t = 2N\pi$. We place Eq. (6.14) in equation (6.8) and exploit the *Jocobi-Anger* identity (4.3), we find the wave function

$$\psi_{p_0}(z, t = 2N\pi) = \frac{1}{\sqrt{2\pi k}} \sum_{m=-\infty}^{\infty} B_m \exp[\frac{i}{k}(p_0 + m k)z] \exp[-\frac{i}{k} p_0^2 N\pi], \quad (6.16)$$

where,

$$b - m = (ie^{-ip_0 N\pi})^n J_m(\frac{\pi \tilde{V}_0}{k} \frac{\sin(N\pi p_0)}{\sin(\pi p_0)} \mathbf{J}_{p_0}(-\lambda)). \quad (6.17)$$

The Fourier transformation

$$\psi_{p_0}(p, t) = \frac{1}{\sqrt{2\pi k}} \int_{\infty}^{\infty} dz \exp[\frac{i}{k} z] \psi_{p_0}(z, t), \quad (6.18)$$

of wave function in Eq. (6.15) represents the wave function in momentum space

$$\psi_{p_0}(p, t) = \frac{1}{\sqrt{2\pi k}} \exp[\frac{i}{k} p_0^2 N\pi] \sum_{m=-\infty}^{infty} b_m(t) \int_{\infty}^{\infty} dz \exp[\frac{i}{k}(p - p_0 - n k)z]. \quad (6.19)$$

Hence the momentum distribution

$$P(p, t) = |\psi_{p_0}(p, t)|^2 = \sum_{m=-\infty}^{\infty} \delta(p - p_0 - m k)|b_m(t)|^2, \quad (6.20)$$

shows that momenta is discrete with probabilities $|b_m(t)|^2$. Eqs. (6.16) and (6.20) give the final analytical expression

$$P(p = p_0 + m k, t = 2\pi N) = J_m^2(\frac{\pi \tilde{V}_0}{k} \frac{\sin(N\pi p_0)}{\sin(\pi p_0)} \mathbf{J}_{p_0}(-\lambda)), \quad (6.21)$$

which gives the momentum distribution of a plane atomic wave in a modulated standing wave field propagating with initial momentum $p_0$ and at time $t = 2N\pi$.

# List of Publications

1. **Muhammad Ayub**, Khalid Naseer, Manzoor Ali and Farhan Saif,
*Atom optics quantum pendulum,*
    J. Rus. Las. Res. **30**, 205 (2009).

2. **Muhammad Ayub**, Khalid Naseer and Farhan Saif,
*Dynamical recurrences based on Floquet theory of nonlinear resonances,*
    Euro Phys. J. D **64**, 491 (2011).

3. **Muhammad Ayub** and Farhan Saif,
*Delicate and robust dynamical recurrences of matter waves in driven optical crystals,*    Phys. Rev. A **85**, 023634, (2012).

4. **Muhammad Ayub** and Farhan Saif,
*Spatio-temporal dynamics of condensate in driven optical lattices,*
    to be submitted.

5. **Muhammad Ayub** and Farhan Saif,
*Quantum chaos with cold atoms in driven optical lattices,*
    submitted.

6. Khalid Naseer, **Muhammad Ayub** and Farhan Saif,
*Atomic bullets: Coherent, non-dispersive, and accelerated matter waves,*
    submitted.

7. **Muhammad Ayub**, Kashif Ammar and Farhan Saif,
*Dynamic localization of Bose-Einstein condensate in optomechanics,*
    submitted.

# Acknowledgements

First and for most, I am thankful to my thesis supervisor Dr. Farhan Saif on many accounts, including his continual encouragement and persistent guidance into this fascinating and encapsulating field of atom optics. I am also greatly indebted to Prof. Dr. Azhar Abbas Rizvi and Dr. Qaiser Abbas Naqvi, Chairman Department of Electronics, for their inspirational teachings, kind and cooperative attitude.

Fruitful discussions with the group fellows, namely, Dr. Rameez, Dr. Shahid, Khalid, Inam, Tasawar, Manzoor, Javed, Jamil, Adnan, Umar, Kashif and Asjad have helped a lot to enrich my conceptual horizons and I submit my sincere gratitude to all of them. I would also like to acknowledge other Ph.D fellows namely, Dr. Ghaffar, Dr. Shakeel, Mustansar, Dr. Ehsan and Arshad for general discussions on research areas other than quantum optics.

I feel strongly indebted to my parents, my family for their love, best wishes, encouragement and unconditional support without which I would have been unable to complete anything worth while.

I am grateful to Higher Education Commission, Government of Pakistan as well for the financial support through Indigious Scholarship Scheme.

**Muhammad Ayub**